\input harvmac
\input rotate
\input epsf
\input xyv2
\def\lroverarrow#1{\raise4.2truept\hbox{$\displaystyle
\leftrightarrow\atop\displaystyle#1$}}


\input psfig
\def\IR{\relax{\rm I\kern-.18em R}}

\def\lp10{l_P^{10}}
\def\lp11{l_P^{11}}
\def\R11{R_{11}}

\let\includefigures=\iftrue
%
%
%
\newfam\black
\input rotate
\input epsf
\input xyv2
\noblackbox
%
%
\includefigures
\message{If you do not have epsf.tex (to include figures),}
\message{change the option at the top of the tex file.}
\def\figin{\epsfcheck\figin}\def\figins{\epsfcheck\figins}
\def\epsfcheck{\ifx\epsfbox\UnDeFiNeD
\message{(NO epsf.tex, FIGURES WILL BE IGNORED)}
\gdef\figin##1{\vskip2in}\gdef\figins##1{\hskip.5in}
\else\message{(FIGURES WILL BE INCLUDED)}%
\gdef\figin##1{##1}\gdef\figins##1{##1}\fi}
\def\DefWarn#1{}
\def\N{{\cal N}}
\def\figinsert{\goodbreak\midinsert}
\def\ifig#1#2#3{\DefWarn#1\xdef#1{fig.~\the\figno}
\writedef{#1\leftbracket fig.\noexpand~\the\figno}%
\figinsert\figin{\centerline{#3}}\medskip\centerline{\vbox{\baselineskip12pt
\advance\hsize by -1truein\noindent\footnotefont{\bf
Fig.~\the\figno:} #2}}
\bigskip\endinsert\global\advance\figno by1}
\else
\def\ifig#1#2#3{\xdef#1{fig.~\the\figno}
\writedef{#1\leftbracket fig.\noexpand~\the\figno}%
\global\advance\figno by1} \fi
\def\yboxit#1#2{\vbox{\hrule height #1 \hbox{\vrule width #1
\vbox{#2}\vrule width #1 }\hrule height #1 }}
\def\fillbox#1{\hbox to #1{\vbox to #1{\vfil}\hfil}}
\def\ybox{{\lower 1.3pt \yboxit{0.4pt}{\fillbox{8pt}}\hskip-0.2pt}}

\def\rightarrowbox#1#2{
  \setbox1=\hbox{\kern#1{${ #2}$}\kern#1}
  \,\vbox{\offinterlineskip\hbox to\wd1{\hfil\copy1\hfil}
    \kern 3pt\hbox to\wd1{\rightarrowfill}}}

\def\half{{1\over 2}}
\def\Tr{{{\rm Tr~ }}}

\def\CA{{\cal A}}

\def\CL{{\cal L}}
\def\CM{{\cal M}}

\def\CO{{\cal O}}

\def\tilde{\widetilde}

\def\II{\relax{I\kern-.10em I}}

\def\pt{pt}
\def\bar{\overline}

\def\IZ{\relax\ifmmode\mathchoice
{\hbox{\cmss Z\kern-.4em Z}}{\hbox{\cmss Z\kern-.4em Z}}
{\lower.9pt\hbox{\cmsss Z\kern-.4em Z}} {\lower1.2pt\hbox{\cmsss
Z\kern-.4em Z}}\else{\cmss Z\kern-.4em Z}\fi}
\def\IB{\relax{\rm I\kern-.18em B}}
\def\IC{{\relax\hbox{$\inbar\kern-.3em{\rm C}$}}}
\def\ID{\relax{\rm I\kern-.18em D}}
\def\IE{\relax{\rm I\kern-.18em E}}
\def\IF{\relax{\rm I\kern-.18em F}}
\def\IG{\relax\hbox{$\inbar\kern-.3em{\rm G}$}}
\def\IGa{\relax\hbox{${\rm I}\kern-.18em\Gamma$}}
\def\IH{\relax{\rm I\kern-.18em H}}
\def\II{\relax{\rm I\kern-.18em I}}
\def\IK{\relax{\rm I\kern-.18em K}}
\def\IN{\relax{\rm I\kern-.18em N}}
\def\IP{\relax{\rm I\kern-.18em P}}

%

\def\inbar{\,\vrule height1.5ex width.4pt depth0pt}

\font\cmss=cmss10 \font\cmsss=cmss10 at 7pt
\def\IR{\relax{\rm I\kern-.18em R}}

\def\underarrow#1{\vbox{\ialign{##\crcr$\hfil\displaystyle
 {#1}\hfil$\crcr\noalign{\kern1pt\nointerlineskip}$\longrightarrow$\crcr}}}
%
\def\hat{\widehat}
\def\tilde{\widetilde}
\overfullrule=0pt
%
\def\tilde{\widetilde}
\def\bar{\overline}

\def\R{{\bf R}}

\font\zfont = cmss10 
\font\litfont = cmr6

\def\bigone{\hbox{1\kern -.23em {\rm l}}}
\def\ZZ{\hbox{\zfont Z\kern-.4emZ}}
\def\half{{\litfont {1 \over 2}}}

\def\CM{{\cal M}}

\Title{hep-th/0312171}
{\vbox{\centerline{Perturbative Gauge Theory   }
\bigskip
\centerline{As A String Theory In Twistor Space   }}}
\smallskip
\centerline{Edward Witten}
\smallskip
\centerline{\it Institute For Advanced Study, Princeton NJ 08540 USA}

\medskip
\input amssym.tex

 \noindent
Perturbative scattering amplitudes in Yang-Mills theory have many
unexpected properties, such as holomorphy of the maximally
helicity violating amplitudes.  To interpret these results, we
Fourier transform the scattering amplitudes from momentum space to
twistor space, and argue that the transformed amplitudes are
supported on certain holomorphic curves.  This in turn is
apparently a consequence of an equivalence between the
perturbative expansion of ${\cal N}=4$ super Yang-Mills theory and
the $D$-instanton expansion of a certain string theory, namely the
topological $B$ model whose target space is the Calabi-Yau
supermanifold $\Bbb{CP}^{3|4}$.

\Date{December, 2003}
\newsec{Introduction}
\nref\pt{S. Parke and T. Taylor, ``An Amplitude For $N$ Gluon Scattering,''
Phys. Rev. Lett. {\bf 56} (1986) 2459.}%
\nref\bg{F. A. Berends and W. T. Giele, ``Recursive Calculations For
Processes With $N$ Gluons,'' Nucl. Phys. {\bf B306} (1988) 759.}%
\nref\texas{B. DeWitt, ``Quantum Theory Of Gravity, III:
Applications Of The Covariant Theory,''
Phys. Rev. {\bf 162} (1967) 1239.}%
\nref\bdkold{Z. Bern, L. Dixon, and D. Kosower, ``Progress In One-Loop
QCD Computations,'' hep-ph/9602280, Ann. Rev. Nucl. Part. Sci. {\bf 46} (1996)
109.  }%
\nref\schwarz{L. Brink, J. Scherk, and  J. H. Schwarz,
``Supersymmetric Yang-Mills Theories,'' Nucl. Phys. {\bf B121}
(1977) 77.}%
\nref\bdknew{C. Anastasiou, Z. Bern, L. Dixon, and D. Kosower,
``Planar Amplitudes In Maximally Supersymmetric Yang-Mills
Theory,'' hep-th/0309040; Z. Bern, A. De Freitas, and L. Dixon,
``Two Loop Helicity Amplitudes For Quark Gluon Scattering In QCD
And Gluino Gluon Scattering In Supersymmetric
Yang-Mills Theory''JHEP 0306:028 (2003), hep-ph/0304168.}%
\nref\penrose{R. Penrose, ``Twistor Algebra,'' J. Math. Phys. {\bf
8} (1967) 345.}%
\nref\pentwo{R. Penrose, ``The Central Programme Of Twistor
Theory,'' Chaos, Solitons, and Fractals
{\bf 10} (1999) 581.}%
\nref\ferber{A. Ferber,
``Supertwistors And Conformal Supersymmetry,'' Nucl. Phys. {\bf B132}
(1978) 55.}%
\nref\witten{E. Witten, ``An Interpretation Of Classical Yang-Mills Theory,''
Phys. Lett. {\bf B77} (1978) 394.}%
\nref\iyg{J. Isenberg, P. B. Yasskin, and P. S. Green,
``Nonselfdual Gauge Fields,'' Phys. Lett. {\bf B78} (1978) 464.}%
\nref\nair{V. Nair,
``A Current Algebra For Some Gauge Theory Amplitudes,'' Phys. Lett.
{\bf B214} (1988) 215.}%
\nref\malda{J. Maldacena, ``The Large $N$ Limit Of Superconformal Field Theories And
Supergravity,'' Adv. Theor. Math. Phys. {\bf 2} (1998) 231, hep-th/9711200.}%
\nref\thooft{G. 't Hooft,
``A Planar Diagram Theory For Strong Interactions,'' Nucl. Phys.
{\bf B72} (1974) 461.}%
\nref\mcpen{M. A. H. MacCallum and R. Penrose, ``Twistor Theory:
An Approach
To The Quantization Of Fields And Space-Time,'' Phys. Rept. {\bf 6C} (1972) 241.}%
\nref\hodges{A. P. Hodges and S. Huggett, ``Twistor Diagrams,''
{\it Surveys In High Energy Physics} {\bf 1} (1980) 333; A.
Hodges, ``Twistor Diagrams,'' Physica {\bf 114A} (1982) 157,
``Twistor Diagrams,'' in {\it The Geometric Universe: Science,
Geometry, and The Work Of Roger Penrose},
eds. S. A. Huggett et. al. (Oxford University Press, 1998).}%
\nref\whodi{J. D. Bjorken and M. C. Chen, ``High Energy Trident
Production With Definite Helicities,'' Phys. Rev. {\bf 154}
(1966) 1335.}%
\nref\whodio{G. Reading Henry, ``Trident Production With Nuclear Targets,''
Phys. Rev. {\bf 154} (1967) 1534.}%
 \nref\whod{F. A. Berends, R. Kleiss, P. De Causmaecker,
R. Gastmans, and T. T. Wu, ``Single Bremsstrahlung Processes In
Gauge Theory,'' Phys. Lett. {\bf 103B} (1981) 124; P. De
Causmaecker, R. Gastmans, W. Troost, and T. T. Wu, ``Helicity
Amplitudes For Massless QED,'' Phys. Lett. {\bf B105} (1981) 215,
Nucl. Phys. {\bf B206} (1982) 53; F. A. Berends, R. Kleiss, P. De
Causmaecker, R. Gastmans, W. Troost, and T. T. Wu, ``Multiple
Bremsstrahlung In Gauge Theories At High-Energies. 2. Single
Bremsstrahlung,'' Nucl. Phys. {\bf B206} (1982) 61.}%
\nref\whone{R. Kleis and W. J. Stirling, ``Spinor Techniques
For Calculating Proton Anti-Proton To $W$ Or $Z$ Plus Jets,''
Nucl. Phys. {\bf B262} (1985) 235.}%
\nref\whonext{J. F. Gunion and Z. Kunzst, ``Improved Analytic
Techniques For Tree Graph Calculations And The ${\rm gg}q\bar q
l\bar l$ Process,''
Phys. Lett. {\bf 161B} (1985) 333.}%
\nref\whoever{F. A. Berends and W-T. Giele, ``The Six Gluon Process As
An Example Of Weyl-van der Waerden Spinor Calculus,''
Nucl. Phys. {\bf B294} (1987) 700.}%
\nref\whot{Z. Xu, D.-H. Zhang, and L. Chang, ``Helicity Amplitudes For
Multiple Bremsstrahlung In Massless Nonabelian Gauge Theories,''
Nucl. Phys. {\bf B291} (1987) 392.}%
\nref\whotwo{G. Chalmers and W. Siegel, ``Simplifying Algebra In
Feynman Graphs, Part I: Spinors,'' Phys. Rev. {\bf D59} (1999)
045012, hep-ph/9708251, ``Part II: Spinor Helicity From The Light
Cone,'' Phys. Rev. {\bf D59} (1999) 045013, hep-ph/9801220, ``Part
III: Massive Vectors,''  Phys. Rev. {\bf D63} (2001) 125027,
hep-th/0101025.}%
\nref\parkrev{M. L. Mangano and S. J. Parke, ``Multiparton Amplitudes In Gauge
Theories,'' Phys. Rept. {\bf 200} (1991) 301.}%
\nref\dixon{L. Dixon, ``Calculating Scattering Amplitudes Efficiently,''
TASI Lectures, 1995, hep-ph/9601359.}%
\nref\lwitten{L. Witten, ``Invariants Of General Relativity And
The Classification Of Spaces,'' Phys. Rev. (2) {\bf 113} (1959)
357.}%
 \nref\penrind{R. Penrose and W. Rindler, {\it
Spinors And Space-Time: Volume 1, Two-Spinor Calculus and
Relativistic Fields}, {\it Volume 2, Spinor And Twistor Methods In
Spacetime Geometry} (Cambridge University Press, 1986). }

 The perturbative
expansion of Yang-Mills theory has remarkable properties that are
not evident upon inspecting the Feynman rules. For example, the
tree level scattering amplitudes that are maximally helicity
violating (MHV) can be expressed in terms of a simple holomorphic
or antiholomorphic function.  This was first conjectured by Parke
and Taylor based on computations in the first few cases \pt; the
general case was proved by Berends and Giele \bg.  (Unexpected
simplicity and selection rules in Yang-Mills and gravitational
helicity amplitudes were first found, as far as I know, by DeWitt
\texas\ for four particle amplitudes.) These unexpected
simplifications have echoes, in many cases, in loop amplitudes,
especially in the supersymmetric case. For a sampling of one-loop
results, see the review \bdkold, and for some recent two-loop
results for the theory with maximal or ${\cal N}=4$ supersymmetry
(which was first constructed in \schwarz), see \bdknew.  ${\cal
N}=4$ super Yang-Mills theory is an important test case for
perturbative gauge theory, since it is the simplest case and, for
example, has the same gluonic tree amplitudes as pure Yang-Mills
theory.

In the present paper, we will offer a new perspective on
explaining these results.  We will study what happens when the
usual momentum space scattering amplitudes are Fourier transformed
to Penrose's twistor space \penrose. We argue that the
perturbative amplitudes in twistor space are supported on certain
holomorphic curves.  Results such as the holomorphy of the
tree-level MHV amplitudes, as well as more complicated (and novel)
differential equations obeyed by higher order
amplitudes, are direct consequences of this.

We interpret these results to mean that the perturbative expansion
of ${\cal N}=4$ super Yang-Mills theory with $U(N)$ gauge group is
equivalent to the instanton expansion of a certain string theory.
The instantons in question are $D$-instantons rather than ordinary
worldsheet instantons.  The string theory is the topological $B$
model whose target space is the Calabi-Yau supermanifold
$\Bbb{CP}^{3|4}$. This is the supersymmetric version of twistor
space, as defined \ferber\ and exploited \witten\ long
ago.\foot{See \iyg\ for the bosonic version of the supersymmetric
construction found in \witten.} From
the string theory, we recover the tree level MHV amplitudes of
gauge theory. They arise from a one-instanton computation that
leads to a formalism similar to that suggested by Nair \nair.

This representation of {\it weakly coupled} ${\cal N}=4$
super-Yang-Mills theory as a string theory is an interesting
counterpoint to the by now familiar description of the {\it
strongly coupled} regime of the same theory via Type IIB
superstring theory on $\rm {AdS}_5\times \Bbb  S^5$ \malda.
However, many aspects of the $B$ model of $\Bbb{CP}^{3|4}$  remain
unclear. One pressing question is to understand the closed string
sector. The closed strings may possibly give some version of
${\cal N}=4$ conformal supergravity, in which case the string
theory considered here is equivalent to super Yang-Mills theory
only in the planar limit (large $N$ with fixed $g^2N$ \thooft), in
which the closed strings decouple.  (If so, as conformal
supergravity is generally assumed to have negative energies and
ghosts, the $B$ model of ${\Bbb{CP}}^{3|4}$ may be physically
sensible only in the planar limit.)

In twistor theory, it has been a longstanding problem
\refs{\mcpen,\hodges} to understand how to use twistors to
describe perturbative field theory amplitudes. Our proposal in the
present paper differs from previous attempts mainly in that we
consider families of holomorphic curves in twistor space, and not
just products of twistor spaces.

\nref\revminus{R. S. Ward and R. O'Neil Jr. Wells,
{\it Twistor Geometry And Field Theory} (Cambridge University
Press, 1991).}%
\nref\revzero{L. P. Hughston, {\it Twistors and Particles}, Lecture Notes
in Physics {\bf 97} (Springer-Verlag, Berlin, 1989).}%
\nref\revone{T. N. Bailey and R. J. Baston, eds.,
{\it Twistors In Mathematics And Physics}, London Mathematical
Society Lecture Notes Series {\bf 156} (1990).}%
\nref\revtwo{S. A. Huggett and K. P. Tod, {\it An Introduction To Twistor Theory},
London Mathematical Society Student Texts {\bf 4}.}%
\nref\nongrav{R. Penrose, ``The Nonlinear Graviton,''  Gen. Rel.
Grav. {\bf 7} (1976) 171.}%
\nref\ward{R. Ward, ``On Self-Dual Gauge Fields,'' Phys. Lett.
{\bf
61A} (1977) 81.}%
 \nref\hetal{M. F. Atiyah, N. J.
Hitchin, and I. M. Singer, ``Self-Duality In Four-Dimensional
Riemannian Geometry,''  Proc. Roy. Soc. London Ser. A  {\bf 362}
(1978) 425.}%
 \nref\atiyah{M. F. Atiyah, {\it Geometry Of
Yang-Mills Fields}, Lezioni Fermiane (Academia Nazionale dei
Lincei and Scuola Normale
Superiore, Pisa, 1979).}%
In section 2, we review Yang-Mills helicity amplitudes and their
description via spinors \refs{\whodi-\whotwo}  and the MHV
amplitudes.  For expositions of this material, see \refs{\parkrev,
\dixon}.  For the use of  spinors in relativity, see
\refs{\lwitten,\penrind}. Then we describe the Fourier transform
to twistor space. For general reviews of twistor theory, see
\refs{\pentwo,\mcpen,\revminus-\revtwo}; for the twistor transform
of the self-dual Einstein and Yang-Mills equations, see
respectively \nongrav\ and \ward, as well as \refs{\hetal,\atiyah}
for the Euclidean signature case. In section 3, we investigate the
behavior of the perturbative Yang-Mills amplitudes in twistor
space.  We demonstrate (in various examples of tree level and
one-loop amplitudes) that they are supported on certain curves in
twistor space, by showing that in momentum space they obey certain
differential equations that generalize the holomorphy of the MHV
amplitudes. In section 4, we argue that the topological $B$ model
of super twistor space $\Bbb{CP}^{3|4}$ gives a natural origin for
these results.

The analysis in section 3 and the proposal in section 4 are
tentative.  They represent the best way that has emerged so far to
organize and interpret the facts, but much more needs to be
understood.

We are left with numerous questions. For example, are there
analogous descriptions of perturbative expansions for other field
theories with less supersymmetry, at least in their planar limits?
What about the ${\cal N}=4$ theory on its Coulomb branch, where
conformal invariance is spontaneously broken?  Can its
perturbative expansion be described by the instanton expansion of
the string theory we consider in this paper, expanded around a
shifted vacuum? In section 3, we also observe that the tree level
MHV amplitudes of General Relativity are supported on curves in
twistor space. Is this a hint that the perturbative expansion of
${\cal N}=8$ supergravity can be described by some string theory?
What would be the target space of such a string theory and would
its existence imply finiteness of ${\cal N}=8$ supergravity?

How do the usual infrared divergences of gauge perturbation theory
arise from the twistor point of view? In the one-loop example that
we consider, the twistor amplitude is finite; the infrared
divergence arises from the Fourier transform back to momentum
space.  Is this  general?  In ${\cal N}=4$ super Yang-Mills
theory, the planar loop amplitudes are known to be simpler than
the non-planar ones. Does including the closed string sector of
the string theory give a model in which the simplicity persists
for non-planar diagrams? Finally, many technical problems need to
be addressed in order to properly define the string theory
amplitudes  and  facilitate their computation.

\newsec{Helicity Amplitudes And Twistor Space}

\subsec{Spinors}

Before considering scattering amplitudes, we will review some
kinematics in four dimensions. We start out in signature $+--\,-$,
but we sometimes generalize to other signatures.  Indeed,  this
paper is only concerned with perturbation theory, for which the
signature is largely irrelevant as the scattering amplitudes are
holomorphic functions of the kinematic variables.  Some things
will be simpler with other signatures or for complex momenta with
no signature specified.

First we recall that the Lorentz group in four dimensions, upon
complexification, is locally isomorphic to $SL(2)\times SL(2)$,
and thus the finite-dimensional representations are classified as
$(p,q)$, where $p$ and $q$ are integers or half-integers.  The
negative and positive chirality spinors transform in the $(1/2,0)$
and $(0,1/2)$ representations, respectively.  We write generically
$\lambda_a$, $a=1,2$, for a spinor transforming as $(1/2,0)$, and
$\tilde \lambda_{\dot a}$, $\dot a=1,2$, for a spinor transforming
as $(0,1/2)$.

Spinor indices of type $(1/2,0)$ are raised and lowered with the
antisymmetric tensor $\epsilon_{ab}$ and its inverse
$\epsilon^{ab}$ (obeying $\epsilon^{ab}\epsilon_{bc}=\delta^a_c$):
$\lambda_a=\epsilon_{ab}\lambda^b$,
$\lambda^b=\epsilon^{bc}\lambda_c$. Given two spinors $\lambda_1,$
$\lambda_2$ both of positive chirality, we can form the Lorentz
invariant $\langle\lambda_1,\lambda_2\rangle
=\epsilon_{ab}\lambda_1^a\lambda_2^b$. From the definitions, it
follows that $\langle \lambda_1,\lambda_2\rangle = -\langle
\lambda_2,\lambda_1\rangle =-
\epsilon^{ab}\lambda_{1\,a}\lambda_{2\,b}$.

Similarly, we raise and lower indices of type $(0,1/2)$ with the
antisymmetric tensor $\epsilon_{\dot a \dot b}$ and its inverse
$\epsilon^{\dot a\dot b}$, again imposing $\epsilon^{\dot a\dot
b}\epsilon_{\dot b\dot c}=\delta^{\dot a}_{\dot b}$.  For two
spinors $\tilde\lambda_1$, $\tilde\lambda_2$ both of negative
chirality, we define $[\tilde\lambda_1,\tilde
\lambda_2]=\epsilon_{\dot a\dot b}\tilde\lambda_1^{\dot a}\tilde
\lambda_2^{\dot b}=-[\tilde\lambda_2,\tilde\lambda_1]$.

The vector representation of $SO(3,1)$ is the $(1/2,1/2)$
representation.  Thus, a momentum vector $p_\mu$, $\mu=0,\dots
,3$, can be represented as a ``bi-spinor'' $p_{a\dot a}$ with one
spinor index $a$ or $\dot a$ of each chirality.  The explicit
mapping  from $p_\mu$ to $p_{a \dot a}$ can be made using the
chiral part of the Dirac matrices.  With signature $+---$, one can
take the Dirac matrices to be \eqn\ymo{\gamma^\mu=\left(\matrix{0
& \sigma^\mu\cr \overline\sigma^\mu & 0 }\right),} where we take
$\sigma^\mu=(1,\vec\sigma)$, $\overline
\sigma^\mu=(-1,\vec\sigma)$, with $\vec\sigma$ being the $2\times
2$ Pauli spin matrices. In particular, the upper right hand block
of $\gamma^\mu$ is a $2\times 2$ matrix $\sigma^\mu_{a\dot a}$
that maps spinors of one chirality to the other.  For any spinor
$p_\mu$, define \eqn\wedef{p_{a\dot a}=\sigma^\mu_{a\dot a}p_\mu.}
Thus, with the above representation of $\sigma^\mu$, we have
$p_{a\dot a}=p_0+\vec \sigma\cdot \vec p$ (where $p_0$ and $\vec
p$ are the ``time'' and ``space'' parts of $p^\mu$), from which it
follows that \eqn\tyup{p_\mu p^\mu=\det(p_{a\dot a}).} Thus a
vector $p^\mu$ is lightlike if and only if the corresponding
matrix $p_{a\dot a}$ has determinant zero.

Any $2\times 2$ matrix $p_{a\dot a}$ has rank at most two, so it
can be written $p_{a\dot a}=\lambda_a\tilde\lambda_{\dot
a}+\mu_a\tilde \mu_{\dot a}$ for some spinors $\lambda,\mu$ and
$\tilde\lambda,\tilde \mu$.  The rank of a $2\times 2$ matrix is
less than two if and only if its determinant vanishes.  So the
lightlike vectors $p^\mu$ are precisely those for which
\eqn\tofo{p_{a\dot a}=\lambda_a\tilde\lambda_{\dot a},} for some
spinors $\lambda_a$ and $\tilde\lambda_{\dot a}$.

If we wish $p_{a \dot a}$ to be real with Lorentz signature, we
must take $\tilde\lambda=\pm \overline\lambda$ (where $\overline
\lambda$ is the complex conjugate of $\lambda$).  The sign
determines whether $p^\mu$ has positive energy or negative energy.

It will also be convenient to consider other signatures. In
signature $++-\,-$, $\lambda$ and $\tilde \lambda$ are
independent, real, two-component objects.  Indeed, with signature
$++-\,-$, the Lorentz group $SO(2,2)$ is, without any
complexification, locally isomorphic to $SL(2,\Bbb R)\times
SL(2,\Bbb R)$, so the spinor representations are real.   With
Euclidean signature $+++\,+$, the Lorentz group is locally
isomorphic to $SU(2)\times SU(2)$; the spinor representations are
pseudoreal.  A lightlike vector cannot be real with Euclidean
signature.

Obviously, if $\lambda$ and $\tilde \lambda$ are given, a
corresponding lightlike vector $p$ is determined, via \tofo. It is
equally clear that if a lightlike vector $p$ is given, this does
not suffice to determine $\lambda $ and $\tilde \lambda$.  They
can be determined only modulo the scaling \eqn\tofog{\lambda \to
u\lambda,~\tilde\lambda \to u^{-1} \tilde\lambda} for $u\in \Bbb
C^*$, that is, $u$ is a nonzero complex number.  (In signature
$+--\,-$, if $p$ is real, we can restrict to $|u|=1$. In signature
$++--$, if $\lambda$ and $\tilde\lambda$ are real, we can restrict
to real $u$.) Not only is there no natural way to determine
$\lambda$ as a function of $p$; there is in fact no continuous way
to do so, as there is a topological obstruction to this. Consider,
for example, massless particles of unit energy; the
energy-momentum of such a particle is specified by the momentum
three-vector $\vec p$, a unit vector which determines a point in
$\Bbb S^2$.  Once $\vec p$ is given, the space of possible
$\lambda$'s is a non-trivial complex line bundle over $\Bbb S^2$
that is known as the Hopf line bundle; non-triviality of this
bundle means that one cannot pick $\lambda$ as a continuously
varying function of $\vec p$.

Once $p$ is given, the additional information that is involved in
specifying $\lambda$ (and hence $\tilde\lambda$)   is equivalent
to a choice of wavefunction for a spin one-half particle with
momentum vector $p$.  In fact, the chiral Dirac equation for a
spinor $\psi^a$ is \eqn\noy{i\sigma^\mu_{a\dot
a}{\partial\psi^a\over\partial x^\mu}=0.}  A plane wave
$\psi^a=\lambda^a\exp(ip\cdot x)$ (with constant $\lambda^a$)
obeys this equation if and only if $p_{a\dot a}\lambda^a=0$.  This
is so if and only if $p_{a\dot a}$ can be written as
$\lambda_a\tilde\lambda_{\dot a}$ for some $\tilde\lambda$.

The formula $p\cdot p =\det (p_{a\dot
a})=\epsilon^{ab}\epsilon^{\dot a\dot b} p_{a\dot a}p_{b\dot b}$
generalizes for any two vectors $p$ and $q$ to $p\cdot
q=\epsilon^{ab}\epsilon^{\dot a\dot b}p_{a\dot a}q_{b\dot b}$.
Hence if $p$ and $q$ are lightlike vectors, which we write in the
form $p_{a\dot a}=\lambda_a\tilde\lambda_{\dot a}$ and $q_{b\dot
b}=\mu_b\tilde \mu_{\dot b}$, then we have \eqn\vuvro{p\cdot
q=\langle \lambda,\mu\rangle [\tilde\lambda,\tilde\mu].}

\subsec{Helicity Amplitudes}

Now we consider scattering amplitudes of massless particles in four dimensions.
We consider $n$ massless particles with momentum vectors $p_1,p_2,\dots,p_n$.
For scattering of scalar particles, the initial and final states are completely
fixed by specifying the momenta.  The scattering amplitude is, therefore, a
function of the $p_i$.  For example, for $n=4$, the only independent
Lorentz invariants
are the usual Mandelstam variables $s=(p_1+p_2)^2$ and $t=(p_1+p_3)^2$, and
the scattering amplitude (being dimensionless) is a function of the ratio $s/t$.

For particles with spin, the scattering amplitude is not merely a
function of the momenta.  For example, in the case of massless
particles of spin one -- the main case that we will consider in
detail in the present paper -- in the conventional description, to
each external particle is associated not just a momentum vector
$p_i^\mu$ but also a polarization vector
$\epsilon_i^\mu$.\foot{Hopefully, the use of $\epsilon_\mu$ or
$\epsilon_{a\dot a}$ for a polarization vector, while the
Levi-Civita tensors for the spinors are called $\epsilon_{ab}$ and
$\epsilon_{\dot a\dot b}$, will not cause confusion. Polarization
vectors only appear in the present subsection.} The polarization
vector obeys the constraint $\epsilon_i\cdot p_i=0$, and is
subject to the gauge invariance \eqn\hurfo{\epsilon_i\to
\epsilon_i+w p_i,} for any constant $w$.

The scattering amplitude is most often introduced in textbooks as
a function of the $p_i$ and $\epsilon_i$, subject to this
constraint and gauge invariance.  However, in four dimensions, it
is more useful to label external gauge bosons by their helicity,
$+1$ or $-1$ or simply $+$ or $-$.\foot{In labeling helicities, we
consider all particles to be outgoing.  In crossing symmetry, an
incoming particle of one helicity is equivalent to an outgoing
particle of the opposite helicity.} If a choice of momentum vector
$p_i$ and helicity $+$ or $-$ enabled us to pick for each particle
a polarization vector, then the scattering amplitude of gauge
bosons would  depend only on the momenta and the choices $\pm$ of
helicities.

However, given a lightlike momentum vector $p$ and a choice of
helicity, there is no natural way to pick a polarization vector
with that helicity.  (There is not even any continuous way to pick
a polarization vector as a function of the momentum; in trying to
do so, one runs into a non-trivial complex line bundle which is
the square of the Hopf bundle.)  Suppose though that instead of
being given only a lightlike vector $p_{a\dot a}$ one is given a
$\lambda$, that is a decomposition $p_{a\dot
a}=\lambda_a\tilde\lambda_{\dot a}$.  Then \refs{\whoever,\whotwo}
 we {\it do} have enough information to determine a polarization
vector, up to a gauge transformation. To get a negative helicity
polarization vector, we pick any positive helicity spinor $\tilde
\mu_{\dot a}$ that is not a multiple of $\tilde\lambda$ and set
\eqn\norskip{\epsilon_{a\dot a}={\lambda_a\tilde\mu_{\dot a}\over
[\tilde\lambda,\tilde \mu]}.} This obeys the constraint
$0=\epsilon_\mu p^\mu=\epsilon_{a\dot a}p^{a\dot a}$, since
$\langle \lambda,\lambda\rangle=0$.  It also is independent of the
choice of $\tilde \mu$ up to a gauge transformation.  To see this,
note that since the space of possible $\tilde\mu$'s is
two-dimensional, any variation of $\tilde\mu$ is of the form
\eqn\orskip{\tilde\mu\to \tilde\mu + \eta \tilde\mu +
\eta'\tilde\lambda,} with some complex parameters $\eta$, $\eta'$.
The $\eta$ term drops out of \norskip, since $\epsilon_{a\dot a}$
is invariant under rescaling of $\tilde\mu$; the $\eta'$ term
changes $\epsilon_{a\dot a}$ by a gauge transformation, a multiple
of $\lambda_a\tilde\lambda_{\dot a}$.

Under $\lambda\to u\lambda$, $\tilde\lambda\to
u^{-1}\tilde\lambda$, $\epsilon_{a\dot a}$ has the same scaling as
$\lambda^2$. This might have been anticipated: since $\lambda$
carries helicity $-1/2$ (as we saw above in discussing the Dirac
equation), a helicity $-1$ polarization vector should scale as
$\lambda^2$.

To determine more directly the helicity of a massless particle
whose polarization vector is $\epsilon_{a\dot a}$, we construct
the field strength $F_{\mu\nu}=\partial_\mu A_\nu -\partial_\nu
A_\mu=-i(p_\mu\epsilon_\nu -p_\nu\epsilon_\mu)$ and verify that it
is selfdual or anti-selfdual. In terms of spinors, the field
strength is $F_{ab\dot a \dot b}=-F_{ba\dot b\dot a}$ and can be
expanded $F_{ab\dot a \dot b}=\epsilon_{ab}\tilde f_{\dot a\dot
b}+f_{ab} \epsilon_{\dot a\dot b}$, where $f$ and $\tilde f$ are
the selfdual and anti-selfdual parts of $F$.  With $p_{a\dot
a}=\lambda_a\tilde\lambda_{\dot a}$ and $\epsilon_{b\dot b}$
defined as above, we find that $f_{ab}\sim\lambda_a\lambda_b$ and
$\tilde f_{\dot a\dot b}=0$. So $F$ is selfdual and the photon has negative helicity.
\foot{In the literature on perturbative QCD, it is conventional that the MHV
amplitude, introduced presently, that is a function of $\lambda$ describes mostly
$+$ helicity scattering.  In the literature on twistor theory
-- at least in the mathematical branch of that literature -- it is conventional
that an instanton
is an anti-selfdual gauge field in spacetime and corresponds to a holomorphic
vector bundle over twistor space.  We will try to follows these two conventions.}

We can similarly make a polarization vector of positive helicity,
introducing an arbitrary negative chirality spinor $\mu_a$ that is
not a multiple of $\lambda_a$ and setting
\eqn\hoffo{\tilde\epsilon_{a\dot a}={\mu_a\tilde\lambda_{\dot
a}\over \langle \mu,\lambda\rangle}.} As one would expect from the
above discussion, under $\lambda\to u\lambda$, $\tilde\epsilon$
has the same scaling as $\lambda^{-2}$.

Although a scattering amplitude of massless gauge bosons cannot be
regarded as a function of the momenta $p_i$, it can be regarded as
a function of the spinors $\lambda_i$ and $\tilde\lambda_i$, as
well as the helicity labels $h_i=\pm 1$, since as we have just
seen this data determines the polarization vectors $\epsilon_i$ up
to a gauge transformation. Thus, instead of writing the amplitude
as $\widehat A(p_i,\epsilon_i)$, where $\epsilon_i$ are the
polarization vectors, we write it as $\widehat
A(\lambda_i,\tilde\lambda_i,h_i)$.
  When formulated in this way, the amplitude
obeys for each $i$ an auxiliary condition
\eqn\ruffo{\left(\lambda^a_i{\partial\over\partial\lambda^a_i}
-\tilde\lambda_i^{\dot a}{\partial\over\partial\tilde
\lambda_i^{\dot a}}\right) \widehat
A(\lambda_i,\tilde\lambda_i,h_i)=-2h_i\widehat
A(\lambda_i,\tilde\lambda_i,h_i),} which reflects the scaling with
$\lambda$ of the polarization vectors. This equation holds for
helicity amplitudes for massless particles of any spin.

The scattering amplitude is of course proportional to a delta
function of energy-momentum conservation, $(2\pi)^4\delta^4(\sum_i
p_i)$, or in terms of spinors
$(2\pi^4)\delta^4(\sum_i\lambda_i^a\tilde\lambda_i^{\dot a})$. The
general form of the scattering amplitude is thus
\eqn\grufo{\widehat A(\lambda_i,\tilde\lambda_i,h_i)=i(2\pi)^4
\delta^4\left(\sum_i\lambda_i^a\tilde\lambda_i^{\dot a}\right)
A(\lambda_i,\tilde\lambda_i,h_i),} where the reduced amplitude $A$
obeys the same equation \ruffo\ as $\widehat A$.  (We often write
$\widehat A$ and $A$ as functions just of $\lambda_i$ and $\tilde\lambda_i$,
with the $h_i$ understood.)

\subsec{Maximally Helicity Violating Amplitudes}

To make this discussion tangible, let us consider the tree level
scattering of $n$ gluons in the simplest configuration.  The
scattering amplitude with $n$ outgoing gluons all of the same
helicity vanishes, as does (for $n>3$) the amplitude with $n-1$
outgoing gluons of one helicity and one of the opposite
helicity.\foot{The amplitude with $n=3$ is exceptional and is
often omitted, but will be discussed in section 3.2.} The
``maximally helicity violating'' or MHV amplitude is the case with
$n-2$ gluons of one helicity and $2$ of the opposite helicity. To
understand the name ``maximally helicity violating,'' recall that
in labeling the helicities, we consider all gluons to be outgoing.
So after allowing for crossing symmetry, the MHV amplitude
describes, for example, a process in which all incoming gluons
have one helicity and all but two outgoing gluons -- the maximal
possible number -- have the opposite helicity.

For $n=4$, the only nonzero tree level amplitude is the MHV
amplitude with helicities some permutation of $++--$, and
similarly for $n=5$, the nonzero amplitudes are MHV amplitudes
such as $++---$ or $+++--$.  These amplitudes dominate two-jet and
three-jet production in hadron colliders at very high energies,
and so are of phenomenological importance.  The lowest order
non-MHV tree level amplitudes are the $n=6$ amplitudes such as
$+++--\,-$.  They enter, for example, in four-jet production at
hadron colliders.

The actual form of the tree-level MHV amplitudes (conjectured by
Parke and Taylor based on results for small $n$ \pt, and proved by
Berends and Giele \bg) is quite remarkable. The reduced amplitude
$A$ can be written as a function only of the $\lambda_i$ or only
of the $\tilde\lambda_i$, depending on whether the outgoing
helicities are almost all $+$ or almost all $-$.  For real momenta
in Minkowski signature, one has $\tilde\lambda_i=\pm\overline
\lambda_i$, and then the MHV amplitudes are holomorphic or
antiholomorphic functions of the $\lambda_i$, depending on whether
the  helicities are mostly $+$ or mostly $-$.

To describe the results more precisely, we take
 the gauge group to be $U(N)$ (for some sufficiently large $N$ as to avoid
accidental equivalences of any traces that we
might encounter). We recall that tree level diagrams in Yang-Mills
theory are planar, and generate a single-trace interaction
\thooft.  In such a planar diagram, the $n$ gauge
bosons are attached to the index loop in a definite cyclic order,
as indicated in figure 1.  If we number the gauge bosons so that
the cyclic order is simply $1,2,3,\dots,n$, then the amplitude
includes a group theory factor $I=\Tr \,T_1T_2\dots T_n$.  It
suffices to study the amplitude with one given cyclic order; the
full amplitude is obtained from this by summing over the possible
cyclic orders, to achieve Bose symmetry.  Gluon scattering
amplitudes considered in this paper are always proportional to the
group theory trace $I$, and this factor is omitted in writing the
formulas.

\ifig\tramp{$n$ external gluons cyclically attached to the
boundary of a disc, representing the group theory structure of a
Yang-Mills tree diagram.} {\epsfbox{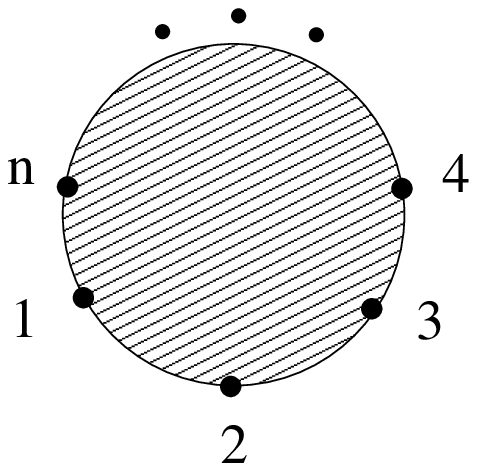}}

 Suppose that gauge bosons
$r$ and $s$ ($1\leq r<s\leq n$) have negative helicity and the
others have positive helicity.  The reduced tree level amplitude
for this process (with the energy-momentum delta function and the
trace $I$ both omitted) is \eqn\hutu{A=g^{n-2}{\langle
\lambda_r,\lambda_s\rangle^4\over \prod_{i=1}^n\langle
\lambda_i,\lambda_{i+1}\rangle}.} (Here $g$ is the gauge coupling
constant, and we
 set $\lambda_{n+1}=\lambda_1$.)  Note that this amplitude has the requisite
homogeneity in each variable.  It is homogeneous of degree $-2$ in
each $\lambda_i$ with $i\not= r,s$, since each $\lambda_i$ appears
twice in the denominator in \hutu.  But for $i=r,s$, it is
homogeneous in $\lambda_i$ with degree $+2$, since in these cases
the numerator is homogeneous of degree four in $\lambda_i$. If
gauge bosons $r$ and $s$ have positive helicity and the others
have negative helicity, the amplitude is instead
\eqn\huttu{A=g^{n-2}{[\tilde \lambda_r,\tilde\lambda_s]^4\over
\prod_{i=1}^n[ \tilde\lambda_i,\tilde\lambda_{i+1}]}.}

In order to write such formulas briefly, one often abbreviates
$\langle \lambda_i,\lambda_j\rangle$ as $\langle i\,\,j\rangle$,
and similarly one write $[\tilde\lambda_i,\tilde\lambda_j]$ as
$[i\,\,j]$. In this way, \hutu\ would be written as
\eqn\hhutu{A=g^{n-2} {\langle
r\,\,s\rangle^4\over\prod_{i=1}^n\langle i\,\, i+1\rangle}.}

Specializing to the case of $n=4$, we may appear to have a
contradiction, since for example the amplitude in which the
helicities in cyclic order are $--++$ is a special case of each
of these constructions.  Via \hutu, we expect \eqn\yutu{A={\langle
\lambda_1,\lambda_2\rangle^4\over \prod_{i=1}^4\langle
\lambda_i,\lambda_{i+1}\rangle},} but via \huttu, we expect
\eqn\yuttu{A={[\tilde \lambda_3,\tilde\lambda_4]^4\over
\prod_{i=1}^4 [\tilde\lambda_i,\tilde\lambda_{i+1}]}.}  How can
the same function be both holomorphic and antiholomorphic? The
resolution is instructive.  From momentum conservation,
$\sum_i\lambda_i^a \tilde\lambda_i^{\dot a}=0$, it follows, upon
taking inner products with $\lambda_y$ and $\tilde \lambda_z$,
that for any $y$ and $z$ we have $\sum_i\langle
\lambda_y,\lambda_i\rangle [\tilde \lambda_i,\tilde\lambda_z]=0$.
Setting, for example, $y=1$, $z=2$, this leads to
\eqn\guyy{{\langle\lambda_1,\lambda_3\rangle\over
\langle\lambda_1,\lambda_4\rangle}
=-{[\tilde\lambda_4,\tilde\lambda_2]
\over[\tilde\lambda_3,\tilde\lambda_2]}.} By repeated use of such
identities, one can show that the two formulas for $A$ are
equivalent, when multiplied by a delta function of energy-momentum
conservation.

\subsec{Conformal Invariance}

A useful next step is to verify the conformal invariance of the
MHV amplitudes.  First we write down the conformal generators in
terms of the $\lambda$ and $\tilde\lambda$ variables.  We consider
conformal generators for a single massless particle; the
corresponding generators for the full $n$-particle system are
obtained simply by summing over the $n$ particles.

Some of the conformal generators are obvious.  For example, the
Lorentz generators are \eqn\tomgo{\eqalign{J_{ab}& = {i\over
2}\left(\lambda_a{\partial\over
\partial\lambda^b}+\lambda_b{\partial\over
\partial\lambda^a}\right)\cr
\tilde J_{\dot a\dot b}& = {i\over 2}\left(\tilde\lambda_{\dot
a}{\partial\over
\partial\tilde \lambda^{\dot b}}+\tilde\lambda_{\dot b}{\partial\over
\partial\tilde\lambda^{\dot a}}\right).\cr}}
The momentum operator is a multiplication operator,
\eqn\trufy{P_{a\dot a}=\lambda_a\tilde\lambda_{\dot a}.} As we can
see from this formula, it is natural to take $\lambda$ and
$\tilde\lambda$ to have dimension 1/2 (so that
$-i[D,\lambda]=\lambda/2$, and similarly for $\tilde\lambda$).
This determines the dilatation operator $D$ up to an additive
constant $k$: \eqn\ormgo{D={i\over
2}\left(\lambda^a{\partial\over\partial\lambda^a}+\tilde\lambda^{\dot
a}{\partial\over\partial\tilde\lambda^{\dot a}}+k\right).} What
about the special conformal generator $K_{a\dot a}$?  As it has
dimension $-1$, it cannot be represented by a multiplication
operator or a first order differential operator; if we try to
represent $K_{a\dot a}$ by a second order differential operator,
the unique possibility (up to a multiplicative constant) is
\eqn\tormgo{K_{a\dot
a}={\partial^2\over\partial\lambda^a\partial\tilde\lambda^{\dot
a}}.}  To verify that these operators do generate conformal
transformations, the non-trivial step is the commutator of
$K_{a\dot a}$ with $P^{b\dot b}$.  A short calculation shows that
the desired relation $[K_{a\dot a},P^{b\dot
b}]=-i(\delta^b_a\tilde J^{\dot b}_{\dot a}+\delta^{\dot b}_{\dot
a}J^b_a+\delta^b_a\delta^{\dot b}_{\dot a}D)$ does arise precisely
if we fix the constant $k$ so that the dilatation operator becomes
\eqn\jormgo{D={i\over
2}\left(\lambda^a{\partial\over\partial\lambda^a}+\tilde\lambda^{\dot
a}{\partial\over\partial\tilde\lambda^{\dot a}}+2\right).}

Now let us verify the conformal invariance of the MHV amplitude,
which as we recall is \eqn\tofgop{\widehat
A=ig^{n-2}(2\pi)^4\delta^4\left(\sum_i\lambda_i^a\tilde\lambda_i^{\dot
a}\right){\langle\lambda_s,\lambda_t\rangle^4\over
\prod_{i=1}^n\langle\lambda_i,\lambda_{i+1}\rangle}.} Lorentz
invariance of this formula is manifest, and momentum conservation
is also clear because of the delta function.  So we really only
have to verify that the amplitude is annihilated by $D$ and by
$K$.

First we consider $D$. The numerator contains a delta-function of
energy-momentum conservation, which in four dimensions has
dimension $-4$, and the factor
$\langle\lambda_s,\lambda_t\rangle^4$, of dimension 4.  So $D$
commutes with the numerator.  We are left acting with $D$ on the
remaining factor \eqn\remfac{{1\over
\prod_{i=1}^n\langle\lambda_i,\lambda_{i+1}\rangle}.} This is
homogeneous in each $\lambda_i$ of degree $-2$, so allowing for
the $+2$ in the definition \jormgo\ of $D$, it is annihilated by
$D$.

Similarly, we can verify that $K_{a\dot a}\widehat A=0$.  We write
$\widehat A=ig^{n-2}(2\pi)^4 \delta^4(P)A(\lambda)$, where
$P^{b\dot b}= \sum_{i=1}^n \lambda_i^b\lambda_{i}^{\dot b}$. Using
the fact that $\partial A/\partial\tilde\lambda=0$, we get, on
using the chain rule, \eqn\weget{\eqalign { K_{a\dot a}\widehat
A=&
\sum_i{\partial^2\over\partial\lambda_i^a\partial\tilde\lambda_i^{\dot
a}}\widehat A\cr
=&ig^{n-2}(2\pi)^4\left(\left(\left(n{\partial\over\partial
P^{a\dot a}}+P^{b\dot b}{\partial^2\over\partial P^{a\dot
b}\partial P^{b\dot
a}}\right)\delta^4(P)\right)A+\left({\partial\over
\partial P^{b\dot a}}\delta^4(P)\right)\sum_i\lambda_i^b{\partial A\over \partial
\lambda_i^a}\right).\cr}}  Since $J_{ab}A=0$, we can replace
$\sum_i\lambda_i^b\partial A/\partial\lambda_i^a$ by $\half
\delta^a_b\sum_i\lambda_i^c\partial
A/\partial\lambda_i^c=-(n-4)\delta^a_bA$. Upon multiplying by a
test function and integrating by parts, we find that the
distribution \eqn\yurmo{P^{b\dot b}{\partial^2\over\partial
P^{a\dot b}\partial P^{b\dot a}}\delta^4(P)} is equal to
\eqn\hurmo{-4{\partial\over
\partial P^{a\dot a}}\delta^4(P).} Combining these statements, the
right hand side of \weget\ vanishes.

\subsec{Fourier Transform to Twistor Space}

The representation of the conformal group that we have encountered
above is certainly quite unusual.  Some generators are represented
by differential operators of degree one, but the momentum operator
is a multiplication operator, and the special conformal generators
are of degree two.

We can reduce to a more standard representation of the conformal
group if (in Penrose's spirit \penrose) we make the transformation
\eqn\umbub{\eqalign{\tilde\lambda_{\dot a} \rightarrow &
i{\partial\over\partial\mu^{\dot a}}\cr
              -i{\partial\over\partial\tilde\lambda^{\dot
              a}}    \rightarrow &\mu_{\dot a}.\cr}}
In making this substitution, we have arbitrarily chosen to
transform $\tilde\lambda$ rather than $\lambda$.  The choice
breaks the symmetry between left and right, and means that
henceforth scattering amplitudes with $n_1$ positive helicity
particles  and $n_2$ of negative helicity will be treated
completely differently from those with $n_1$ and $n_2$ exchanged.
Our choice, as we will see in detail in section 3, causes
amplitudes with an arbitrary number of $+$ helicities and only a
fixed number of $-$ helicities to be treated in a relatively
uniform way, while increasing the number of $-$ helicities makes
the description more complicated. With the opposite choice for the
Fourier transform, the roles of the two helicities would be
reversed. Each amplitude can be studied in either of the two
formalisms, and hence potentially obeys two different sets of
differential equations, as we see in detail in section 3.

Upon making the substitution \umbub, the momentum and special
conformal operators become first order operators,
\eqn\uvu{\eqalign{P_{a\dot
a}&=i\lambda_a{\partial\over\partial\mu^{\dot a}}\cr K_{a\dot
a}&=i\mu_{\dot a}{\partial\over \partial\lambda^a}.\cr}} The
Lorentz generators are unchanged in form,
\eqn\tomgot{\eqalign{J_{ab}& = {i\over
2}\left(\lambda_a{\partial\over
\partial\lambda^b}+\lambda_b{\partial\over
\partial\lambda^a}\right)\cr
\tilde J_{\dot a\dot b}& = {i\over 2}\left(\mu_{\dot
a}{\partial\over
\partial\mu^{\dot b}}+\mu_{\dot b}{\partial\over
\partial\mu^{\dot a}}\right).\cr}}
Finally, the dilatation generator becomes a {\it homogeneous}
first order operator, as the $+2$ in \jormgo\ disappears:
\eqn\formgo{D={i\over
2}\left(\lambda^a{\partial\over\partial\lambda^a}-\mu^{\dot
a}{\partial\over\partial\mu^{\dot a}}\right).}

This representation of the four-dimensional conformal group is a
much more obvious one than the representation that we described
above in terms of $\lambda$ and $\tilde\lambda$. Consider the
four-dimensional space $\Bbb T$ (called twistor space by Penrose)
spanned by $\lambda^a$ and $\mu^{\dot a}$. It is a copy of $\Bbb
R^4$ if we are in signature $++-\,-$ and can consider $\lambda$
and $\mu$ to be real. Otherwise, we must think of $\Bbb T$ as a
copy of $\Bbb C^4$.  At any rate, the traceless four by four
matrices acting on $\lambda$ and $\mu$ generate a group which is
$SL(4,\Bbb R)$ in the real case, or $SL(4,\Bbb C)$ in the complex
case.  In fact,  the conformal group in four dimensions is a real
form of $SL(4)$ (namely $SL(4,\Bbb R)$, $SU(2,2)$, or $SU(4)$ for
signature $++-\,-$, $+--\,-$, or $+++\,+$), and the conformal
generators found in the last paragraph simply generate the natural
action of $SL(4)$ on $\Bbb T$.

The ``identity'' matrix on $\Bbb T$ also has a natural meaning in
this framework.  In \ruffo, we found that the scattering amplitude
for a particle of helicity $h$ obeys a condition that in terms of
$\lambda $ and $\mu$ becomes
\eqn\bruffo{\left(\lambda^a{\partial\over\partial\lambda^a}
+\mu^{\dot a}{\partial\over\partial\mu^{\dot a}}\right) \tilde
A(\lambda_i,\tilde\lambda_i,h_i)=(-2h-2)\tilde
A(\lambda_i,\tilde\lambda_i,h_i).} (We write $\tilde A$ for the
scattering amplitude in twistor space; its proper definition will
be discussed momentarily.  For now, we just formally make the
transformation from $\tilde\lambda$ to $\partial/\partial \mu$ in
\ruffo.) This equation means that the scattering amplitude of a
massless particle of helicity $h$ is best interpreted not as a
function on $\Bbb T$ but as a section of a suitable line bundle
${\cal L}_h$ over the projective space $\Bbb{PT}$ whose
homogeneous coordinates are $\lambda $ and $\mu$. $\Bbb{PT}$ is a
copy of $\Bbb{RP}^3$ or $\Bbb{CP}^3$ depending on whether we
consider $\lambda$ and $\mu$ to be real or complex.  $\Bbb{PT}$ is called
projective twistor space, but when confusion with $\Bbb{T}$ seems unlikely,
we will just call it twistor space.

In the complex case, ${\cal L}_h={\cal O}(-2h-2)$, which is defined
as the line bundle whose sections are functions homogeneous of
degree $-2h-2$ in the homogeneous coordinates. In the real case, we
can give a description in real differential geometry.  If we let
$Z^I$, $I=1,\dots,4$, be the coordinates of $\Bbb T$ (thus
combining together $\lambda^a$ and $\mu^{\dot a}$), then on
$\Bbb{PT}$ there is a natural volume form
$\Omega=\epsilon_{IJKL}Z^I dZ^J dZ^K dZ^L$ of degree four.
Multiplying by $\Omega$ converts a function homogeneous of degree
$-4$ into a three-form or measure.  So functions homogeneous of
degree $-2h-2$ can be described as $\half(1+h)$-densities.  The
scattering amplitudes, in the case of signature $(2,2)$, can thus
be interpreted as fractional densities on $\Bbb{PT}$ of these
weights.

In signature $++-\,-$, where $\tilde\lambda$ is real, the
transformation from scattering amplitudes regarded as functions of
$\tilde\lambda$ to scattering amplitudes regarded as functions of
$\mu$ is made by a simple Fourier transform that is familiar in
quantum mechanics. Any function $f(\tilde\lambda)$  is transformed
to \eqn\yubu{ \tilde f(\mu)=\int {d^2\tilde \lambda\over
(2\pi)^2}\exp(i\mu^{\dot a}\tilde \lambda_{\dot
a})f(\tilde\lambda).}  Starting with the momentum space scattering
amplitude $\widehat A(\lambda_i,\tilde\lambda_i)$ and making this
Fourier transform for each particle, we get the twistor space
scattering amplitude $\tilde A(\lambda_i,\mu_i)$.

\bigskip\noindent{\it Analog With Euclidean Signature}

\nref\penroseb{R. Penrose, ``Twistor Quantization And Curved
Spacetime,'' Int. J. Theor. Phys. {\bf 1} (1968) 61.}%
 For spacetime signatures
other than $++-\,\,-$, it is unnatural to interpret $\lambda $ and
$\tilde\lambda$ as real variables. If the variables are complex
variables, we can optimistically attempt to proceed in the same
way, interpreting the integral in \yubu\ as a contour integral.
This will make sense when a suitable contour exists. When the
twistor variables are complex, an alternative and more systematic
approach to defining the transformation from $\tilde\lambda$ to
$\tilde\mu$ can be used, modeled on Penrose's description
\penroseb\ of the particle wavefunctions in complex twistor space.
This alternative approach uses the weightier mathematical
machinery of $\bar\partial$ cohomology (or sheaf cohomology, the
alternative most exploited by Penrose).   The description of the
amplitudes using $\bar\partial$ cohomology will not be needed in
section 3, but is useful background for the string theory
discussion in section 4.   I will here describe this approach in a
rather naive way, taking the starting point to be Euclidean
signature in spacetime.   Though we start with Euclidean
signature, since we get a description with complex variables
$\lambda$, $\tilde\lambda$ or $\lambda,\mu$, the result makes
sense for computing scattering amplitudes with arbitrary complex
momenta. (Hopefully, a reader of this article will be able to give
a less naive description, and perhaps a formulation that is more
directly related to Lorentz signature.)

In $+++\,+$ signature, the spinor representations of the Lorentz
group are pseudo-real, so in particular $\tilde\lambda$ and its
complex conjugate $\overline{\tilde\lambda}$ transform in the same
way.  We simply interpret $\mu$ as $\overline{\tilde\lambda}$, or
equivalently $\tilde\lambda$ as $\overline \mu$.  Then we write
down the same formula as in \yubu\ except that we omit to do the
integral.  For any function $f(\tilde\lambda)$, we  define
\eqn\hutty{\hat f= {d^2\overline\mu\over
(2\pi)^2}\exp(i\mu\overline\mu)f(\bar\mu).} We interpret this as a
$(0,2)$-form, which is obviously $\bar\partial$-closed as $\mu$
has only two components.   We interpret the $\bar\partial$
cohomology class represented by this form as the twistor version
of the helicity scattering amplitude. Granted this, the
scattering amplitude for a massless particle of helicity $h$ can
be interpreted in twistor space as an element of the sheaf
cohomology group $H^2(\Bbb {PT}',{\cal O}(-2h-2))$, where the
homogeneity was determined above.

The reason that we have written here $\Bbb{PT}'$, and not
$\Bbb{PT}$, is that, as with many twistor constructions, one
really should not work with all of $\Bbb{PT}$ but with a suitable
open set thereof.  In fact, $H^2(\Bbb {PT},{\cal O}(-2h-2))=0$. For
scattering of plane waves, one can  take $\Bbb{PT}'$ to be
the subspace of $\Bbb{PT}$ in which the $\lambda_a$ are not both
zero.\foot{This statement reflects the fact that plane waves are
regular throughout $\Bbb{R}^4$, but have a singularity upon
conformal compactification to $\Bbb{S}^4$.  It is standard in
twistor theory that omitting the point at infinity in Euclidean
spacetime corresponds to omitting the subspace of twistor space in
which the $\lambda$'s vanish. We will give an illustration of the idea behind this
statement at the end of the present subsection, and more information in the
appendix.}
Understanding exactly what open set $\Bbb{PT}'$  and what
kind of $\bar\partial$ cohomology to use in  a given physical
problem is a large part
of more fully understanding the twistor transform with complex variables.

\bigskip\noindent{\it Pairing With External Wavefunctions}

Physical particles are not normally in momentum eigenstates.
Initial and final states of interest might be arbitrary solutions
of the free wave equation. If a momentum space scattering
amplitude $\widehat A(p_1,\dots,p_n)$ is known (for simplicity we
consider scalar particles, so that the amplitude depends only on
the momenta), the amplitude for scattering with initial and final
states $\phi_1(x),\dots,\phi_n(x)$ (each of which obeys the
appropriate free wave equation) is  obtained by taking a suitable
convolution.  If $\phi_i(x)$ is expressed in terms of momentum
space wavefunctions $a_i(p)$ by $\phi_i(x)=\int d^4p \,e^{ip\cdot
x}\delta(p^2)a_i(p)$ (where the factor $\delta(p^2)$ ensures that
the momentum space wavefunction is supported at $p^2=0$), then the
scattering amplitude with external states $\phi_i$ is
\eqn\jumbo{A(\phi_1,\dots,\phi_n)=\prod_{i=1}^n\int
d^4p_i\delta(p_i^2)a_i(p_i) \widehat A(p_1,\dots,p_n).} In other
words, the amplitude with specified external states is obtained
from the momentum space scattering amplitude by multiplying by the
momentum space wavefunctions and integrating over momentum space.

Similarly, to go from the twistor space amplitude to an amplitude
with specified external states, we must multiply by twistor space
wavefunctions and integrate over twistor space. To carry this out,
we need the twistor description of the initial and final state
wavefunctions. This description was originally developed
 by Penrose \penroseb\ for complex twistor
space $\Bbb{CP}^3$.  This is the most useful case for most purposes and is
reviewed in the appendix. There is also an analog
(explained by Atiyah in \atiyah, section VI.5) for real
twistor space $\Bbb{RP}^3$. We consider first the real case.  (The following
discussion is useful background for the computation of tree level MHV amplitudes
in section 4.7, but otherwise is not needed for the rest of the paper.)

For signature $++-\,-$, a massless
field of helicity $h$ corresponds to a $\half(1-h)$-density over
${\bf RP}^3$.   As we have seen above, the scattering amplitude
for a massless particle of helicity $h$
 is a $\half(1+h)$ density.  When these
are multiplied, we get a density on ${\bf RP}^3$, which can be
integrated over ${\bf RP}^3$ to get a number.  We interpret this
number as the scattering amplitude for the given initial and final
states. (This process of integration over ${\bf RP}^3$ is
analogous to the momentum space integral in \jumbo\ and has to be carried out
separately for each initial and final particle.)

Instead, for the complex case of the twistor transform, the
wavefunction of an external particle is, according to the usual
complex case of the Penrose transform, an element of the
$\bar\partial$ cohomology group $H^1(\Bbb{PT}', {\cal O}(-2-h))$.
Upon taking the cup product of such an element with the scattering
amplitude, which as we have argued above is
 an element of $H^2(\Bbb{PT}',{\cal O}(h-2))$, we get an element of $H^3(\Bbb{PT}',
{\cal O}(-4))$.  As ${\cal O}(-4)$ is the canonical bundle of
$\Bbb{CP}^3$, an element of $H^3(\Bbb{PT}',{\cal O}(-4))$ can be
interpreted as a $(3,3)$-form on $\Bbb{PT}'$; upon integrating
this form over $\Bbb{PT}'$ (and repeating this process for each
external particle) one gets the desired scattering amplitude for
the given initial and final states.  This procedure is implemented
for MHV tree amplitudes in eqn. (4.54).

\bigskip\noindent{\it Physical Interpretation Of Twistor Space Amplitudes}

Finally, let us discuss the physical meaning of the twistor space
scattering amplitude $\tilde A(\lambda_i,\mu_i)$.  We carry out this discussion
in signature $++-\,-$, where the definition of $\tilde A$ is more elementary.
 The scattering amplitude as a
function of $p_{a\dot a}=\lambda_a\tilde \lambda_{\dot a}$ has a
clear enough meaning: it is the amplitude for scattering a
particle whose wavefunction is $\exp(ip\cdot x)$. What is the
wavefunction of a particle whose scattering is described by a
function of $\lambda$ and $\mu$?

The twistor space scattering amplitude $\tilde A(\lambda,\mu)$ has
been obtained from the momentum space version $\widehat
A(\lambda,\tilde\lambda)$ by multiplying by $\exp(i\tilde
\lambda_{\dot a} \mu^{\dot a})$ and integrating over
$\tilde\lambda$. This describes a scattering process in which the
initial and final states are not plane waves but have the wave
function \eqn\ploon{\int {d^2\tilde\lambda\over
(2\pi)^2}\exp(ix^{a\dot a}\lambda_a\tilde \lambda_{\dot
a})\exp(i\tilde\lambda_{\dot a}\mu^a).} The result of the integral
is simply \eqn\nonon{\delta^2(\mu_{\dot a}+x_{a\dot a}\lambda^a).}
In other words, the scattering amplitude as a function of $\lambda
$ and $\mu$ describes the scattering of a particle whose wave
function is a delta function supported on the subspace of
Minkowski space given by the equation $\mu_{\dot a}+x_{a\dot
a}\lambda^a=0$. This is the two-dimensional subspace of Minkowski
space that is associated by Penrose \penrose\ with the point in
$\Bbb{PT}$ whose homogeneous coordinates are $\lambda,\mu$.

From the equation $\mu_{\dot a}+x_{a\dot a}\lambda^a$, we see that
for $x\to\infty$, $\lambda$ must vanish.  (As $\lambda,\mu$ are
homogeneous coordinates, $\lambda\to 0$ is equivalent to $\mu\to
\infty$.) This illustrates the idea that not allowing
 $x\to\infty$ is related to omitting the
set in twistor space with $\lambda=0$.

\subsec{Extension With ${\cal N}=4$ Supersymmetry}

Here, along lines suggested by Nair \nair, we will describe the
generalization of the MHV amplitudes to ${\cal N}=4$ super
Yang-Mills theory. Then we will transform to super twistor space.
(This discussion is useful background for section 4, but it is not
really needed for section 3.)

We describe each external particle by the familiar (commuting)
spinors $\lambda_a,\tilde\lambda_{\dot a}$, and also by a
spinless, anticommuting variable $\eta_A$, $A=1,\dots ,4$.
$\eta_A$ will have dimension zero and transforms in the
${\overline 4}$ representation of the $SU(4)_R$ symmetry of ${\cal
N}=4$ Yang-Mills.  We take the helicity operator to be
\eqn\into{h=1-\half \sum_A\eta_A{\partial\over\partial\eta_A}.}
Thus, a term in the scattering amplitude that is of $k^{th}$ order
in $\eta_{iA}$ for some $i$ describes a scattering process in
which the $i^{th}$ particle has helicity $1-k/2$. Notice that in
adopting this formalism, we are breaking the symmetry between
positive and negative helicity, choosing the term in the
scattering amplitude with no $\eta$'s to have helicity $1$
instead of $-1$. This choice is well adapted to describing MHV
amplitudes in which most external particles have helicity $1$;
one would make the opposite choice to give a convenient
description of the other MHV amplitudes.

The form \into\ of the helicity operator means that the relation
\ruffo\ between the helicity and the homogeneity of the scattering
amplitude $\widehat A$ becomes
\eqn\incob{\left(\lambda_i^a{\partial\over\partial\lambda_i^a}
-\tilde\lambda_i^{\dot
a}{\partial\over\partial\tilde\lambda_i^{\dot a}}
-\eta_{iA}{\partial\over\partial\eta_{iA}}+2\right)\widehat A =
0.} If we let $P_{a\dot a}=\sum_i\lambda_{ia}\tilde\lambda_{i\dot
a}$, and $\Theta_{b A}=\sum_i\lambda_{ib}\eta_{iA}$, then the MHV
scattering amplitude  (with the gauge theory factor $\Tr
T_1T_2\dots T_n$ suppressed as usual) is \eqn\tyco{\widehat
A=ig^{n-2}(2\pi)^4\delta^4(P)\delta^8(\Theta) \prod_{i=1}^n{1\over
\langle \lambda_i,\lambda_{i+1}\rangle}.} We recall that for a
fermion $\psi$, one defines $\delta(\psi)=\psi$.

Now let us describe the action of the superconformal group
$PSU(2,2|4)$ in this formalism.  The conformal group $SU(2,2)$
commutes with $\eta$ and acts on $\lambda$ and $\tilde\lambda$
exactly as we have described above.  The $SU(4)$ $R$-symmetry
group is generated by
\eqn\bogus{\eta_A{\partial\over\partial\eta_B}-{1\over
4}\delta^B_A \eta_C{\partial\over\partial\eta_C}.} Note that we
consider only ``traceless'' generators here, as the ``trace''
generator $\eta_C({\partial/\partial\eta_C})$ is not contained in
$PSU(2,2|4)$ and is not a symmetry of ${\cal N}=4$ super
Yang-Mills theory.  Nonetheless, it will play an important role
later.

Finally,  $PSU(2,2|4)$ has 32 fermionic generators, which act as
follows. Half of them are first order differential operators
\eqn\firsthalf{\tilde\lambda^{\dot
a}{\partial\over\partial\eta_A},~~
\eta_A{\partial\over\partial\tilde\lambda^{\dot a}}.} One quarter
are multiplication operators \eqn\normgo{\lambda^a \eta_A,} and
the remainder are second order differential operators
\eqn\formgo{{\partial^2\over\partial\lambda^a\partial\eta_A}.} It
is not hard to verify that these generate $SU(2,2|4)$ (but this
result is more transparent after the transformation to super
twistor space that we make presently).

Next, let us verify the superconformal invariance of the MHV
amplitudes. Apart from dilatations, the bosonic and fermion
 generators that act by first order differential
operators are manifest symmetries.  The bosonic and fermionic
generators that act as multiplication operators are conserved
because of the delta functions in the scattering amplitude.
Dilatation invariance  can be verified exactly as in the previous
discussion of the pure Yang-Mills case. Special conformal symmetry
follows via closure of the algebra if  the scattering amplitude is
annihilated by the fermionic second order differential operators
\formgo. So we need only check those symmetries, which can be
established by a procedure similar to the one that was  used to
show special conformal invariance of the Yang-Mills amplitudes.
For brevity, we set $B=ig^{n-2}(2\pi)^4$, $\delta_1=\delta^4(P)$,
$\delta_2=\delta^8(\Theta)$, and $S=1/\prod_{i=1}^n
\langle\lambda_i,\lambda_{i+1}\rangle$.  So the scattering
amplitude is $\widehat A=B\delta_1\delta_2S$.  Now we compute that
\eqn\gono{B^{-1}\sum_i{\partial^2\widehat
A\over\partial\lambda_i^a\partial
\eta_{iA}}=\delta_1{\partial\delta_2\over
\partial\Theta^b_A}\sum_i\lambda_i^b {\partial S\over
\partial\lambda_i^a}+\delta_1\cdot n{\partial\delta_2
\over\partial\Theta^a_A}S-\delta_1\Theta^b_C{\partial^2\delta_2\over
\partial\Theta^b_A\partial\Theta^a_C}S
+P^{a\dot b}{\partial\delta_1\over\partial P^{b\dot b}}{\partial\delta_2\over
\partial\eta^a_A}S.}
This vanishes, upon using the following facts that are analogous
to the facts that we used in examining the special conformal
symmetries: (i) $\sum_i\lambda_i^a{\partial
S\over\partial\lambda_i^b}=-n\delta^a_bS$; (ii) $P^{a\dot
b}(\partial\delta_1/\partial P^{b\dot b})=-2\delta^a_b \delta_1$;
and finally (iii)
$\Theta^b_C(\partial^2\delta_2/\partial\Theta^b_A
\partial\Theta^a_C)=-2(\partial\delta_2/\partial \Theta^a_A).$

\bigskip\noindent{\it Transform To Super-Twistor Space}

The next step should not be hard to guess.  We make the action of
$PSU(2,2|4)$ more transparent by a supersymmetric extension of the
same transformation as in the bosonic case:
\eqn\pilpo{\eqalign{\tilde\lambda_{\dot a}& \to
i{\partial\over
\partial\mu^{\dot a}}\cr
-i{\partial\over\partial\tilde\lambda^{\dot a}}&\to
\mu_{\dot a}\cr
    \eta_A & \to i{\partial\over\partial\psi^A}\cr
    -i{\partial\over\partial\eta_A}&\to \psi^A.\cr}}
Now all $PSU(2,2|4)$ generators become first order differential operators.

Moreover, the representation of $PSU(2,2|4)$ that we get this way
is easy to describe.  We introduce a space $\hat {\Bbb T}=\Bbb
C^{4|4}$ with four bosonic coordinates $Z^I=(\lambda^a,\mu^{\dot
a})$ and four fermionic coordinates $\psi^A$.   $\hat {\Bbb T}$ is
the supersymmetric extension of (nonprojective) twistor space
(with ${\cal N}=4$ supersymmetry). The full supergroup of linear
transformations of $\hat {\Bbb T}$ is called $GL(4|4)$. If we take
the quotient of this group by its center -- which consists of the
nonzero multiples of the identity -- we get the quotient group
$PGL(4|4)$. This is the symmetry group of the projectivized
super-twistor space $\hat{\Bbb {PT}}$ that has $Z^I$ and $\psi^A$
as homogeneous coordinates.  In other words, $\hat{\Bbb {PT}}$ is
parameterized by $Z^I$ and $\psi^A$ subject to the equivalence
relation $(Z^I,\psi^A)\sim (t Z^I,t \psi^A)$ for
nonzero complex $t$.    If the $Z^I$ are taken to be
complex, then $\hat {\Bbb{PT}}$ is a copy of the supermanifold
$\Bbb{CP}^{3|4}$. It also makes sense, and is natural in signature
$++-\,-$, to take the $Z^I$ to be real. We cannot take the
$\psi^A$ to be real, as this would clash with the $SU(4)$
$R$-symmetry. But we can do the next best thing: we simply do not
mention $\bar\psi$ and consider only functions that depend only on
$Z^I$ and $\psi^A$. (As functions of $\psi^A$ are automatically
polynomials, there are no choices to make of what kind of
functions to allow.)
 This version of $\hat {\Bbb{PT}}$ might be called
$\Bbb{RP}^{3|4}$.

When confusion seems unlikely because the context clearly refers to the
supersymmetric case, we will sometimes omit the hats from
$\hat{\Bbb T}$ and $\hat{\Bbb{PT}}$.

If we further introduce the object
\eqn\indop{\Omega_0=dZ^1dZ^2dZ^3dZ^4d\psi^1d\psi^2d\psi^3d\psi^4,}
then the subgroup of $PGL(4|4)$ that preserves $\Omega_0$ is
called $PSL(4|4)$.  Finally, the desired superconformal group is a
real form of $PSL(4|4)$.  In Lorentz signature, it is
$PSU(2,2|4)$. The object $\Omega_0$, in contrast to our experience
in the bosonic case, is best understood as a measure (in the
holomorphic sense) on $\hat {\Bbb{T}}$, or a section of the
Berezinian of the tangent bundle of $\hat {\Bbb T}$, rather than
as a differential form.

After the transform to twistor variables, the homogeneity
condition for the scattering amplitudes becomes simply
\eqn\onxon{\left(Z_i^I{\partial\over\partial
Z_i^I}+\psi_i^A{\partial\over
\partial\psi_i^A}   \right) \tilde A = 0.}
In other words, the scattering amplitude is homogeneous in the
 twistor coordinates of each external particle.

In the case of signature $++-\,-$, rather as in the bosonic case,
the transformation to twistor variables is made by an ordinary
Fourier transform from $\tilde\lambda$ to $\mu$ together with a
Fourier transform from $\eta$ to $\psi$.  In this signature
$\tilde\lambda$ is real (or at least real modulo nilpotents), so
the Fourier transform from $\tilde\lambda$ to $\mu$ makes sense;
fermions are infinitesimal so the Fourier transform from $\eta$ to
$\psi$ is really an algebraic operation that has no analytic
difficulties.  The twistor transformed scattering amplitude is
thus in the signature $++-\,-$ case a function on the twistor
superspace $ {\Bbb {RP}}^{3|4}$.  In contrast to $\Bbb{RP}^3$,
${\Bbb{RP}}^{3|4}$ has a natural measure, associated with the
object \eqn\imopo{\Omega={1\over
4!^2}\epsilon_{IJKL}Z^IdZ^JdZ^KdZ^L\epsilon_{ABCD} d\psi^A d\psi^B
d\psi^Cd\psi^D.} In the real version of the twistor transform, the
external particle wavefunctions are just functions on
$\Bbb{RP}^{3|4}$; to compute a scattering amplitude with specified
initial and final states, one multiplies the external
wavefunctions by the twistor space scattering amplitude and
integrates it over twistor space, using the measure \imopo.

For other signatures, one must take twistor space to be the
complex manifold $\hat{\Bbb{PT}}={\Bbb{CP}}^{3|4}$. The Fourier
transform in the fermions does not introduce any special
subtleties, but we must treat the complex bosons just as we did in
the absence of supersymmetry.  So the transform to twistor
variables produces in the complex case  a scattering amplitude
that for each external particle is an element of the sheaf
cohomology group $H^2(\hat{\Bbb{PT}}\,',{\cal O})$. As before,
$\hat{\Bbb{PT}}\,'$ is $\hat{\Bbb{PT}}$ with the set $\lambda^a=0$
omitted. Also, ${\cal O}$ is the trivial line bundle, which is the
right one since the twistor wave function according to \onxon\ is
homogeneous of degree zero in $Z^I$ and $\psi^A$. The
wavefunctions for external particles with specified quantum states
are elements of the sheaf cohomology group
$H^1(\hat{\Bbb{PT}}\,',{\cal O})$. To compute a scattering
amplitude with specified initial and final states, one takes the
cup product of the external wave functions with the scattering
amplitudes, to get for each particle an element of
$H^3(\hat{\Bbb{PT}}\,',{\cal O})$.  This can then be integrated,
using the section $\Omega$ of the Berezinian of the tangent
bundle, to get a number, the scattering amplitude.

Even when $\lambda$ and $\mu$ are complex and one is doing
$\bar\partial$ cohomology, there is no need to introduce a complex
conjugate of $\psi^A$.  Though it may be inevitable to consider
not necessarily holomorphic functions of $\lambda$ and $\mu$,
there is no need to consider non-holomorphic functions of
$\psi^A$.  (In section 4, we will see that the complex conjugate
of $\psi^A$ is unavoidable for closed strings, but can be avoided
for open strings.)

\newsec{Scattering Amplitudes In Twistor Space}

After the Fourier transform to twistor space, each external particle
in an $n$-particle scattering process is labeled by a point $P_i$ in twistor
space.  The homogeneous coordinates of $P_i$ are
$Z^I_i=(\lambda_i^a,\mu_i^{\dot a})$.
The scattering amplitudes are functions of the $P_i$, that is, they are functions
defined on the product of $n$ copies of twistor space.

In this section, we make an empirical study of scattering
amplitudes in twistor space.  We work in signature $++-\,-$, and
therefore in real twistor space $\Bbb{RP}^3$. We consider gluon
scattering and will (for the most part) make no attempt to make
supersymmetry manifest, so we can use $\Bbb{RP}^3$ rather than its
supersymmetric extension $\Bbb{RP}^{3|4}$. The advantage of
signature $++-\,-$ is that the transform to twistor space is an
ordinary Fourier transform, and the scattering amplitudes in real
twistor space are ordinary functions on $(\Bbb{RP}^3)^n$, one copy
of twistor space for each external particle. With other
signatures, we would have to use  $\bar\partial$ cohomology and a
weightier mathematical machinery.

The goal is to show that the twistor version of the $n$ particle
scattering amplitude is nonzero only if the points $P_i$ are all
supported on an algebraic curve in twistor space.  This algebraic
curve has degree $d$ given by \eqn\ibob{d=q-1+l,} where $q$ is the
number of negative helicity gluons in the scattering process, and
$l$ is the number of loops. It is not necessarily connected.  And
its genus $g$ is bounded by the number of loops, \eqn\nibob{g\leq
l.}

Our goal in the remainder of this section is to explore the
hypothesis that twistor amplitudes are supported on the curves
described in the last paragraph.
 We will consider in this light  various tree
level Yang-Mills scattering amplitudes (which are not sensitive to
supersymmetry) and one example of a one-loop scattering amplitude
in ${\cal N}=4$ super Yang-Mills theory. In each case, we verify
our conjecture. (Along the way, a few unexplained properties also
appear, showing that there is much more to understand.)
These examples, though certainly far short of proving or
even determining the general structure, do give a good motivation
for seeking a string theory whose instanton expansion might
reproduce the perturbation expansion of super Yang-Mills theory
and explain the conjecture. We make a proposal for such a string
theory in the next section.

We conclude this section with
a peek at General Relativity, showing that the tree level MHV
amplitudes are again supported on curves.  Unfortunately, I do not know of any string
theory whose instanton expansion might reproduce the perturbation expansion of
General Relativity or supergravity.

\bigskip\noindent{\it A Note On Algebraic Curves}

An algebraic curve $\Sigma$ in $\Bbb{RP}^3$ is a curve defined
as the zero set of a collection of polynomial equations with real coefficients
in the homogeneous coordinates $Z^I$ of $\Bbb{RP}^3$.  The degree and genus of
such a curve are defined by complexifying the $Z^I$, whereupon $\Sigma$ becomes
a Riemann surface in $\Bbb{CP}^3$, whose degree and genus are defined in the usual
way.

The simplest type of algebraic curve is a ``complete
intersection,'' obtained by setting to zero two homogeneous
polynomials $F(Z^I)$ and $G(Z^I)$, of degrees (say) $d_1$ and
$d_2$.\foot{In the more general case, one must define a given
curve $C$ by the vanishing of more than two polynomials, all of
which vanish on $C$ but any two of which vanish on additional
branches other than $C$.  Curves of genus zero and degree three
give an example; we briefly comment on their role at the end of section 3.4.}
 The degree of such a
complete intersection is $d=d_1d_2$. Curves of degree one and two
are of this type, with $(d_1,d_2)=(1,1)$ and $(1,2)$,
respectively.

\subsec{MHV Amplitudes}

We begin with the $n$ gluon tree level MHV amplitude with two
gluons of negative helicity and $n-2$ of positive helicity.  As we
reviewed in section 2.3, it can be written \eqn\torbo{\widehat
A(\lambda_i,\tilde\lambda_i)=
ig^{n-2}(2\pi)^4\delta^4\left(\sum_i\lambda_i^a \tilde\lambda_i^{\dot
a}\right)f(\lambda_i),} where $f(\lambda_i)$ is a function of only the
$\lambda$'s and not the $\tilde\lambda$'s. The details of
$f(\lambda)$ need not concern us here. Using a standard
representation of the delta function (used in discussing MHV
amplitudes by Nair \nair), we can rewrite the amplitude as
\eqn\forbo{\widehat A(\lambda_i,\tilde\lambda_i) =ig^{n-2}\int
d^4x \,\exp\left(ix_{a\dot a}\sum_{i=1}^n\lambda_i^a
\tilde\lambda_i^{\dot a}\right)f(\lambda_i).}

To transform to twistor space, we simply carry out a Fourier
transform with respect to all of the $\tilde\lambda$ variables.
The twistor space amplitude is hence \eqn\jorbo{\tilde
A(\lambda_i,\mu_i)=ig^{n-2}\int d^4x \int {d^2\tilde\lambda_1
\over (2\pi)^2}
\dots{d^2\tilde\lambda_n\over (2\pi )^2}
\,\,\exp\left(i\sum_{i=1}^n\mu_{i\,\dot a} \tilde\lambda_i^{\dot
a}\right) \,\exp\left(ix_{a\dot a}\sum_{i=1}^n\lambda_i^a
\tilde\lambda_i^{\dot a}\right)f(\lambda_i).} The $\tilde \lambda$
integrals can be done trivially, with the result \eqn\vorbo{\tilde
A(\lambda_i,\mu_i)=ig^{n-2}\int d^4x \prod_{i=1}^n
\delta^2(\mu_{i \dot a}+x_{a\dot a}\lambda_i^a)f(\lambda_i).}

Let us now interpret this result.  For every (real) $x_{a\dot a}$,
the pair of equations \eqn\porbo{\mu_{\dot a}+x_{a\dot
a}\lambda^a=0, ~~\dot a=1,2} defines a real algebraic curve $C$ in
$\Bbb{RP}^3$, or if we complexify the variables, in $\Bbb{CP}^3$.
This curve is a complete intersection, since it is defined by
vanishing of a pair of homogeneous polynomials; it is of degree 1
since the polynomials are linear (thus $d_1=d_2=1$ in the notation
used at the end of section 3.1).  Moreover, $C$ has genus zero.
Indeed, the equations \porbo\ can be solved for $\mu_{\dot a}$ as
a function of $\lambda^a$, so the $\lambda^a$ serve as homogeneous
coordinates for $C$, which therefore is a copy of $\Bbb{RP}^1$ or
$\Bbb{CP}^1$ depending on whether the variables are real or
complex.  Conversely, if one allows limiting cases with
$x\to\infty$, the degree one, genus zero curves are all of this
type.

The integral $\int d^4 x$ in \vorbo\ is thus an integral over the
moduli space of real, degree one, genus zero curves in
$\Bbb{RP}^3$.  The delta function means that the amplitude
vanishes unless all $n$ points $P_i=(\lambda_i^a,\mu_{i\,\dot a})$
are contained on one of these curves.  The MHV amplitudes with
mostly plus helicities are thus supported, in twistor space, on
configurations of $n$ points that all lie on a curve in ${\Bbb
{RP}}^3$ of degree one and genus zero.  This is a basic example of
our proposal.

As long as we are working in the real version of twistor space,
which is appropriate to $++-\,-$ signature, this curve can be
described more intuitively as a straight line.  Indeed, throwing
away, for example, the set $\lambda_1=0$ in ${\Bbb {RP}}^3$, we
can describe the rest of ${\Bbb {RP}}^3$ by the affine coordinates
$x=\lambda_2/\lambda_1$, $y=\mu_1/\lambda_1$, $z=\mu_2/\lambda_2$.
$x,y,$ and $z$ parameterize a copy of ${\Bbb R}^3$ and the curve
$C$ is simply a straight line in ${\Bbb R}^3$.  Thus (figure 2),
the MHV amplitude with these helicities is supported for points
$P_i$ that are collinear in ${\Bbb R}^3$.  The conformal symmetry
group (in $++-\,-$ signature) is the $SL(4,\Bbb{R})$ symmetry of
${\Bbb {RP}}^3$; it does not preserve a metric on ${\Bbb R}^3$,
but it maps straight lines to straight lines.

\ifig\collinear{The MHV amplitude for gluon scattering is
associated with a collinear arrangement of points in $\Bbb R^3$.}
{\epsfbox{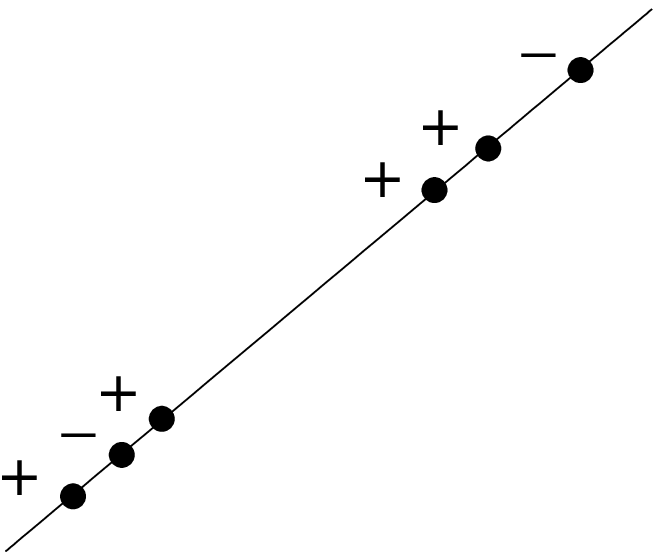}}

\bigskip\noindent{\it Supersymmetric Extension}

Here, by Fourier-tranforming the supersymmetric MHV amplitude
\tyco, we will obtain the supersymmetric extension of the above
result.  This is the only example where we will study the twistor
transform of a manifestly supersymmetric amplitude.

We write the fermionic delta function in \tyco\ as
\eqn\florry{\delta^8(\Theta)=\int
d^8\theta_a^A\,\exp\left(i\theta_a^A\sum_i\eta_{iA}\lambda_i^a\right).}
Using this and the familiar representation of the bosonic delta
function in \tyco, the supersymmetric MHV amplitude becomes
\eqn\plook{\hat A=ig^{n-2}\int
d^4x\,d^8\theta\,\exp\left(ix_{a\dot
a}\sum_i\lambda_i^a\tilde\lambda_i^{\dot
a}\right)\exp\left(i\theta_a^A\sum_i\eta_{iA}\lambda_i^a\right)
\prod_{i=1}^n{1\over\langle\lambda_i,\lambda_{i+1}\rangle}.} The
Fourier transform is therefore straightforward.  The amplitude in
super twistor space is \eqn\tyno{\eqalign{\tilde
A(\lambda^a_i,\mu^{\dot a}_i,\psi^A_i)&=\int {d^2\tilde\lambda_1
d^4\eta_{1}\over (2\pi)^2}\dots {d^2\tilde\lambda_n
d^4\eta_{n}\over
(2\pi)^2}
\exp\left(i\sum_i\mu_i^a\tilde\lambda_{i\,a}+i\sum_i\psi^A_i\eta_{i\,A}\right)\hat
A\cr & = ig^{n-2}\int
d^4x\,d^8\theta_a^A\,\,\prod_{i=1}^n\delta^2\left(\mu_{i\,\dot
a}+x_{a\dot
a}\lambda_i^a\right)\delta^4\left(\psi_i^A+\theta^A_a\lambda_i^a\right)
\prod_{i=1}^n{1\over\langle\lambda_i,\lambda_{i+1}
\rangle}.\cr}}

The interpretation is very much as in the bosonic case. For any
given $x$ and $\theta$, the equations
\eqn\ononp{\eqalign{\mu_{\dot a}+x_{a\dot a}\lambda^a &=  0\cr
\psi^A+\theta_a^A\lambda^a&=0\cr}} determine a curve in the
supersymmetric extension of twistor space.  The equations can be
used to solve for $\mu$ and $\psi$ in terms of $\lambda$, so this
curve has the $\lambda^a$ as homogeneous coordinates and is a copy
of $\Bbb{CP}^1$.  It has degree one; its moduli  are $x$ and
$\theta$. In the real version of super twistor space, in affine coordinates,
this curve is a straight line.  The delta functions in \tyno\ mean that the
supersymmetric MHV amplitudes, when transformed to twistor space,
vanish unless all external particles are inserted at collinear
points in supertwistor space.

\subsec{Degree Minus One And Degree Zero}  We considered first the
MHV amplitudes with two positive helicity gluons, with the aim of
giving the reader a first orientation about our basic conjecture
that twistor space scattering amplitudes are supported on the
curves described in \ibob. However, there are a few cases to
consider that in a sense are more primitive.

 The conjecture actually gives a new perspective
on the vanishing of the tree level $n$ gluon scattering amplitudes
with all or all but one gluons of positive helicity.  A tree
amplitude with all gluons of positive helicity has $q=l=0$ in
\ibob, leading to $d=-1$.  There are no algebraic curves of degree
$-1$, so such amplitudes vanish.  If all gluons but one have
positive helicity, we get $q=1$, $l=0$, whence $d=0$.  A curve of
degree zero is collapsed to a point, so amplitudes of this type
are supported in twistor space by configurations in which all
gluons are attached at the same point $P=(\lambda,\mu)$.  In
particular, $\lambda_i=\lambda_j$ for all particles $i$ and $j$.
For the corresponding momenta $p_i$ and $p_j$, we have $p_i\cdot
p_j=\langle\lambda_i,\lambda_j\rangle[\tilde\lambda_i,\tilde\lambda_j]=0$.
This is impossible for a non-trivial scattering amplitude with
$n\geq 4$ particles, since such amplitudes depend on non-trivial
kinematic invariants $p_i\cdot p_j$ (such as the Mandelstam
variables for $n=4$).  Hence $n$ gluon tree amplitudes with all
but one gluon of positive helicity must vanish for $n\geq 4$.

The $n=3$ case is exceptional, because here it is true that
$p_i\cdot p_j=0$ for all $i$ and $j$.  Indeed, if $p_1+p_2+p_3=0$
and $p_i^2=0$ for $i=1,2,3$, then for example $p_1\cdot
p_2=(p_1+p_2)^2/2=p_3^2/2=0$.  For real momenta in Lorentz
signature, the condition $p_i\cdot p_j=0$ implies that the $p_i$
are collinear, whence the amplitude (written below) vanishes and
the phase space also vanishes.  Hence this rather degenerate case
is often omitted in discussing the Yang-Mills scattering
amplitudes.  However, the tree level three point function makes
sense with other signatures or with complex momenta, so we will
consider it here (albeit with some difficulty as will soon
appear).

Since $p_i\cdot p_j=\langle \lambda_i,\lambda_j\rangle
[\tilde\lambda_i,\tilde\lambda_j]$, it follows that for each
$i,j$, either $\langle\lambda_i,\lambda_j\rangle=0$ or
$[\tilde\lambda_i,\tilde\lambda_j]=0$.  The first condition
implies that $\lambda_i$ and $\lambda_j$ are proportional, and the
second condition implies that $\tilde\lambda_i$ and
$\tilde\lambda_j$ are proportional.  Since at least one of these
conditions is satisfied for each pair $i,j$, it follows that
either all $\lambda_i$, $i=1,2,3$ are proportional, or that all
$\tilde\lambda_i$ are proportional.

The momentum space amplitude for $-++$ is \eqn\hutob{\widehat
A(\lambda_i,\tilde\lambda_i)=ig\delta^4\left(\sum_{i=1}^3
\lambda_i^a\tilde\lambda_i^{\dot a}\right)
{[\tilde\lambda_1,\tilde\lambda_2]^4
\over\prod_{i=1}^3[\tilde\lambda_i,\tilde\lambda_{i+1}]}.} This
can be read off from the Yang-Mills Lagrangian, or it can be
regarded as a special case of the general MHV amplitude \hhutu\
with two positive helicities and an arbitrary number (which in
this case we take to be one) of negative helicities.  Since this
amplitude vanishes if the $\tilde\lambda_i$ are all proportional
(for such a configuration, the numerator has a higher order zero
than the denominator), it is supported on configurations for which
the $\lambda_i$, $i=1,2,3$, are proportional.  $SL(4,\Bbb R)$
invariance then implies that the $(\lambda_i,\mu_i)$ are all
proportional, so that the points $P_i\in \Bbb{PT}$ at which the
gluons are inserted all coincide, as predicted above.

However, it does not appear possible to justify this statement by
Fourier transforming the amplitude \hutob. Actually, trying to
define the twistor amplitudes by an ordinary Fourier transform
rests on the claim in section 2 that in signature $--++$, the
twistor amplitudes are ordinary functions in the real form of
$\Bbb{PT}$ and can be defined by an ordinary Fourier transform
without need of $\bar\partial$ cohomology. Possibly this claim is
too naive in the degenerate case of the $-++$ helicity amplitude.
The string theory proposed in section 4 does lead in the complex
form of $\Bbb{PT}$ to a local twistor space three-point function
for the $-++$ amplitude in terms of $\bar\partial$ cohomology, as
the argument given above predicts. (It is described in section
4.3.) Apparently, though I do not really know why a problem arises
for this one case, the mapping of the $\bar\partial$ cohomology
classes on complex twistor space to functions (or real densities)
on real twistor space does not work for this particular amplitude.

\subsec{First Look At Degree Two Curves And Differential
Equations}

The next case to consider is that the number of negative helicity
gluons is three.
 Amplitudes with many positive helicity gluons and
three of negative helicity are quite complicated; no general
formula for them is known.  We will analyze only the first two
cases that the number of positive helicity gluons is two or three.

The five-particle amplitudes with three negative and two positive
helicity gluons are MHV amplitudes with the opposite
``handedness'' from those that we have just considered.  Hence, if
instead of Fourier transforming from $\tilde \lambda$ to $\mu$, we had made the
opposite Fourier transform from $\lambda$ to $\tilde\mu$,
the above analysis would apply, showing
that the amplitudes with helicities $---++$ (or permutations
thereof) are supported on straight lines in what Penrose calls the
``dual twistor space.'' But we do not want to go over to this dual
twistor space where life would be easier.  We want to understand
the $---++$ amplitudes in the ``original'' twistor space with
homogeneous coordinates $\lambda,\mu$.  In the language of twistor
theorists, we want to understand the ``googly'' description of the
$---++$ amplitudes.  (The term is borrowed from cricket and refers
to a ball thrown with the opposite of the natural spin.)

At any rate, because they are MHV amplitudes, the amplitudes with
two negative and three positive helicities are simple, as we
reviewed in section 2.3: \eqn\torfgo{\widehat
A=ig^3(2\pi)^4\delta^4\left(\sum_i\lambda_i^a\tilde\lambda_i^{\dot
a}\right) {[\tilde\lambda_a,\tilde\lambda_b]^4\over
\prod_{i=1}^5[\tilde\lambda_i, \tilde\lambda_j]}.}  The gluons $a$
and $b$ are the ones with negative helicity; up to a cyclic
permutation of the five gluons, there are two cases, namely
$---++$ and $--+-+$.  The two cases need to be treated separately,
but the arguments below turn out to work for each.

In principle, we would now like to compute the Fourier transform
of this amplitude with respect to the $\tilde\lambda$'s and show
that it is supported on curves of genus zero and degree two in
twistor space.  How to actually compute that Fourier transform is
not clear. Happily, a short cut is available. The property we want
to prove can be proved without actually computing the Fourier
transform. It is equivalent to certain differential equations
obeyed by the momentum space amplitudes $\widehat A$.

First, let us describe what genus zero, degree two curves look like.
We simply impose a linear and a quadratic equation in the homogeneous
coordinates of $\Bbb{RP}^3$ or $\Bbb{CP}^3$:
\eqn\bumgo{\eqalign{\sum_{I=1}^4a_IZ^I&=0\cr
              \sum_{I,J=1}^4 b_{IJ}Z^IZ^J&=0.\cr}}
Here $a_I$ and $b_{IJ}$ are generic real parameters.
The degree of the zero set $C$ of these equations is the product of the degrees
of the equations, or $1\cdot 2=2$.  The complex Riemann surface in $\Bbb{CP}^3$
defined by these equations has genus zero.  Conversely, any genus zero, degree
two curve is of this form.  The second equation in \bumgo\
 can be simplified by solving
the first for one of the coordinates.  For example, generically
$a_4\not= 0$, in which case we can solve the first for $Z_4$.
Eliminating $Z_4$ from the second equation gives an equation
\eqn\tumgo{\sum_{I,J=1}^3c_{IJ}Z^IZ^J=0} with some new
coefficients $c_{IJ}$. Since only three homogeneous coordinates
enter, this equation describes a curve in $\Bbb{RP}^2$.  As we did
at the end of section 3.1 for curves of degree one, we can give a
more elementary description of the curve defined by this equation
 if we go to affine coordinates.
We throw away the subset of $\Bbb{RP}^2$ with (for example)
$Z_1=0$, and introduce affine coordinates $x=Z^2/Z^1$, $y=Z^3/Z^1$
that parameterize a real two-plane $\Bbb{R}^2$.    Then \tumgo\
becomes a quadratic equation in $x$ and $y$, so the solution set,
in terminology possibly familiar from elementary algebra, is a
conic section of the plane.  Thus, we may refer to the degree two
curves as conics or conic sections, while the degree one curves,
in the same sense, are straight lines.

We would like to prove that the Fourier transform of the amplitude
\torfgo\ is supported on configurations of five points $P_i
=(\lambda_i,\mu_i)$ in $\Bbb{RP}^3$ that obey the equations in
\bumgo\ for some set of the coefficients. Since the zero set of
the first equation is an $\Bbb{RP}^2\subset\Bbb{RP}^3$, we need to
prove first of all that the $P_i$ are contained in a common
$\Bbb{RP}^2$. {\it A priori}, we also need to prove that the five
points are contained in a common conic section of $\Bbb{RP}^2$.
However, for the five point function this is trivial.  It is
always possible to pick the six coefficients $c_{IJ}$ to obey the
five linear equations \eqn\umgo{\sum_{I,J=1}^3c_{IJ}Z_i^IZ_i^J=0,
~~ i=1,\dots,5.} Since the equations are homogeneous, it follows
that for generic $P_i$, they uniquely determine the $c_{IJ}$ up to
an overall scaling. Such a scaling does not affect the curve
defined in \tumgo, so
 a generic set of five points
in $\Bbb{RP}^2$ (or $\Bbb{CP}^2$) is contained in a unique conic.
So for the five particle amplitudes, we only need to verify that
the five points are contained in an $\Bbb{RP}^2$.

It may still seem that we need to compute the Fourier transform of
the amplitude, but this can be avoided as follows.  Consider any
four points $Q_\sigma$ in $\Bbb{RP}^3$ with homogeneous
coordinates $Z_\sigma^I$, $\sigma=1,\dots,4$.
  The $Q_\sigma$ are contained in an $\Bbb{RP}^2$ if and
only if (regarding $Z_\sigma^I$, for fixed $\sigma$, as the
coordinates of a vector in $\Bbb{R}^4$) the vectors $Z_\sigma^I$
are linearly dependent. The condition for this is that the matrix
$Z_\sigma^I$ of the coefficients of these four vectors has
determinant zero.  The condition, in other words, is that the
twistor space amplitude is supported at $K=0$, where $K$ is the
determinant of the $4\times 4$ matrix with entries $Z_\sigma^I$:
\eqn\tomcon{K=\epsilon_{IJKL}Z_1^IZ_2^JZ_3^KZ_4^L.}

If we take the four points $Q_\sigma$ to be the points
$P_i,P_j,P_k$, and $P_l$ associated with four of the external
gluons, $K$ becomes a function of the twistor coordinates of the
gluons that we will call $K_{ijkl}$. We want to show that the
twistor space amplitude $\hat A$ is supported where each of the
functions $K_{ijkl}$ vanishes. In fact we will show that
\eqn\bomcon{K_{ijkl}\widehat A=0.}

Since the Fourier transform to twistor space is difficult, we want
to evaluate this condition in momentum space, where the twistor
coordinates $Z^I=(\lambda,\mu)$ are represented  as
$Z^I=(\lambda^a,-i\partial/\partial\tilde\lambda^{\dot a})$. So
$K$ can be interpreted as a differential operator in $\lambda$ and
$\tilde\lambda$.  In fact, this operator is homogeneous of degree
two in the $\lambda$'s and of degree two in the $\tilde\lambda$';
it is \eqn\boncon{K={1\over
4}\sum_{\sigma_1,\dots,\sigma_4}\epsilon_{\sigma_1\dots\sigma_4}
\langle\lambda_{\sigma_1},\lambda_{\sigma_2}\rangle \epsilon^{\dot
a\dot b} {\partial^2\over \partial\tilde\lambda_{\sigma_3}^{\dot
a}\partial \tilde\lambda_{\sigma_4}^{\dot b}}.}

So we do not need to actually compute the Fourier transform.  It
suffices to show that  the differential operators $K$ annihilate
the momentum space scattering amplitudes $\widehat A$ of helicites
$--+++$ and $-+-++$.

This assertion is true, and is simple enough that it can be proved
by hand, in contrast with a variety of other statements  made
later whose verification required computer assistance.

\bigskip\noindent{\it A Preliminary Simplification}

In doing so, it is very helpful to make a preliminary
simplification that we can describe informally as using conformal
invariance to set $P_1=(1,0,0,0)$ and $P_2=(0,1,0,0)$.  The full
procedure is a little more elaborate.  We will go through the
steps in detail as this procedure will be useful in other examples
as well.

First of all, the conformal group $SL(4,\Bbb R)$ contains an
$SL(2,\Bbb R)$ subgroup that acts on the $\lambda$'s.   We can use
this group plus overall scaling of the homogeneous coordinates
to set \eqn\uvu{\lambda_1=(1,0),~~\lambda_2=(0,1).} The
conformal group also contains a subgroup which in Minkowski space
is the group of translations of the spatial coordinates $x^{a\dot
a}$, and which  on the twistor variables is generated by
$\lambda^a\partial/\partial\mu^{\dot a}$. It acts as $\mu_{\dot
a}\to \mu_{\dot a}+x_{a\dot a}\lambda^a$.  We can  use this to set
\eqn\guvu{\mu_1=\mu_2=0.} Of course, \uvu\ and \guvu\ together are
equivalent to  \eqn\ycy{P_1=(1,0,0,0),~~ P_2=(0,1,0,0).} The
twistor amplitude with $\mu_1=\mu_2=0$ is \eqn\untu{\tilde
A(\lambda_i,\mu_i)=i\int {d\tilde\lambda_1\over (2\pi)^2}\dots
{d\tilde\lambda_n\over
(2\pi)^2}\exp\left(i\sum_{j=3}^n\mu_{j\,\dot
a}\tilde\lambda_j^{\dot
a}\right)(2\pi)^4\delta^4\left(\sum_{i=1}^n\lambda_i^a\tilde\lambda_i^{\dot
a}\right)A(\lambda_i,\tilde\lambda_i).} The integrals over
$\tilde\lambda_1$ and $\tilde\lambda_2$ can be done with the aid
of the delta function.  The delta function sets
\eqn\bunto{\eqalign{\tilde\lambda_1^{\dot
a}&=-\sum_{j=3}^n\lambda_j^1\tilde\lambda_j^{\dot a}\cr
\tilde\lambda_2^{\dot
a}&=-\sum_{j=3}^n\lambda_j^2\tilde\lambda_j^{\dot a}.\cr}} After
eliminating $\tilde\lambda_1$ and $\tilde\lambda_2$, the  twistor
amplitude reduces to a Fourier transform with respect to the
remaining $\tilde\lambda$'s: \eqn\zunot{\tilde A'(\lambda_i,\mu_i)
=i\int {d\tilde\lambda_3\over (2\pi)^2}\dots
{d\tilde\lambda_n\over
(2\pi)^2}\exp\left(i\sum_{j=3}^n\mu_{j\,\dot
a}\tilde\lambda_j^{\dot a}\right)A'(\lambda,\tilde\lambda).} Here
the symbol $ \tilde A'$ refers to $\tilde A$ with \uvu\ and \guvu\
imposed, while $A'$ is $A$ with \uvu\ and \bunto\ imposed.

Now suppose that we want to determine whether the twistor
amplitude $\tilde A$ vanishes when multiplied by some polynomial
function $X(\lambda,\mu)$.  The amplitude $\tilde A$ is $SL(4,\Bbb
R)$ invariant.  The one example we have so far of a polynomial
that should annihilate a twistor amplitude is the polynomial $K$
defined in  \tomcon; it is $SL(4,\Bbb R)$ invariant. In some later
examples, we will meet polynomials $X$ that should annihilate an
amplitude and are not $SL(4,\Bbb R)$ invariant. Even when this
occurs, the family of functions $X_i$ that should annihilate the
amplitude is $SL(4,\Bbb R)$-invariant.

To justify a claim that $X\hat A=0$, where $X$ and $\hat A$ are
$SL(4,\Bbb R)$-invariant, it suffices to show that this claim is
valid when \uvu\ and \guvu\ are imposed.  The same is true for
justifying a claim that $X_i\hat A=0$, where $\hat A$ is
$SL(4,\Bbb R)$-invariant, and $X_i$ ranges over an $SL(4,\Bbb
R)$-invariant family of polynomials.

We can use equation \zunot\ to evaluate $X\hat A$ (or $X_i\hat A$)
on the locus with $P_1=(1,0,0,0)$ and $P_2=(0,1,0,0)$.  On the
right hand side of \zunot, $\mu_j$, $j>2$, is equivalent to
$-i\partial/\partial\tilde\lambda_j$. So,  once we eliminate
$\tilde\lambda_i, $ $i=1,2$ by using \bunto, we can convert $X$ to
a differential equation $ X'$ acting on $A'$ by naively setting
$\mu_1=\mu_2=0$ and
\eqn\ixono{\mu_j=-i{\partial\over\partial\tilde\lambda_j},~~j>2.}
The condition $X\hat A=0$ for the twistor amplitude is equivalent
to the differential equation $ X'  A'=0$ for the reduced and
restricted amplitude $A'$ in momentum space.

What have we gained in this process? We have reduced the number of
variables, and obtained a simpler differential equation that is
equivalent to the original one.  The one subtlety in this
procedure, and the reason that we have described it at some
length, is that to arrive at this result, one must regard
$\tilde\lambda_i$, $i=1,2$ as functions of the $\tilde\lambda_j$,
$j>2$, via  \bunto. This renders more complicated the reduced
function $A'$ and the action of the derivatives
$\partial/\partial\tilde \lambda_j, $ $j>2$.

\bigskip\noindent{\it Verification}

We will now verify by the above procedure that (for example)
$K_{1234}\hat A=0$.  Upon setting $P_1=(1,0,0,0)$ and
$P_2=(0,1,0,0)$, $K_{1234}$ reduces to
\eqn\unco{K_{34}=\epsilon^{\dot a\dot b}{\partial^2\over\partial
\tilde\lambda_3^{\dot a}\partial\tilde\lambda_4^{\dot b}}.} Since
$\tilde\lambda_1$ and $\tilde\lambda_2$ have been eliminated, the
reduced momentum space amplitude $ A'$ is a function only of the
$\tilde\lambda_i$ with $i\geq 3$ (as well as the $\lambda$'s).
Because of $SL(2,\Bbb R)$ symmetry acting on the
$\tilde\lambda$'s, the dependence on the $\tilde\lambda$'s is only
via $a=[\tilde\lambda_3,\tilde\lambda_4]$,
$b=[\tilde\lambda_3,\tilde\lambda_5]$, and
$c=[\tilde\lambda_4,\tilde\lambda_5]$. Moreover, $A'$ is
homogeneous in $a,b,$ and $c$ of degree $-1$:
\eqn\funco{\left(a{\partial\over\partial
a}+b{\partial\over\partial b}+c{\partial\over\partial c}\right)
A'=- A'.}  This follows directly from the homogeneity of the full
momentum space amplitude $\widehat A$ in \torfgo\ as well as the
fact that the equations used to solve for $\tilde\lambda_1$ and
$\tilde\lambda_2$ are homogeneous.

A short computation using the chain rule shows that acting on any
function $F(a,b,c)$, \eqn\mimbo{K_{34}F=-2{\partial F\over\partial
a}-a{\partial^2 F\over\partial a^2}-b{\partial^2F\over \partial
a\partial b}-c{\partial^2F\over\partial a\partial c}.}    The
right hand side can be written \eqn\zimbo{-{\partial\over\partial
a}\left(a{\partial F\over\partial a}+b{\partial F\over\partial
b}+c{\partial F\over\partial c}+F\right),} and so vanishes for any
function that obeys \funco.

Thus, we have demonstrated that $K_{1234}\widehat A=0$.  Nothing
essentially new is needed to show that $K_{ijkl}\widehat A=0$ for
all $i,j,k,l$; one just uses conformal invariance to set (for
example) $P_i=(1,0,0,0)$ and  $P_j=(0,1,0,0)$, and then proceeds
as above.

\subsec{The Six Gluon Amplitude With Three Positive And Three Negative Helicities}

\nref\mpxsix{M. Mangano, S. Parke, and Z. Xu, ``Duality And Multi-Gluon
Scattering,'' Nucl. Phys. {\bf B298} (1988) 653. }%
Continuing our study of tree amplitudes associated with curves of
degree two, the next case is the six gluon amplitudes with three
positive and three negative helicities. These were first computed
by Mangano, Parke, and Xu \mpxsix\
 and by Berends
and Giele \whoever\ and are quite complicated.  There are three
essentially different cases, namely helicities $+++-\,-\,-$,
$++--+-$, or $+-+-+-$. These amplitudes can all be written
\eqn\longamp{\eqalign{A=8g^4&\left[{\alpha^2\over
t_{123}s_{12}s_{23}s_{45}s_{56}}+{\beta^2\over
t_{234}s_{23}s_{34}s_{56}s_{61}}\right.\cr&  \left.+{\gamma^2\over
t_{345}s_{34}s_{45}s_{61}s_{12}}+{t_{123}\beta\gamma+t_{234}\gamma\alpha+
t_{345}\alpha\beta \over
s_{12}s_{23}s_{34}s_{45}s_{56}s_{61}}\right].\cr}} with
$s_{ij}=(p_i+p_j)^2$, $t_{ijk}=(p_i+p_j+p_k)^2$. The functions
$\alpha$, $\beta$, and $\gamma$ are different for the different
helicity orderings.  They are presented in the table.

$$\vbox{\hsize=4.5in\noindent
Table 1.  Coefficients for six gluon amplitudes with three
helicities of each type (table from \mpxsix).  The symbol $\langle
I|T|J\rangle $ is here short for $[IT]\langle TJ\rangle$; for
$T=T_1+T_2+T_3$, a sum over the $T_i$ is understood. The notation
$\langle i\,j\rangle $ is used for
$\langle\lambda_i,\lambda_j\rangle$, and  $[i\,j]$ for
$[\tilde\lambda_i,\tilde\lambda_j]$.
\smallskip
\halign to 4.5in{#\hfil\hfil\tabskip=5em plus5em
minus5em&\hfil#\hfil\tabskip=5em plus5em
minus5em&\hfil#\hfil\tabskip=5em plus5em
minus5em&\hfil#\hfil\tabskip=0pt\cr
\noalign{\medskip\hrule\smallskip\hrule\medskip}
&$1^{+}2^{+}3^{+}4^{-}5^{-}6^{-}$&
$1^{+}2^{+}3^{-}4^{+}5^{-}6^{-}$&$1^{+}2^{-}3^{+}4^{-}5^{+}6^{-}$\cr
&$X=1+2+3$&$Y=1+2+4$&$Z=1+3+5$\cr \noalign{\medskip\hrule\medskip}
$\alpha$&0&$-[12]\langle56\rangle\langle 4 \vert Y \vert
3\rangle$&$[13]\langle46\rangle\langle5 \vert Z \vert 2\rangle$\cr
$\beta$&$[23]\langle56\rangle\langle 1 \vert X \vert
4\rangle$&$[24]\langle56\rangle\langle 1 \vert Y \vert 3\rangle$
&$[51]\langle24\rangle\langle 3 \vert Z \vert 6\rangle$\cr
$\gamma$&$[12]\langle45\rangle\langle 3 \vert X \vert 6\rangle$
&$[12]\langle35\rangle\langle 4 \vert Y \vert 6\rangle$
&$[35]\langle62\rangle\langle 1 \vert Z \vert 4\rangle $\cr
\noalign{\medskip\hrule} }}
$$
\vskip 1cm

Our conjecture says again that these amplitudes should be
supported on configurations in which all six points $P_i$ labeling
the external particles lie on a common genus zero degree two curve
or conic in $\Bbb{RP}^3$.  First of all, to show that the six
points are contained in an $\Bbb{RP}^2$ subspace, we must establish
that the amplitudes are annihilated by the differential operator
$K$ defined in \boncon, where the $Q_\sigma$, $\sigma=1,\dots,4$,
may be any of the six points $P_i$.  This was verified with some
computer assistance, after simplifying the problem as in section
3.3 by using conformal symmetry to set $P_1=(1,0,0,0)$ and
$P_2=(0,1,0,0)$.

Next, we need to show that the six points are contained not just
in an $\Bbb{RP}^2$ but in a conic section therein. This means that
it must be possible to pick the coefficients $c_{IJ}$ in \tumgo\
so that the equations \eqn\zumgo{\sum_{I,J=1}^3c_{IJ}Z_i^IZ_i^J=0,
~~ i=1,\dots,6} are obeyed.  In contrast to the five gluon case
that we considered in section 3.3, here we have six homogeneous
equations for six unknowns, so for a generic set of points $P_i$,
a nonzero solution for the $c_{IJ}$ does not exist. Existence of a
nonzero solution is equivalent to vanishing of the determinant of
the $6\times 6$ matrix of coefficients in this equation. With
$(Z^1,Z^2,Z^3)=(\lambda^1,\lambda^2,\mu^1)$, this determinant is
\eqn\xumgo{\widehat V=\det\left(\matrix{
(\lambda_1^1)^2&\lambda_1^1\lambda_1^2&(\lambda_1^2)^2&\lambda_1^1\mu_1^1&
\lambda_1^2\mu_1^1 & (\mu_1^1)^2\cr
(\lambda_2^1)^2&\lambda_2^1\lambda_2^2&(\lambda_2^2)^2&\lambda_2^1\mu_2^1&
\lambda_2^2\mu_2^1 & (\mu_2^1)^2\cr
(\lambda_3^1)^2&\lambda_3^1\lambda_3^2&(\lambda_3^2)^2&\lambda_3^1\mu_3^1&
\lambda_3^2\mu_3^1 & (\mu_3^1)^2\cr
(\lambda_4^1)^2&\lambda_4^1\lambda_4^2&(\lambda_4^2)^2&\lambda_4^1\mu_4^1&
\lambda_4^2\mu_4^1 & (\mu_4^1)^2\cr
(\lambda_5^1)^2&\lambda_5^1\lambda_5^2&(\lambda_5^2)^2&\lambda_5^1\mu_5^1&
\lambda_5^2\mu_5^1 & (\mu_5^1)^2\cr
(\lambda_6^1)^2&\lambda_6^1\lambda_6^2&(\lambda_6^2)^2&\lambda_6^1\mu_6^1&
\lambda_6^2\mu_6^1 & (\mu_6^1)^2\cr
                               }\right).}
(The subscripts $1,\dots ,6$ label the six gluons, while the
superscripts refer to the component $a$  or $\dot a$ of
$\lambda^a$ or $\mu^{\dot a}$ for each gluon.) Upon interpreting
$\mu$ as $-i\partial/\partial\tilde\lambda$, $\widehat V$ becomes
a fourth order differential operator that should annihilate the
six gluon amplitudes with three positive helicities. This
statement appears too complicated to check by hand and was
verified with computer assistance.  A preliminary simplification
was again made by using
 conformal invariance to fix the point $P_1$
to have coordinates $(1,0,0,0)$ and $P_2$ to have coordinates
$(0,1,0,0)$. Upon doing so, $\widehat V$ reduces to the
determinant of a $4\times 4$ matrix
\eqn\pumgo{V=\det\left(\matrix{ \lambda_3^1\lambda_3^2 &
\lambda_3^1\mu_3^1 & \lambda_3^2\mu_3^1 & (\mu_3^1)^2\cr
\lambda_4^1\lambda_4^2 & \lambda_4^1\mu_4^1 & \lambda_4^2\mu_4^1 &
(\mu_4^1)^2\cr \lambda_5^1\lambda_5^2 & \lambda_5^1\mu_5^1 &
\lambda_5^2\mu_5^1 & (\mu_5^1)^2\cr \lambda_6^1\lambda_6^2 &
\lambda_6^1\mu_6^1 & \lambda_6^2\mu_6^1 & (\mu_6^1)^2\cr
}\right).} $V$ again is interpreted via
$\mu_j=-i\partial/\partial\tilde\lambda_j$, $j>2$, as a fourth
order differential operator that should annihilate the reduced
momentum space amplitudes $ A'$. As the computer program was
unreasonably slow, the vanishing of $V A'$ was verified as a
function of $\tilde\lambda_i^1$, $i=3,\dots,6$, with the other
variables set to randomly selected values.

\bigskip\noindent{\it Remaining Six Gluon Amplitudes}

By now, we have shown that, in accord with our general conjecture,
the six gluon tree level amplitudes with two negativeive
helicities are supported on lines, and those with three negative
helicities are supported on conics.  The remaining six gluon
amplitudes are those with four negative helicity gluons.  Our
conjecture asserts that these amplitudes should be supported on
curves of genus zero and degree three, which are called twisted
cubic curves. This statement is trivial, however, as any six
points in $\Bbb{RP}^3$ lie on some twisted cubic.  A specific
string theory proposal (such as we will make in section 4) may
lead to a new way to understand the $----++$ amplitudes, but there
is no content in merely saying that they are supported on twisted
cubics. Since a generic set of seven points does not lie on a
twisted cubic, the seven gluon tree amplitude with helicities
$----+++$ should obey interesting differential equations related
to twisted cubics. (Seven gluon tree amplitudes have been computed
in \ref\otherber{F. A. Berends, W. T. Giele, and H. Kuijf, ``Exact
And Approximate Expressions For Multi-Gluon Scattering,'' Nucl.
Phys. {\bf B333} (1990) 120.}.) This question will not be
addressed here.

\subsec{$---++$ and $--+-+$ Amplitudes Revisited}

By now, we have obtained what may seem like a tidy story for the
five and six gluon amplitudes with three negative helicities.
However, further examination, motivated by the string theory
proposal in section 4 as well as the preliminary examination of
one-loop amplitudes that we present in section 3.6, has shown that
the full picture is more elaborate and involves disconnected
instantons. Here we will re-examine the $---++$ and $--+-+$ tree
level amplitudes to consider such contributions. (It would be
desireable to similarly re-examine the six gluon amplitudes, but
this will not be done here.)

We so far interpreted these five gluon amplitudes in terms of genus zero
curves of degree two.  In string theory, these curves will be interpreted as
instantons. The action of an instanton of degree two is precisely twice
the action of a degree one instanton.  It therefore has precisely the
same action as a pair of separated degree one instantons.  Might
the $---+++$ and $--+-+$ amplitudes receive contributions from configurations
with two separated instantons of degree one?

\ifig\twoar{In part (a), we depict two different straight lines in
$\Bbb R^3$, representing two disjoint curves of genus zero and
degree one.  A twistor field, represented by a curved dotted line
which we call the internal line, is exchanged between them.
Various points on the two lines, including the endpoints of the
internal line, are labeled by $+$ or $-$ helicity.  There are two
$-$ helicities on each line. (b) Here we give a complex version of
the same picture.  The lines of part (a) are replaced by
two-spheres, and the internal line becomes a thin tube connecting
them.  The whole configuration is topologically a two-sphere. }
{\epsfbox{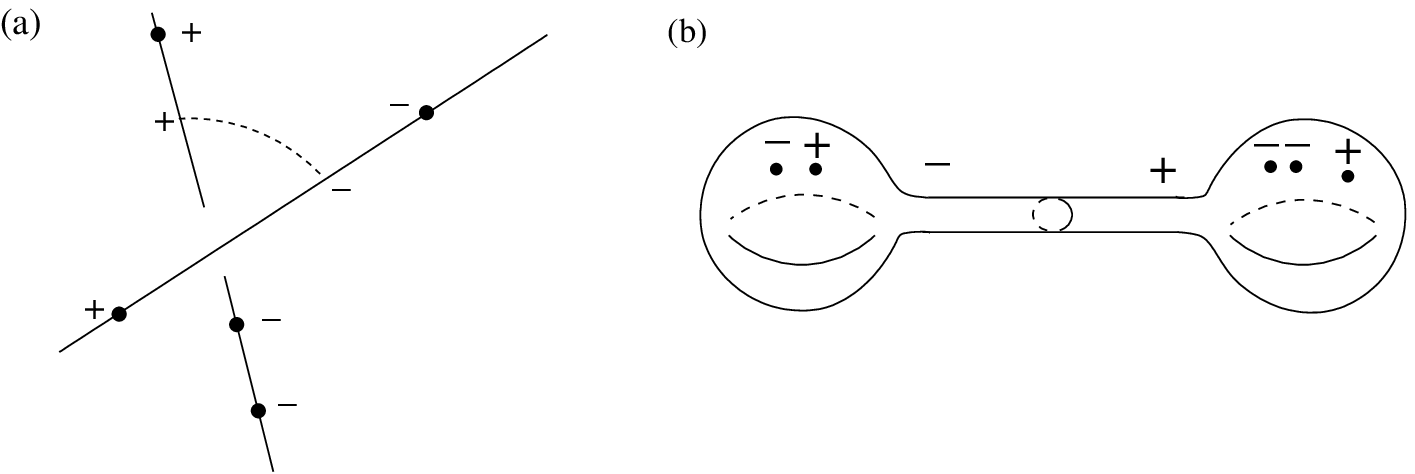}} In figure 3, we sketch two different
pictures of a configuration with two widely separated instantons
of degree one.  Figure 3a contains a view of this situation in
real twistor space.  The degree one instantons are represented as
straight lines.  We have attached the five external gluons to the
two instantons. In the example sketched, we attached helicities
$-\,+$ to one side and $--+$ to the other. Since we are trying to
construct a connected amplitude, we also assume that a twistor
space field of some kind is exchanged between the two instantons.
What it might be will be clearer in section 4. We assume that this
field carries negative helicity at one end and positive helicity
at the other. (The helicities are reversed between the two ends
because we consider all fields attached to either instanton to be
outgoing and because crossing symmetry relates an incoming gluon
of one helicity to an outgoing gluon of opposite helicity.)
Propagation of the twistor space field between the two instantons
is shown in figure 3a by connecting them via an internal line
(shown as a dotted line in the figure). A degree one instanton
must have exactly two $-$ helicities attached to it. In figure 3a,
after distributing the external particles between the two
instantons, we labeled
 the ends of the internal line such that this
condition is obeyed.

Note that the number of internal lines must be precisely one or we
would end up with too many $-$ helicities on one instanton or the
other. In figure 3b, we try to give another explanation of what
this means. Here we consider the instantons as curves in complex
twistor space, so a degree one curve of genus zero is represented
as a $\Bbb{CP}^1$, or topologically a two-sphere.  We also assume
that the internal line connecting the two $\Bbb{CP}^1$'s
represents a collapsed limit of a cylinder (exchange of a closed
string -- presumably a $D$-string in the proposal of section 4).
So in figure 3b, we have drawn two $\Bbb{S}^2$'s connected by a
single narrow tube, corresponding to the internal line in figure
3a. Two two-spheres joined this way make a surface of  genus zero,
and we regard this as a degenerate case of a Riemann surface of
genus zero that can contribute to a tree level scattering
amplitude. If we connect the two instantons with more than one
narrow tube, we get a Riemann surface of genus one or higher,
which should be considered as a contribution to a one-loop
scattering amplitude.

The reasoning in the last two paragraphs is certainly not meant to
be rigorous, but hopefully it will encourage the reader to follow
along with us in contemplating differential equations that the
$---++$ and $--+-+$ amplitudes might obey reflecting their support
on configurations like that in figure 3a. The salient aspect of
figure 3a is that three of the external particles are attached to
the same straight line, or degree one curve.  In other words, of
the five twistor points representing the external particles, three
are collinear.

In figure 3, we assume that the field represented by the internal line
 transforms in the adjoint
representation of the gauge group.  To get a single trace
amplitude whose group theory factor is $\Tr \,T_1T_2T_3T_4T_5$,
the particles must be divided between the two instantons in a way
that preserves the cyclic order -- for example, 12 on one side and
345 on the other or 34 on one side and 512 on the other. The color
flow is then as shown in figure 4.

\ifig\colorflow{This diagram is intended to show the color flow in
figure 3.  We suppose that in a dual Yang-Mills or open string
description, each line (or genus zero, degree one curve) in figure
3 corresponds to a disc with gluons cyclically attached on the
boundary.  Moreover, the internal line becomes a small strip
connecting the two discs.  The overall figure is thus
topologically a disc.  It can contribute to a single trace
amplitude (such  as a Yang-Mills tree amplitude) if the internal
line is in the adjoint representation of the gauge group.  The
amplitude is proportional to $\Tr \,T_1T_2\dots T_n$, where $T_i$
is a $U(N)$ generator of the $i^{th}$ particle, and the particles
attached to either disc are consecutive with respect to the cyclic
order. } {\epsfbox{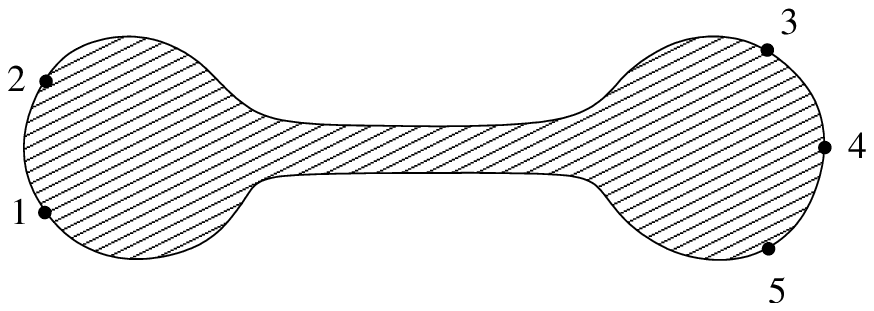}}

We will now construct a differential operator $F_{ijk}$ that
annihilates amplitudes in which the points $P_i$, $P_j$, and $P_k$
are collinear. Once this is done, we can construct an operator
that annihilates amplitudes in which {\it any} three consecutive
points are collinear by simply forming the product \eqn\huv{\hat
F=F_{123}F_{234}F_{345}F_{451}F_{512}.}  Thus, the operator $\hat
F$ should annihilate any amplitude that arises from the sort of
configuration sketched in figure 3.

If $Q_\sigma$, $\sigma=1,2,3$ are three points in twistor space
with homogeneous coordinates $Z_\sigma^I$, then the condition that
the $Q_\sigma$ are collinear is that\foot{For example, if we set
$L=4$ and introduce affine coordinates $x^I=Z^I/Z^4$, $I=1,2,3$,
then three points $Q_\sigma\in \Bbb{R}^3,$ $\sigma=1,2,3$, with
coordinates $x_\sigma^I$, are collinear if and only if
$\epsilon_{IJK}(x_1-x_2)^I(x_2-x_3)^J=0$.  This can be rewritten
as $\epsilon_{IJK}(x_1^Ix_2^J+x_2^Ix_3^J+x_3^Ix_1^J)=0$, and in
that form is readily compared to the following equation in
homogeneous coordinates. }
\eqn\hyfcon{\epsilon_{IJKL}Z_1^IZ_2^JZ_3^K=0,~~~L=1,\dots,4.}
Setting
$Z=(\lambda,\mu)=(\lambda,-i\partial/\partial\tilde\lambda)$, the
expressions on the left hand side of \hyfcon\ become, in the usual
way, differential operators that act on the momentum space
amplitudes. For example, if we set $L=4$, then we get an operator
\eqn\nuv{F_{ijk}=\langle\lambda_i,\lambda_j\rangle{\partial\over\partial\tilde
\lambda_k^1}+\langle\lambda_k,\lambda_i\rangle{\partial\over\partial\tilde
\lambda_j^1}+\langle\lambda_j,\lambda_k\rangle{\partial\over\partial\tilde
\lambda_i^1}} that annihilates amplitudes in which the points
$P_i$, $P_j$, and $P_k$ are collinear. Inserting this definition
of $F_{ijk}$ in \huv, we get a differential operator $\hat F$ that
should annihilate amplitudes in which any three consecutive points
are collinear.

It is not hard to verify (by computer) that the $---++$ and $--+-+$
amplitudes are indeed annihilated by
$\hat F$.  (The computation is again simplified by using $SL(4)$
to set $P_1=(1,0,0,0)$ and $P_2=(0,1,0,0)$.) One may  wonder if
these amplitudes are annihilated by a simpler operator obtained by
omitting some of the five factors in $\hat F$.  A little
experimentation reveals that the $--+-+$ amplitude is not
annihilated by the product of any four of the five factors in
$\hat F$, while the $---++$ amplitude is annihilated by
$F_{234}F_{345}F_{451}F_{512}$.  In other words, we can omit the
factor $F_{123}$ from $\hat F$ and get an operator that still
annihilates the $---++$ amplitude.  This fact has a simple
interpretation.  The role of $F_{123}$ is to annihilate
contributions in which points $P_1,$ $P_2$, and $P_3$ are attached
to the same degree one curve.  But for helicities precisely
$---++$ in that order, these particular contributions vanish
anyway since, as in this case the gluons attached at $P_1$, $P_2$,
and $P_3$ all have negative helicity, this configuration has too
many negative helicity gluons attached to a curve of degree one.

\bigskip\noindent{\it Interpretation}

At this point, we really should pause to discuss the interpretation of
these results.  We have found that the same five gluon amplitudes are annihilated
both by an operator $K$ associated with connected curves of degree two
and by an operator $\hat F$ associated with disconnected pairs of degree one
curves.  Does this indicate a duality wherein the same amplitude can be
computed either using connected degree two curves or using the disconnected
pairs?

I believe that actually the amplitude is a sum of the two types of
contribution, and that the reality is  more mundane.  The
amplitudes $\hat A$ that we are exploring have various
singularities, and hence one should consider the possibility  of
delta function contributions in $K\hat A$ and $\hat F\hat A$. I
suspect that $K\hat A$ is not quite zero but contains delta
functions that are annihilated by $\hat F$, and vice-versa. I will
not try to justify this statement here, but I will give a simple
example of a singularity that leads to a delta function
contribution. Consider the expression \eqn\ufox{\epsilon^{\dot
a\dot b}{\partial^2\over\partial\tilde\lambda_i^{\dot
a}\partial\tilde\lambda_j^{\dot b}}\left({1\over
[\tilde\lambda_i,\tilde\lambda_j]}\right),} which is a simplified
example of the sort of contribution we meet in evaluating $K\hat
A$. According to the formulas  used at the end of section 3.3 in
proving that $K\widehat A=0$, this appears to vanish, but actually
it is a multiple of $\delta^2(\tilde \lambda_i)\delta^2(\tilde\lambda_j)$.  In
fact, we can regard the four components of
$\tilde\lambda_i^{\dot a}$ and $\tilde\lambda_j^{\dot a}$
as coordinates of $\Bbb R^4$.  The differential operator
$L=\epsilon^{\dot a\dot
b}{\partial^2\over\partial\tilde\lambda_i^{\dot
a}\partial\tilde\lambda_j^{\dot b}}$ is then the Laplacian of
$\Bbb R^4$, endowed with a suitable metric $g$ of signature
$++-\,-$. Thus, if we combine together the $\tilde\lambda_i^{\dot a}$
and $\tilde\lambda_j^{\dot a}$ to coordinates $x^\alpha$,
$\alpha=1,\dots,4$ of $\Bbb R^4$, $L$ becomes the Laplacian
$g^{\alpha\beta}\partial^2/\partial x^\alpha \partial x^\beta$. In
the same notation, $1/[\tilde\lambda_i,\tilde\lambda_j]$ becomes
$1/g_{\alpha\beta}x^\alpha x^\beta$, which is the usual propagator
or Green's function of the Laplacian of $\Bbb{R}^4$.  So in acting
on $1/[\tilde\lambda_i,\tilde\lambda_j]$, the differential
operator $L$ produces a delta function supported at the origin. It
seems plausible that analogous delta function terms appear in more
carefully evaluating $K\widehat A$ or $\widehat F \widehat A$.

\subsec{A Few One-Loop Amplitudes}

To conclude this exploration of some perturbative Yang-Mills amplitudes,
we would like to at least glimpse a few of the simplest issues concerning some
one-loop amplitudes.  The obvious one-loop amplitudes to look at first
are the planar MHV amplitudes with precisely two negative helicity gluons in
${\cal N}=4$ super Yang-Mills theory.  General and relatively simple
formulas are known for these amplitudes \ref\simpamp{Z. Bern, L. Dixon, D. Dunbar,
and D. A. Kosower, ``One-Loop $n$-Point Gauge Theory Amplitudes, Unitarity
And Collinear Limits,'' hep-ph/9403226.}.

Until this point, it has not generally mattered if we  contemplate pure
Yang-Mills theory or a supersymmetric extension thereof.  The
reason is that so far we have mainly limited ourselves to tree amplitudes
in which the external particles are gluons.  In such diagrams, the
internal particles (in gauge theory or its supersymmetric
extensions) are also gluons and supersymmetry simply does not
matter. For loop diagrams, supersymmetry definitely does matter as
any particle can propagate in the loop.  We will consider the
amplitudes with maximal supersymmetry, expecting them to be the
most likely ones to lead to a simple theory.

 Also, we consider planar amplitudes because
computations show that they are simpler; in fact, the analysis in
section 4 suggests that to get an equally simple result from
non-planar diagrams, one should modify the ${\cal N}=4$ theory to
include closed string contributions (which are not yet
understood).

The formula \ibob\ says that a one-loop amplitude with two gluons
of negative helicity will be associated with curves of degree two
in twistor space, since  $q=2,$ $l=1$ leads to $d=2$. Moreover,
with $l=1$, the genus of these curves will be bounded by $g\leq
1$. However, there are no curves in twistor space of genus one and
degree two. So these amplitudes will actually come from curves of
genus zero and degree two. There are two kinds of curves to
consider, both of which we have already encountered in studying
tree diagrams:

(1) There are connected curves of genus zero and degree two, consisting of conics
located in some $\Bbb{RP}^2\subset\Bbb{RP}^3$.

(2) There are disconnected curves, consisting
of a pair of degree one curves or lines.  Generically these lines are ``skew,''
not contained in any plane or $\Bbb{RP}^2$.

\ifig\twoint{The two configurations that contribute to the five
gluon amplitude in one-loop order.  In (a), we have a degree two
curve of genus zero -- the two bulges are meant merely as a
reminder that the degree is two.  An internal line -- representing
propagation of a twistor field -- connects the curve to itself.
The ends of the internal lines are labeled by $+$ or $-$ helicity,
as are the five points at which external gluons are attached. In
(b), we consider instead a configuration of two disjoint degree
one curves connected by two internal lines, whose ends are again
labeled along with the points at which external gluons are
attached.  If internal lines are replaced by thin tubes, both
configurations become topologically equivalent to Riemann surfaces
of genus one.} {\epsfbox{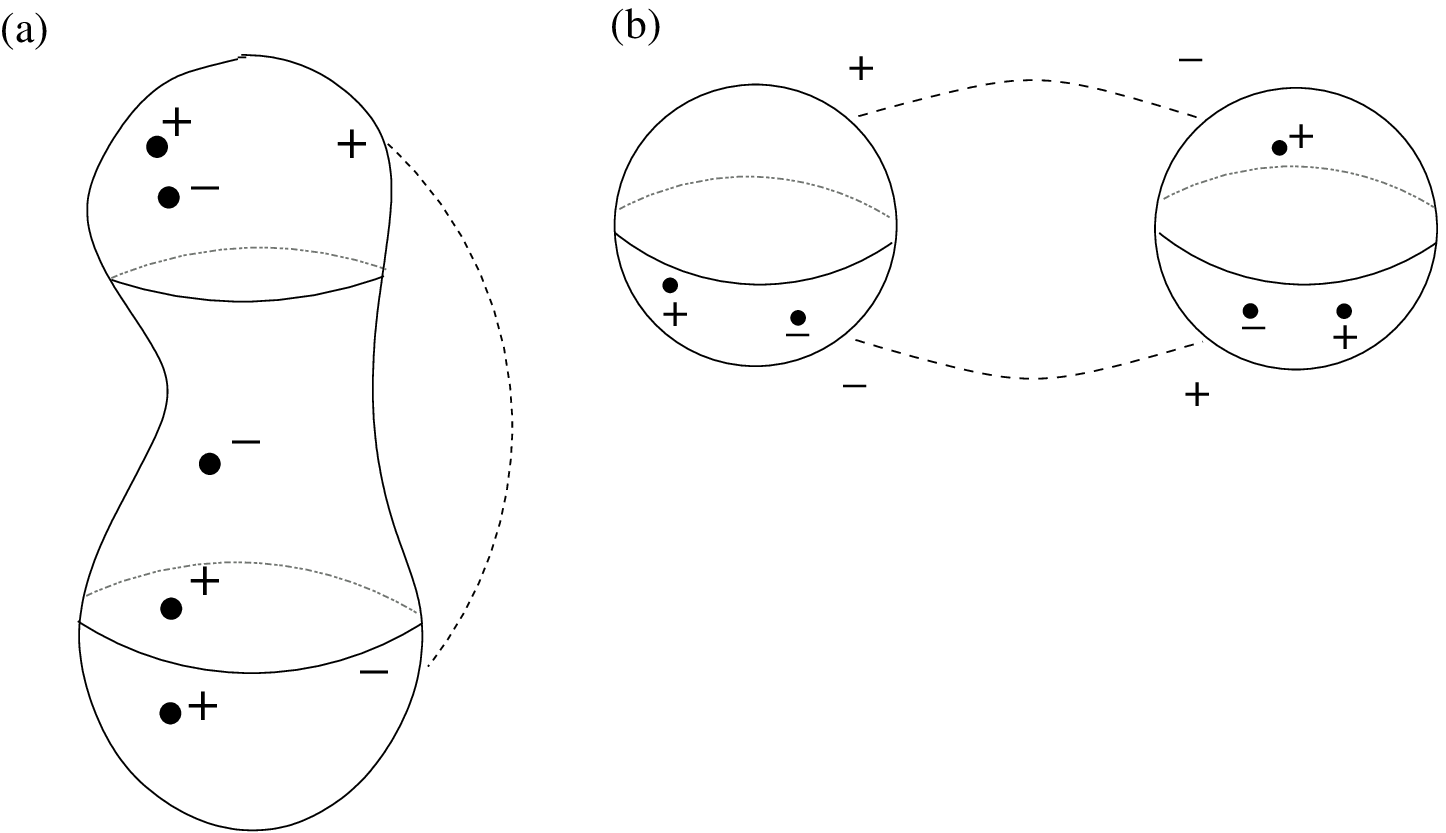}}

In figure 5, we sketch in what sense these two kinds of curve
represent degenerate cases of configurations of genus one.  In
figure 5a, we consider a degree two curve with gluons of
helicities $--+++$ attached to it, while in addition some twistor
space field (of a type that will be clearer in section 4) is
exchanged between two points on this curve. This is represented by
the ``internal line'' in the figure, which connects the curve to
itself.
 Labeling the ends of the internal line by helicities $+$ and $-$, there are a total
of three $-$ helicities on the curve in figure 5a.  As we have
seen, this is the right number for a curve of genus zero and
degree two.  In addition, if we think of the internal line as
representing a very thin tube, then the configuration in figure 5a
represents a degenerate case of a Riemann surface of genus one.

In figure 5b, the external particles have been distributed in some
way between the two degree one curves, which are also connected by
two internal lines.  Labeling the ends of the internal lines by
$+$ and $-$, we can ensure that each degree one curve has two $-$
helicities attached to it, which is the correct number.  Each
degree one curve is topologically ${\Bbb S}^2$; when we interpret
the internal lines as two very thin tubes connecting the two
${\Bbb S}^2$'s, we arrive, again, at a degenerate case of a
Riemann surface of genus one.  The number of internal lines should be exactly two
both to ensure the right number of negative helicity insertions on each side
and to get a configuration of genus one.

The simplest one-loop amplitude in the ${\cal N}=4$ theory is the
four gluon amplitude, inevitably with two positive and two
negative helicities (as the other cases vanish).  This amplitude,
however, is too simple for our purposes. The reason is that any
four points are trivially contained pairwise in two skew lines, so
we cannot expect to derive from the configurations of figure 5 any
differential equation obeyed by the four gluon amplitude at one
loop. (However, a string theory, such as the one proposed in
section 4, may lead to a new way to calculate these amplitudes.)

We move on, therefore, to the five gluon amplitudes with
helicities $--+++$ or $-+-++$.  (It would be desireable to
consider the one-loop MHV amplitudes with any number of positive
helicity gluons, but this will not be done in the present paper.)
We recall that for these MHV configurations, the tree level
amplitudes $\hat A_0$ are ``holomorphic,'' that is, they are
functions only of $\lambda$ and not $\tilde\lambda$ (times a delta
function of energy-momentum conservation). Denoting the
corresponding one-loop amplitudes as $\hat A_1$, the relation
between them is \eqn\hucon{\hat A_1=g^2\hat A_0(L_1+L_2),} with
\ref\bdkx{Z. Bern, L. Dixon, and D. Kosower, ``One Gluon Corrections To
Five Gluon Amplitudes,'' Phys. Rev. Lett. {\bf
70} (1993) 2677.}
\eqn\bucon{\eqalign{L_1 & =-{1\over
\epsilon^2}\sum_{i=1}^5\left(\mu^2\over -s_{i,i+1}\right)^\epsilon
\cr
                    L_2 & = \sum_{i=1}^5\ln\left(-s_{i,i+1}\over -s_{i+1,i+2}\right)
                    \ln\left(-s_{i+2,i+3}\over -s_{i-2,i-1}\right)+{5\pi^2\over 6}
 .\cr}}
Here $s_{i,j}=(p_i+p_j)^2$. These amplitudes have been computed
with dimensional regularization in $4-2\epsilon$ dimensions; the
pole in $L_1$ at $\epsilon=0$ is the usual infrared divergence of
the one-loop diagram.

We have at our disposal the operator $K$ which annihilates any
amplitude that has delta function support on configurations
sketched in figure 5a, and the operator $\hat F$ which annihilates
any amplitude that has delta function support on configurations
sketched in figure 5b. Note that $K$ and $\hat F$ trivially
commute with $\hat A_0$, since they contain derivatives only with
respect to $\tilde\lambda$, while $\hat A_0$, apart from the delta
function, is a function of $\lambda $ only.\foot{$K$ and $\hat F$
commute with the delta function of energy-momentum conservation
since they express geometrical relations that are invariant under
translations.  More explicitly, our usual method of implementing
differential operators such as $K$ and $\hat F$ is to make a
preliminary simplification in which we use conformal invariance to
set $P_1=(1,0,0,0)$ and $P_2=(0,1,0,0)$.  In the process, the
delta function of energy-momentum conservation is used to
eliminate $\tilde\lambda_1$ and $\tilde\lambda_2$, and the
operators such as $K$ and $\hat F$ are expressed in a way that
only involves derivatives with respect to the other
$\tilde\lambda_i$'s.  Thus, in this procedure $K$ and $\hat F$ are
reduced to simpler operators which still  involve derivatives only  with
respect to $\tilde\lambda$, and the energy-momentum delta
function is eliminated.}
 Hence
$K$ and $\hat F$ will act only on the $L$'s.

By inspection, one can see that the amplitude $L_1$ is annihilated
by $\hat F$. Indeed, $L_1$ is a sum of terms each of which depends
on the coordinates of two particles only.  But $\hat F$ is a
product of five operators, each of which contains only derivatives
acting on the coordinates of three adjacent particles. Each term
in $L_1$ is thus annihilated by one factor in $\hat F$, and hence
$\hat F L_1=0$.  So it is reasonable to interpret $L_1$ as arising
from the configurations of figure 5b with two disconnected curves
of degree one.

This observation, together with the fact that the one-loop
infrared divergence is contained entirely in $L_1$, leads to an
interesting thought.  If we specify the location in twistor space
of five points, three of which are collinear, then the choice of
the two degree one curves containing them is uniquely determined;
one passes through the three collinear points and the other is the
unique straight line through the remaining two points. There are
no moduli to integrate over. We do have to integrate over the
positions at which the internal lines in figure 5b are attached.
But these integrations run over compact parameter spaces (choices
of points on the degree one curves in the figure), and divergences
appear quite unlikely. (One can worry about short distance
divergences in twistor space, but they would not be expected in
the string model considered in section 4.) All of this strongly
suggests that at least in this example, the one-loop twistor
amplitude is free of infrared divergences, and the  usual infrared
problem comes from the Fourier transform back to momentum space.
Perhaps twistor space amplitudes are free of infrared divergences
in general.

If $L_1$ comes from the disconnected curves, perhaps $L_2$ is the contribution
of the connected curves of degree two.  This thought motivates the question
of whether $K$ annihilates $L_2$.  A small amount of computer-based inquiry
reveals that $KL_2\not=0$, but
\eqn\grify{K^2L_2=0.}
Here we should recall that $K$ is really a collection of operators $K_{ijkl}$.
The statement in \grify\ is that the product of any two of these operators
 annihilates $L_2$.

What does it mean if an amplitude is annihilated not by $K$ (or
one of the other differential operators that have appeared in our
investigation), but by its square? These operators are all
polynomial in the $\mu$'s (which are interpreted in momentum space
as $-i\partial/\partial \tilde\lambda$). Consider the simplest
case of an operator linear in $\mu$. In fact, consider the
operator $W$ of multiplication by $\mu$ (that is, by one of the
components of $\mu$ for one of the external particles). $W$
annihilates the distribution $\delta(\mu)$.  Now what is
annihilated by $W^2$ but not by $W$? The answer to this question
is that multiplication by $\mu^2$ annihilates the distribution
$\delta'(\mu)$ which is not annihilated by $\mu$. So a
distribution annihilated by $W^2$ is supported on the same set as
a distribution annihilated by $W$, but in general, rather than
delta function support, it has ``derivative of a delta function
support'' in the normal directions.  This is the general
situation, for any operator that (like all the differential operators we have
considered) is in  twistor space  a multiplication operator by
some polynomial $P(\lambda,\mu)$.  The distributions annihilated
by $P^2$ are those that have ``derivative of a delta function
support'' on the zero set of $P$. (We make this reasoning more explicit in discussing
General Relativity in  section 3.7.)

 So that is the meaning of \grify: the $L_2$ term
is supported on configurations coming from connected curves of
degree two, but has ``derivative of a delta function support''
rather than delta function support on the space of configurations
of this type.

It is not difficult to prove by hand that no power of $K$
anihilates $L_1$. So the full amplitude $\hat A_1$ is not
supported on configurations contained in an $\Bbb{RP}^2$.  But
since  $\hat FL_1=0$, the amplitude $\hat A_1=\hat A_0(L_1+L_2)$
is a sum of contributions supported on the two types of
configuration in figure 5. We can write a differential equations
that expresses this fact: \eqn\urify{0=\hat FK^2\hat A_1.}
(Again, $K^2$ refers to the product of any two components of $K$.)

That it is necessary to combine $\hat F$ and $K$ in this way should not
come as a surprise -- it is what one would guess from the
existence of the two configurations of figure 5.  What does remain
surprising is the rather different result of section 3.5 that,
modulo possible delta function terms,  the tree level $---++$
amplitude is annihilated by $\hat F$ and $K$
separately, while one might have expected it to be annihilated
only by the product $\widehat F K$. Furthermore, one would like to
understand, perhaps using the proposal in section 4, why the one-loop amplitude
that we have examined
is annihilated precisely by $\hat F K^2$, rather than the
minimal operator that annihilates it involving some other powers
of $\hat F$ and $K$.  These questions remain open.

\bigskip\noindent{\it A Note On Nonsupersymmetric Amplitudes}

It is fascinating to ask whether in some theories with reduced
supersymmetry, or no supersymmetry at all, a version of the
structure we have found may persist. This question is much too
ambitious to be tackled in the present paper.  However, we will
make one simple observation here.

The formula \ibob\ for the degree of a curve from which a given
amplitude should be derived implies that at the one loop level,
scattering amplitudes with four or more external gluons that all
have the same helicity must vanish.  Indeed, if we set $q=0$,
$l=1$, we get $d=0$, which implies that the support of the
amplitude must collapse to a point in twistor space.  As we
explained in section 3.3, a non-trivial scattering amplitude with
at least four external particles cannot have this property.

However, in gauge theories in general, the one-loop diagrams with
all gluons of the same helicity are non-zero in general. Indeed,
they are given by a simple formula that was conjectured by Bern, Dixon, and
Kosower
\ref\beretal{Z. Bern, L. Dixon, and D. A. Kosower, {\it Strings
1993}, ed. M. B. Halpern et. al. (World Scientific, 1995),
hep-th/0311026; Z. Bern, G. Chalmers, L. Koxon, and D. A. Kosower,
Phys. Rev. Lett. {\bf 72} (1994) 2134.} and proved by Mahlon
\ref\mahlon{G. Mahlon, ``Multigluon Helicity Amplitudes Involving
A Quark Loop,'' Phys. Rev. {\bf D49} (1994) 4438,
hep-ph/9312276.}, following some early computations in special
cases \nref\mahlon{G. Mahlon and T.-M. Yan, ``Multi-Photon
Production At High Energies In The Standard Model: I,'' Phys. Rev.
{\bf D47}
(1993) 1776, hep-ph/9210213.}%
\nref\bk{Z. Bern and D. Kosower, ``The Computation Of
Loop Amplitudes In Gauge Theories,'' Nucl. Phys. {\bf B379} (1992)
451.}%
\refs{\mahlon-\bk}.  For a useful summary, see \ref\bddk{Z. Bern,
L. Dixon, D. C. Dunbar, and D. A. Kosower, ``One-Loop Gauge Theory
Amplitudes With An Arbitrary Number Of External Legs,''
hep-ph/9405248.}.  Interestingly, these amplitudes are polynomial
(and in fact quartic) in $\tilde\lambda$, and hence are
supported on a degree one curve of genus zero, by the same
argument that we give momentarily for General Relativity. This is
not in accord with the most naive extension of our conjecture to
non-supersymmetric theories, but it does suggest the possibility
of some sort of generalization.

\subsec{A Peek At General Relativity}

Although we mainly focus on Yang-Mills theory in this paper, it is
hard to resist taking a peek at General Relativity. Tree level $n$
graviton amplitudes in General Relativity vanish if more than
$n-2$ gravitons have the same helicity.  The maximally helicity
violating amplitudes are thus, as in the Yang-Mills case, those
with $n-2$ gravitons of one helicity and two of the opposite
helicity.  These have been computed by Berends, Giele, and Kuijf
\ref\bgk{F. A. Berends, W. T. Giele, and H. Kuijf, ``On Relations
Between Multi-Gluon And Multi-Graviton Scattering,'' Phys. Lett.
{\bf 211B} (1988) 91.}; the four particle case was first computed by
DeWitt \texas.  The salient features are as follows.

If we factor out the delta function of energy-momentum
conservation via $\widehat
A(\lambda_i,\tilde\lambda_i)=i(2\pi)^4\delta^4\left(\sum_i\lambda_i^a\tilde
\lambda_i^{\dot a}\right)A(\lambda_i,\tilde\lambda_i) $, then for
Yang-Mills MHV amplitudes, $A$ is actually a function of $\lambda$
only.  In section 3.1, we deduced from this that the twistor
transform of those amplitudes is supported on genus zero curves of
degree one.

The formulas in \bgk\ show that for tree level MHV scattering in
General Relativity, the reduced amplitudes $A$, although not
independent of $\tilde\lambda$, are polynomial in $\tilde\lambda$.
This has the following result.  In Yang-Mills theory, when we carry
out the Fourier transform of MHV amplitudes
from momentum space to twistor space,
 we meet  an integral \eqn\ufro{\int {d^2\tilde\lambda_1\over
(2\pi)^2}\dots {d^2\tilde\lambda_n\over
(2\pi)^2}\,\exp\left(i\sum_i\tilde\lambda_i^{\dot a}(\mu_{i\,\dot
a}+x_{a\dot a}\lambda_i^a)\right).} But in General Relativity,  we must instead
evaluate integrals of the form \eqn\cufro{\int{
d^2\tilde\lambda_1\over (2\pi)^2}\dots {d^2\tilde\lambda_n\over
(2\pi)^2}\,\exp\left(i\sum_i\tilde\lambda_i^{\dot a}(\mu_{i\,\dot
a}+x_{a\dot a}\lambda_i^a)\right)P(\tilde\lambda_i^{\dot a}),}
where $P$ is a polynomial. Since this can be written
\eqn\dufro{P(-i\partial/\partial\mu_{i\dot a})\int
{d^2\tilde\lambda_1\over (2\pi)^2}\dots {d^2\tilde\lambda_n\over
(2\pi)^2}\,\exp\left(i\sum_i\tilde\lambda_i^{\dot a}(\mu_{i\,\dot
a}+x_{a\dot a}\lambda_i^a)\right),} the result of the integral is
simply \eqn\kufro{P(-i\partial/\partial\mu_{i\dot a})\prod_{i=1}^n
\delta^2(\mu_{\dot a}+x_{a\dot a}\lambda^a).} Thus, the twistor
transform of the gravitational MHV amplitudes is
supported on the same degree one curves as in the Yang-Mills case,
 but now with ``multiple derivative of a delta function''
behavior in the normal directions, roughly as we found in
supersymmetric Yang-Mills theory at the one-loop level. It would
certainly be interesting to know if such behavior persists for
other amplitudes in General Relativity.

\newsec{Interpretation As A String Theory}

\nref\wittrev{E. Witten, ``Mirror Manifolds And Topological Field
Theory,'' hep-th/9112056, in S.-T. Yau, ed., {\it Mirror
Symmetry}.}%
\nref\hori{K. Hori et. al., ed., {\it Mirror Symmetry} (American
Mathematical
Society, 2003).}%
\nref\wittopen{E. Witten, ``Chern-Simons Gauge Theory As A String
Theory,'' Prog. Math. {\bf 133} (1995) 637, hep-th/9207094.}%
\nref\dv{R. Dijkgraaf and C. Vafa, ``Matrix Models, Topological
Strings, And Supersymmetric Gauge Theories,'' Nucl. Phys. {\bf
B644} (2002) 3, hep-th/0206255.}%
 \nref\tsih{M. Aganagic, R. Dijkgraaf, A. Klemm, M. Marino,
and C. Vafa, ``Topological Strings And Integrable Hierarchies,'' hep-th/0312085.}%
In this section, we will propose a string theory that gives a
natural framework for understanding the results of section 3. This
is the topological $B$ model whose target space is the Calabi-Yau
supermanifold $\Bbb{CP}^{3|4}$.  We begin by outlining how this
model is constructed, and then explore its properties, culminating
with a computation of the supersymmetric MHV tree amplitudes. As
we will have to summarize many things (and because various points
are not yet clear),  the present section will not be as nearly
self-contained as the rest of the paper.  The reader will probably
find it helpful to have more  familiarity that we can convey here
with the topological $B$ model and its extension to open strings.
 For reviews of the $B$ model, see \refs{\wittrev,\hori}, for the extension to
 open strings see  \wittopen, and for  some recent
applications of the $B$ model, see \refs{\dv,\tsih}.
 Some of the basics about the Penrose transform that are needed for
this analysis are explained in an appendix, but the reader will
probably find it helpful to consult \penroseb\ or the reviews
cited in the introduction.

\subsec{Construction Of The Model}

To construct the ordinary $\Bbb{CP}^{M-1}$ model in two
dimensions, we introduce complex fields $Z^I$, $I=1,\dots,M$, with
a (constant) hermitian metric $g_{I\bar J}$, and a $U(1)$ gauge
field $B$ and auxiliary field $D$. The $Z^I$ all have charge one
with respect to $B$, so their covariant derivative is
$DZ^I=dZ^I+iBZ^I$. We work on a two-dimensional surface $C$ with
coordinates $x^\alpha$, $\alpha=1,2$ and metric
$\gamma_{\alpha\beta}$. The action is taken to be \eqn\morfgo{
I=\int_C
\,d^2x\sqrt{\gamma}\left(\gamma^{\alpha\beta}g_{I\overline J}{D
Z^I\over Dx^\alpha}{D\bar Z{}^{\bar J}\over Dx^\beta}
+D\left(g_{I\bar J}Z^I\bar Z{}^{\bar J}-r\right)\right),} where
$r$ is a positive constant. The Lagrange multiplier imposes the
constraint \eqn\impcon{g_{I\overline J}Z^I\bar Z{}^{\bar J}=r.} To
divide by the gauge group $U(1)$, we must impose the gauge
equivalence relation \eqn\impo{Z^I\to e^{i\alpha}Z^I,\,\,\alpha\in
[0,2\pi].}

 Assuming that the hermitian form $g_{I\bar J}$ is positive definite,
 the solution space of \impcon\ subject to the equivalence relation \impo\
 is a copy of $\Bbb{CP}^{M-1}$.
This statement is not completely trivial, since in section 2 we
used a different definition of $\Bbb{CP}^{M-1}$. According to this
definition,  $\Bbb{CP}^{M-1}$  is parameterized by $M$ complex
variables $Z^I$, not all zero, subject to the scaling $Z^I\to
t Z^I$, for $t\in \Bbb C^*$. Writing $t=\rho
e^{i\alpha}$, with $\rho$ real and positive, the scaling by $\rho$
can be used in a unique fashion to obey \impcon, and then the
scaling by $e^{i\alpha}$ is the gauge equivalence \impo.

  Since
the gauge field $B$ has no kinetic energy in \morfgo, it is an
``auxiliary field,'' like the Lagrange multiplier $D$.  We can
solve for $B$ in terms of $Z$ using its equation of motion, which
says that $g_{I\bar J}\bar Z{}^{\bar J}D_\alpha Z^I=0$, leading to
$B=-ir^{-1}g_{I\bar J}\bar Z{}^{\bar J}d Z^I$.  This formula says
that $B$ is the natural $U(M)$-invariant connection on the Hopf
bundle over $\Bbb{CP}^{M-1}$ (or more precisely, on the pullback of
this to $C$ via the map $Z:C\to \Bbb{CP}^{M-1}$).

The parameter $r$ determines the Kahler class of $\Bbb{CP}^{M-1}$.
However, as we will ultimately be studying the topological $B$
model, which is independent of the Kahler class, the choice of $r$
will be irrelevant.  Likewise, although we have to pick some
$g_{I\bar J}$ to write the action, the topological $B$ model is
independent of the choice of $g_{I\bar J}$, and our amplitudes
will really be invariant under the complexification $GL(M,\Bbb C)$
of the unitary group $U(M)$.

To extend this construction to a sigma model in which the target
space is a supermanifold $\Bbb {CP}^{M-1|P}$, we make the same
construction, except that we replace $Z$ by an extended set of
coordinates ${\cal Z}=(Z^I,\psi^A),$ $I=1,\dots,M$, $A=1,\dots,P$,
where $Z^I$ are as before and the $\psi^A$ are fermionic and of
charge one with respect to the $U(1)$ gauge field $B$. The
components of ${\cal Z}$ span a complex supermanifold
$\Bbb{C}^{M|P}$.  We endow this space with a hermitian form $G$
which we may as well take to be block diagonal:
\eqn\ndkn{G=\left(\matrix{g_{I\bar J}& 0 \cr 0 & g_{A\bar
B}\cr}\right).} It is invariant under a supergroup $U(M|P)$. The
action is the obvious extension of \morfgo\ to include the
$\psi^A$: \eqn\norfgo{ I=\int_C
\,d^2x\sqrt{\gamma}\gamma^{\alpha\beta}\left[g_{I\overline J}{D
Z^I\over Dx^\alpha}{D\bar Z{}^{\bar J}\over
Dx^\beta}+g_{A\overline B}{D\psi^A\over
Dx^\alpha}{D\bar\psi{}^{\bar B}\over Dx^\beta}
+\half\gamma_{\alpha\beta}D\left(g_{I\bar J}Z^I\bar Z{}^{\bar
J}+g_{A\overline B}\psi^A\bar\psi^{\bar B}-r\right)\right].} The
constraint and gauge equivalence become \eqn\reforma{g_{I\bar
J}Z^I\bar Z{}^{\overline J}+g_{A\bar B}\psi^A\overline
\psi{}^{\bar B}=r,} and \eqn\triforma{Z^I\to
e^{i\alpha}Z^I,~~\psi^A\to e^{i\alpha}\psi^A.} The constraint and
gauge equivalence turn the model into a sigma model with target
space $\Bbb{CP}^{M-1|P}$. Their combined effect is the same as
taking the space of all $Z^I$ and $\psi^A$, with the $Z^I$ not all
zero, and dividing by $(Z^I,\psi^A)\to (t Z^I,t\psi^A)$, $t\in\Bbb
C^*$. That was the definition of $\Bbb{CP}^{M-1|P}$ in section
2.6.

\bigskip\noindent{\it World-Sheet Supersymmetry}

\nref\aa{A. D'Adda, P. Di Vecchia, and M. Luscher,
``Confinement And Chiral Symmetry
Breaking In $\Bbb{CP}^{N-1}$
Models With Quarks, Nucl. PHys. {\bf B152} (1979) 125.}%
\nref\bb{E. Witten, ``Instantons, The Quark Model, And The $1/N$
Expansion,'' Nucl. Phys. {\bf B149} (1979) 285.}%

 The next step is to introduce world-sheet
supersymmetry. Because $\Bbb{CP}^{M-1|P}$ is a Kahler manifold, a
supersymmetric sigma model with this target space (first
constructed and studied in \refs{\aa,\bb} in the case $P=0$) will
automatically have $N=2$ worldsheet supersymmetry.  We thus
replace $C$ with a super Riemann-surface with $N=2$ supersymmetry,
the fermionic coordinates being a complex spinor $\theta^\alpha$
on $C$ and its complex conjugate $\bar\theta{}^\alpha$.  The $Z^I$
and $\psi^A$ are promoted to chiral superfields $\hat
Z^I(x,\theta)$ and $\hat\psi{}^A(x,\theta)$.  The gauge field $B$
and auxiliary field $D$ combine as part of a vector multiplet in
superspace, whose field strength is a ``twisted chiral
superfield'' $\Sigma$. This means in particular that $Z^I$ and
$\psi^A$ have partners of opposite statistics, \eqn\hundo{\hat
Z^I=Z^I+i\theta^\alpha\chi^I_\alpha+\dots,~\hat\psi^A=\psi^A+
i\theta^\alpha b_\alpha^A+\dots,}
 where $\chi^I_\alpha$ is
fermionic and $b_\alpha^A$ is bosonic. The superspace action is
\eqn\goofy{I=\int d^2x d^2\theta d^2\bar\theta \left(g_{I\bar
J}\bar Z{}^{\bar J} Z^I+g_{A\bar B}\bar\psi{}^{\bar
B}\psi^A\right) +{r\over 2} \left(\int d^2x
d\theta^+d\overline\theta{}^- \Sigma+\int d^2xd\theta^-
d\overline\theta{}^+\bar \Sigma\right).} For simplicity, we took
$C$ to be flat; otherwise, we would need two-dimensional
supergravity to make this construction. \goofy\ might look
mysterious at first sight because of the absence of derivatives
with respect to the $x^\alpha$.  In sigma models such as this one
with four supersymmetries, the derivatives appear \ref\zumino{B.
Zumino, ``Supersymmetry and Kahler Manifolds,'' Phys. Lett. {\bf
B87} (1979) 203.} upon performing the $\theta$ integrals, which
convert \goofy\  into a supersymmetric extension of the
bosonic action \norfgo.

\bigskip\noindent{\it The Calabi-Yau Condition}

So far, $M$ and $P$ are arbitrary.  For reasons that will appear,
however, we want to impose a Calabi-Yau condition. The
supermanifold $\Bbb{CP}^{M-1|P}$ is a Calabi-Yau supermanifold if
and only if $M=P$.  Indeed, the holomorphic measure
\eqn\jurdo{\Omega_0={1\over M!P!}\epsilon_{I_1I_2\dots
I_M}\epsilon_{A_1A_2\dots A_P}dZ^{I_1}dZ^{I_2}\dots
dZ^{I_M}d\psi^{A_1}d\psi^{A_2}\dots d\psi^{A_P} } on
$\Bbb{C}^{M|P}$ is invariant under the $U(1)$ gauge transformation
\triforma\ if and only if $M=P$.\foot{In verifying this statement,
one must recall that for fermions, $\psi$ and $d\psi$ transform
oppositely, so if $\psi\to e^{i\alpha}\psi$, then $d\psi\to
e^{-i\alpha}d\psi$.  This relation is compatible with the defining
property $\int d\psi\,\psi=1$ of the Berezin integral for
fermions.}  When it is $U(1)$-invariant, $\Omega_0$ descends to a
holomorphic measure $\Omega$ on $\Bbb{CP}^{M-1|P}$, ensuring that
$\Bbb{CP}^{M-1|P}$ is a Calabi-Yau manifold for $M=P$.  Of course,
the objects $\Omega_0$ and $\Omega$ were introduced in section
2.6, for closely related reasons.

As a Calabi-Yau supermanifold, $\Bbb{CP}^{M-1|M}$ should have a
Ricci-flat Kahler metric. This in fact is simply the Fubini-Study
metric -- the one we obtain starting with the flat metric on $\Bbb
C^{M|M}$ and imposing the constraint and gauge invariance that
were described above.  It does not take any computation to show
the Ricci-flatness.   The Ricci tensor of $\Bbb{CP}^{M-1|P}$ is completely
determined up to a multiplicative constant by the $SU(M|P)$
symmetry; the multiplicative constant is proportional to the first
Chern class of $\Bbb{CP}^{M-1|P}$, which is
$M-P$ times a generator of the second cohomology group of this space.
For $M=P$, the Ricci tensor therefore vanishes.
(Concretely, the Riemann tensor of $\Bbb{CP}^{M-1|P}$ is non-zero
 and is given by a natural generalization of what it is in the
bosonic case.  To construct the Ricci tensor, we must take a
supertrace of the Riemann tensor on two of its indices; when we do
this, the fermions contribute with opposite sign from the bosons,
giving Ricci-flatness for $M=P$.)

This means that, for $M=P$, the supersymmetric sigma model with
action \goofy\ is conformally invariant.  More important for our
present purposes, the Calabi-Yau condition means that we can
introduce a twisted version of the model which is a topological
field theory called the $B$ model.

\bigskip\noindent{\it The $B$ Model}

The two-dimensional nonlinear sigma model with any Kahler manifold
$X$ as the target space has a vector-like $R$-symmetry which acts
on the worldsheet coordinates $\theta^\alpha$ as $\theta^\alpha\to
e^{i\gamma}\theta^\alpha$; its action on the component fields in
\hundo\ can be deduced from this. The classical theory also has an
axial or parity-violating $R$-symmetry, acting by $\theta^+\to
e^{i\gamma}\theta^+$, $\theta^-\to e^{-i\gamma}\theta^-$, where
$\theta^+$ and $\theta^-$ have positive and negative chirality. We
write $K$ for the generator of this symmetry. In the quantum
theory, $K$ is anomaly-free if and only if $X$ is a Calabi-Yau
manifold.

The $B$ model is defined by ``twisting'' by $K$, so it can only be
defined when $X$ is Calabi-Yau. The twisting operation means, if
the theory is formulated on a flat worldsheet $C\cong \Bbb R^2$,
that one defines a new action of the two-dimensional Poincar\'e
group in which the translation operators $P_i$ are unchanged but
the rotation generator $J$ is replaced by $J'=J+K/2$.  This does
give a representation of the Poincar\'e Lie algebra, since
$[K,P_i]=0$. The twisting shifts the spin of every field by $K/2$.
All fermions have integer spin in the twisted theory and two of
the supercharges, say $Q_1$ and $Q_2$, have spin zero. They obey
$Q_1^2=Q_2^2 =\{Q_1,Q_2\}=0$, and their cohomology classes are
regarded as the physical states of the twisted model. This
construction on flat $\Bbb R^2$ can be generalized to an arbitrary
curved two-dimensional surface in such a way that $Q_1$ and $Q_2$
are still conserved.

When we get to open strings, only one linear combination of $Q_1$
and $Q_2$ is conserved; we call this combination $Q$.  Even for
closed strings, it will be adequate for our purposes to describe
the action of $Q$.

We will briefly describe the field content and transformation laws
of the $B$-model.  Let $\phi^i$, $i=1,\dots,{\rm dim}_{\Bbb C}X$,
be a set of fields representing local complex coordinates on $X$.
(In our example of $\Bbb{CP}^{3|4}$, we can take the $\phi^i$ to
be $Z^I/Z^1$, $I>1$, and $\theta^A/Z_1$.) The superpartners of the
$\phi^i$  are as follows (as one learns by considering the
expansion \hundo\ in the twisted theory):  $\eta^{\bar i}$ is a
zero-form on $C$  that  transforms\foot{To be more precise, $\eta$
is a section of $\phi^*\Omega^{0,1}(X)$, where $\Omega^{0,1}(X)$
is the space of $(0,1)$-forms on $X$, and $\phi^*\Omega^{0,1}(X)$
is its pullback to $C$ via the map determined by the fields
$\phi^i$.  A similar remark holds for $\theta$ and $\rho$.} as a
$(0,1)$-form on $X$; $\theta_i$ is a zero-form on $C$ that
transforms as a section of the holomorphic tangent bundle of $X$;
and $\rho^i$ is a one-form on $C$ that transforms as a
$(1,0)$-form on $X$.  The BRST transformation laws of the fields,
that is, the transformation laws under the symmetry generated by
$Q$, are \eqn\ononpp{\eqalign{\delta \phi^i & = 0 \cr
                     \delta \bar\phi^{\bar i}& = i\alpha\eta^{\bar i}\cr
                     \delta\eta^{\bar i}&=\delta\theta_i = 0\cr
                     \delta\rho^i & = -\alpha \,d\phi^i.\cr}}
($\alpha$ is an infinitesimal anticommuting parameter.)
 The space
of physical states is obtained by taking the cohomology of $Q$ in
the space of local functions of these fields (a local function is
a functional of the fields and their derivatives up to some finite
order, polynomial in the derivatives, evaluated at some given
point in $C$). In fact, the cohomology classes can all be
represented by operators that are functions only of
$\phi,\,\bar\phi,\,\theta$, and $\eta$ without any derivatives.
Such operators take the form $V_\alpha= \alpha(\phi,\bar
\phi)_{\bar i_i\bar i_2\dots \bar i_p}{}^{j_1j_2\dots
j_q}\eta^{\bar i_1}\eta^{\bar i_2} \dots \eta^{\bar
i_p}\theta_{j_1}\theta_{j_2}\dots \theta_{j_q}$.  Upon
interpreting $\eta^{\bar i}$ as $d\bar\phi^{\bar i}$, $V_\alpha$
can be associated with an object $\alpha=d\bar \phi^{\bar
i_1}d\bar\phi ^{\bar i_2}\dots d\bar\phi^{\bar i_p}\alpha_{\bar
i_i\bar i_2\dots \bar i_p}{}^{j_1j_2\dots j_q}$ that we interpret
as a
 $(0,p)$-form on $X$ with values in $\wedge^q TX$,
the $q^{th}$ antisymmetric power of the holomorphic tangent bundle
$TX$ (or $T^{1,0}X$) of $X$.  With this interpretation, $Q$ can be
identified as the $\bar\partial$ operator on the space of such
forms. The space of physical states is hence the direct sum over
$p$ and $q$ of the $\bar\partial$ cohomology groups
$H^p(X,\wedge^q TX)$.   For compact $X$ and more generally for the
type of examples familiar in critical string theory, these
cohomology groups are finite-dimensional.  The richness of twistor
theory comes partly from the fact that for $X$ a suitable region
in twistor space, the $\bar\partial$ cohomology groups are
infinite-dimensional and can be identified with solution spaces
of wave equations in Minkowski spacetime.  For a  brief
explanation of this, see the appendix.

Other physical quantities in the $B$ model are likewise naturally
described in terms of complex geometry of $X$.  For example, for
$C$ of genus zero, the  $B$ model correlation functions (which for
$X$ a Calabi-Yau threefold are important in heterotic and Type II
superstring theory) are expressed as follows in terms of the wedge
products of classes in $H^p(X,\wedge^qX)$.\foot{In this
discussion, we only consider the case of a bosonic Calabi-Yau
manifold.  The extension to a Calabi-Yau supermanifold involves
some technical issues that have not been addressed yet, reflecting
the fact that on a supermanifold, what can be integrated is not a
differential form but an ``integral form.''  A similar issue would
arise for open strings on $\Bbb{CP}^{3|4}$ if we used
space-filling branes.  That is why we will use branes that are not
quite space-filling.  See \ref\mov{M. Movshev and A. Schwarz, ``On
Maximally Supersymmetric Yang-Mills Theory,'' hep-th/0311132.} for
construction of integral forms on certain complex supermanifolds
associated with Yang-Mills theory.} Let $\alpha_1,\dots,\alpha_s$
be elements of $H^{p_i}(X,\wedge^{q_i}X)$, with
$\sum_ip_i=\sum_jq_j=n$, where $n={\rm dim}_{\Bbb C} X$. Each
$\alpha_i$ corresponds to a vertex operator $V_i$, as explained in
the last paragraph. The wedge product of the $\alpha_i$ is
naturally an element of $H^n(X,\wedge^nTX)$.  To define the $B$
model, one must pick a holomorphic $n$-form $\Omega$ on $X$.  By
multiplying by $\Omega^2$, one can map $H^n(X,\wedge^nTX)$ to
$H^{n,n}(X)$, the space of $(n,n)$-forms.  Such a form can be
integrated over $X$ to obtain the genus zero correlation
functions: \eqn\ononon{\langle V_1\dots V_s\rangle =
\int_X\alpha_1\wedge\alpha_2\wedge \dots\wedge \alpha_s\,
\Omega^2.} For $C$ of  genus greater than zero, $B$ model
observables involve more sophisticated invariants of the complex
geometry. For  genus one, one encounters analytic torsion, and for
higher genus one meets less familiar invariants, to whose study
powerful methods including mirror symmetry and the holomorphic
anomaly have been applied
 \ref\cec{M. Bershadsky,
S. Cecotti, H. Ooguri, and C. Vafa, ``Holomorphic Anomalies In
Topological Field Theories,'' Nucl. Phys. {\bf B405} (1993) 279,
hep-th/9302103, ``Kodaira-Spencer Theory Of Gravity And Exact
Results For Quantum String Amplitudes,'' Commun. Math. Phys. {\bf
B165} (1994) 311, hep-th/9309140.}.

\nref\roc{M. Grisaru,
M. Rocek, and W. Siegel, ``Superloops 3, Beta 0: A Calculation
In $N=4$ Yang-Mills Theory,'' Phys. Rev. Lett. {\bf 45} (1980) 1063.}%
\nref\tar{L. V. Avdeev and O. V. Tarasov,
``The Three Loop Beta Function In The $N=1$, $N=2$,
$N=4$ Supersymmetric Yang-Mills Theories,''
Phys. Lett. {\bf B112} (1982) 356.}%
\nref\mandelstam{S. Mandelstam, ``Light Cone Superspace And The Ultraviolet
Finiteness Of
The $N=4$ Model,'' Nucl. Phys. {\bf B213} (1983) 149.}%

In our example of $\Bbb{CP}^{M-1|P}$, we want to take $M=4$, because
 it is $\Bbb{CP}^3$ (and its supersymmetric
extensions), and not some other $\Bbb{CP}^{M-1}$, that is related
to four-dimensional Minkowski spacetime by the Penrose
transform.\foot{Moreover, the twistor transform does not have a
very close analog in Minkowski spacetimes of other dimensions,
though some properties can be generalized, as discussed recently
in \ref\kras{K. Krasnov, ``Twistors, CFT, And Holography,''
hep-th/0311162.}.  For example, the conformal symmetry of
Minkowski spacetime of $n$ dimensions is $SO(2,n)$ while the
symmetry of $\Bbb{CP}^{M-1}$ is $SL(M)$; for one of these to be a
real form of the other, we set $n=M=4$.  (The case $n=1$, $M=2$
does not seem useful.)   The closest analog of twistor space in a
different dimension is probably the ``mini-twistor space,'' a
complex line bundle over $\Bbb{CP}^1$ that is used to solve the
equations for BPS monopoles in three dimensions \ref\ah{M. F.
Atiyah and N. Hitchin,  {\it The Geometry And Dynamics Of Magnetic
Monopoles} (Princeton University Press, 1988).}.  This
construction is naturally obtained by dimensional reduction from
the twistor correspondence in four dimensions.}
 Once we set $M=4$, we also need
$P=4$, for the Calabi-Yau condition.

\bigskip\noindent{\it Symmetries Of The $B$ Model}

In general,  the symmetries of the
$B$ model are the transformations of the target space
 that preserve its complex structure and also
act trivially on the holomorphic measure $\Omega$. The reason for
this last requirement is visible in \ononon: the correlation
functions are proportional to $\Omega^2$, so a symmetry of $X$
that acts nontrivially on $\Omega^2$ is not a symmetry of the $B$
model.   For the open string version that we introduce in section
4.2, the analogous formula (see eqn. (4.15)) is linear in
$\Omega$, so symmetries must act trivially on $\Omega$.

The group of symmetries of $\Bbb{CP}^{3|4}$ that act trivially on the holomorphic
measure $\Omega$
was determined in section 2.6. The group is $PSL(4|4)$, a real form
of which is the symmetry group $PSU(2,2|4)$ of $\N=4$ super
Yang-Mills theory in Minkowski space.

Among the supersymmetric Yang-Mills theories, the ${\cal N}=4$
theory is special, as it has the maximal possible supersymmetry
\schwarz.  Among the pure supersymmetric Yang-Mills theories (with
only the fields of the super Yang-Mills multiplet), it is also
special in being conformally invariant, which makes it a natural
candidate for being encoded in twistor space, where the conformal
symmetries are built in.  We have found here a different
explanation for what is special about ${\cal N}=4$: it cancels the anomalies in the
$B$ model of super twistor space.

Transformations  that preserve the complex structure of the target
space of the $B$ model but act non-trivially on the holomorphic measure $\Omega$
 are also
interesting.  They are not symmetries of $B$ model amplitudes, but
they can still be used to constrain these amplitudes in an
interesting way.  To jump ahead of our story a bit, we will argue
that the relation \ibob\ between the helicities in a Yang-Mills
scattering amplitude and the degree and genus of a holomorphic
curve on which its twistor transform is supported arise from such
an anomalous symmetry of the $B$ model of $\Bbb{CP}^{3|4}$.

Indeed, $\Bbb{CP}^{3|4}$ has a $U(1)$ symmetry which does {\it
not} leave invariant the holomorphic measure $\Omega$.  This is
the transformation that rotates the fermions by a phase,
\eqn\reftrans{S:\psi^A\to e^{i\beta}\psi^A,} while leaving the
bosonic coordinates $Z^I$ invariant.  Under the transformation
$S$, we have $\Omega_0\to e^{-4i\beta}\Omega_0$, and (since the
transformation commutes with the scaling by which we descend to
$\Bbb{CP}^{3|4}$) likewise $\Omega\to e^{-4i\beta}\Omega$.  So
$\Omega$ has $S=-4$.

\subsec{Open String Sector Of The $B$ Model}

$\Bbb{CP}^{3|4}$ is a Calabi-Yau supermanifold whose bosonic
reduction is of complex dimension three.  Before trying to
describe its $B$ model, it is well to begin by recalling the $B$
model on an ordinary Calabi-Yau threefold $X$.

To define open strings while preserving the topological symmetry
of the $B$ model, one needs a boundary condition that preserves a
linear combination of the fermionic symmetries of the model. As
explained in \wittopen, the simplest boundary conditions that do
this are Neumann boundary conditions.  We introduce Chan-Paton
factors of the gauge group $GL(N,\Bbb C)$ (a real form of which is $U(N)$).
In modern language, this
construction amounts to \ref\polchinski{J. Polchinski, ``Dirichlet
Branes And Ramond-Ramond Charges,'' Phys. Rev. Lett. {\bf 75}
(1995) 4724.} introducing $N$ space-filling $D$-branes wrapped on
$X$. The branes are endowed by a vector bundle $E$ with structure
group $GL(N,\Bbb C)$. As is also explained in \wittopen, the only physical
open string field in this model is a field $A$ that is the $(0,1)$
part of a connection on $E$. It is subject to the gauge invariance
\eqn\gauginv{\delta A=\bar\partial \epsilon +[A,\epsilon],} for
any zero-form $\epsilon$ with values in the Lie algebra of $GL(N,\Bbb C)$.

The basic idea of the derivation is that, among the worldsheet fields
described
 in the closed
string case in section 4.1, the open string boundary conditions
are such that  $\theta$ and one component of $\rho$ vanish on the
boundary of $C$. Open string vertex operators are local functions
of the fields evaluated at a point on the boundary of $C$, so they
depend only on $\phi,\bar\phi,\eta$, and the surviving component
of $\rho$.   The cohomology of $Q$ in this space can be
represented by vertex operators that are functions of just
$\phi,\bar\phi$, and $\eta$. Vertex operators $V_\alpha=
\eta^{\bar i_1}\eta^{\bar i_2} \dots \eta^{\bar i_p}\alpha_{\bar
i_1\bar i_2\dots \bar i_p}(\phi,\bar\phi)$ correspond to
$(0,p$)-forms $\alpha=d\bar\phi^{\bar i_1}\dots d\bar\phi^{\bar
i_p}\alpha_{\bar i_1\dots\bar i_p}(\phi,\bar\phi)$. The open
string $B$ model (for this type of space-filling brane) can thus
be described in terms of a set of fields that are $(0,p)$-forms on
$X$, for $0\leq p\leq n$.  (These fields are all $N\times N$
matrices because of the Chan-Paton factors.) However, for $X$ a
Calabi-Yau threefold, the important such field is the $(0,1)$-form
$A$.  The others can be interpreted in the low energy effective
field theory as ghosts that enter in the quantization of $A$.  The
BRST operator $Q$ acts as the $\bar\partial$ operator on $A$ and
the other $(0,p)$-forms.

 $A$ is a complex field (as is the gauge parameter $\epsilon$).
Its complex conjugate would be $\overline A$, the $(1,0)$ part of
the connection. However, the topological sector of the theory can
be described without ever mentioning $\overline A$. Rather, the
action is a holomorphic function of $A$: \eqn\nolfun{I={1\over
2}\int_X\Omega\wedge \Tr\left(A\bar\partial A+{2\over 3}A\wedge
A\wedge A\right). } Here $\Tr\left(A\bar\partial A+{2\over
3}A\wedge A\wedge A\right)$ is the Chern-Simon $(0,3)$-form
constructed from $A$.  There is no need to introduce explicitly a
string coupling constant in \nolfun, as this can be absorbed in a
scaling of the holomorphic three-form $\Omega$. The classical
equations of motion derived from \nolfun\ assert  simply the
vanishing of the curvature $(0,2)$-form $F=\bar\partial A+A\wedge
A$.  This means that a classical solution defines a holomorphic
vector bundle on $X$. This Chern-Simons action is very special; it
is the unique local action that depends only on the complex
structure and holomorphic volume-form of $X$ and is invariant
under complex gauge transformations of $A$.

The quantum theory is described by a path integral, which (if for
simplicity we omit gauge-fixing and ghosts) is roughly of the form
$\int DA\exp(-I)$. Since $A$ is a complex variable and the action
is a holomorphic function, one must try to understand this path
integral as a contour integral for each mode of $A$.  (See
\dv\  for matrix models based on such
contour integrals, and \ref\laz{C. Lazaroiu, ``Holomorphic Matrix
Models,'' hep-th/0303008.} for a thorough discussion of the
contours in that context.)  In general, the result of the path
integral may depend on the choice of contour.

How to make sense of the path integral as a contour integral can
be made explicit in perturbation theory. To construct
perturbation theory, one must expand around a classical solution,
that is, around a field $A$ that defines a holomorphic vector
bundle on $X$.  In the case of an isolated and nondegenerate
bundle (no zero modes for $A$), to construct perturbation theory
one merely needs to know how to integrate a Gaussian function or a
Gaussian times a polynomial. For example, for a single variable
$\phi$, $\int d\phi\,\exp(-\lambda\phi^2)=\sqrt{\pi/\lambda}$. One
can pick a contour in the complex plane that justifies this
Gaussian integral; the same contour will suffice to construct
perturbation theory.

More interesting is the case that one is expanding around a moduli
space ${\cal Y}$ of flat connections.  In this case, choosing a
contour entails picking a suitable middle-dimensional real
homology cycle in ${\cal Y}$. For example, in a favorable
situation, $X$ may have a $\Bbb Z_2$ symmetry $\tau$ that reverses
its complex structure. Such a symmetry defines what is called a
real structure of $X$. If so, the $\tau$-invariant subspace of
${\cal Y}$, if non-empty, is a suitable real cycle to integrate
over. Perturbation theory can be constructed by integrating over
this real slice of ${\cal Y}$ and constructing perturbation theory
in the normal directions. Conceivably, different perturbative
series could be constructed using other cycles in ${\cal Y}$.

If $H^3({\cal Y},\Bbb R)\not= 0$, then because of the behavior of
the Chern-Simons form under gauge transformations that are not
homotopic to the identity, the action integral \nolfun\ is
well-defined only modulo certain periods of the holomorphic volume
form $\Omega$. This does not affect perturbation theory, but it is
certainly important nonperturbatively.  A primary application of
holomorphic Chern-Simons theory at the moment is to computing
certain chiral amplitudes in physical string theory (for a recent
dramatic example, see \dv).
In that context, there is a closed
string field, the two-form field or $B$-field, that must be
included in establishing invariance under disconnected gauge
transformations. In our study below of string theory on
$\Bbb{CP}^{3|4}$, this problem does not arise, since
$H^3(\Bbb{CP}^{3|4},\Bbb R)=0$. In any event, in this paper, we
will be treating the open string fields perturbatively, in order
to compare to perturbative Yang-Mills theory in spacetime.

Since we will be expanding around the trivial solution $A=0$, we
will not need to construct such a real cycle in
a  moduli space ${\cal Y}$ of bundles on $\Bbb
{CP}^{3|4}$.  But we will encounter a somewhat analogous moduli
space ${\cal M}$ of holomorphic curves in $\Bbb{CP}^{3|4}$
(representing $D$-instanton configurations), and we will have to
pick a real cycle in ${\cal M}$, which we will do by using a real
structure on $\Bbb{CP}^3$.

\subsec{Extension to $\Bbb{CP}^{3|4}$}

Let us now consider the analog of this in $\Bbb{CP}^{3|4}$. The
$D$-branes that we will consider are not quite
space-filling.\foot{It is conceivable that similar results would
arise from space-filling branes, but the necessary formalism
appears more complicated as noted in a previous footnote.}
They are defined by the condition that
on the boundary of an open string, $\bar\psi=0$, while $\psi$ is
free. The analog of this condition would not make sense for
bosons; it would not make sense for a complex field $\Phi$ to say
that $\bar\Phi=0$ on the boundary while $\Phi$ is unrestricted.
But this condition does make sense for fermions: it merely means
that the vertex operators that can be inserted on the boundary are
functions of $\psi$ (as well as of the bosonic coordinates $Z$ and
$\bar Z$) and not of $\bar\psi$. Again we introduce $GL(N,\Bbb C)$
Chan-Paton factors, or in other words, we consider a stack of $N$
such $D$-branes.

We write $Y$ for the world-volume of the branes.  $Y$ is thus the
subspace of $\Bbb{CP}^{3|4}$ parameterized by $Z,\bar Z,\psi$ with
$\bar\psi=0$.

If we repeat the derivation in \wittopen, we find that the
physical states are now described by a field ${\cal A}=d\bar
Z^{\bar I}\CA_{\bar I}$, where the  $\CA_{\bar I}$ depend, of
course, on $\psi$ as well as $Z$ and $\bar Z$.  (For space-filling
branes, we would have had an extra term in the expansion, namely
$d\bar\psi_A \CA^A$, and the functions would all depend on
$\bar\psi$ as well as the other variables.)  We can describe this
by saying that $\CA$  is a $(0,1)$-form on $\Bbb{CP}^3$ that
depends on $\psi$ as well as on the coordinates of $\Bbb{CP}^3$.
We can expand $\CA$ in terms of ordinary $(0,1)$-forms
$(A,\chi_A,\phi_{AB},\tilde\chi^A,G)$ with values in various line
bundles: \eqn\duggo{\eqalign{{\cal A}(Z,\bar Z,\psi)=d\bar
Z{}^{\bar I}&\left(A_{\bar I}(Z,\bar Z)+\psi^A\chi_{\bar I\,A}
(Z,\bar Z)+{1\over 2!}\psi^A\psi^B \phi_{\bar
I\,AB}(Z,\bar Z)\right.\cr &\left. +{1\over
3!}\epsilon_{ABCD}\psi^A\psi^B\psi^C\tilde\chi_{\bar I}^D(Z,\bar
Z)+{1\over 4!}\epsilon_{ABCD}\psi^A\psi^B\psi^C\psi^DG_{\bar
I}(Z,\bar Z)\right).\cr}} The gauge invariance is
\eqn\dosoggo{\delta\CA=\bar \partial \epsilon+[\CA,\epsilon].}
where now $\epsilon$ depends on $\psi$ as well as $Z$ and $\bar
Z$.

Since the fermionic homogeneous coordinates $\psi$ of
$\Bbb{CP}^{3|4}$ take values in the holomorphic line bundle
$\CO(1)$ over $\Bbb{CP}^3$, a field that multiplies $\psi^k$ must
take values in $\CO(-k)$. So $A,\chi,\phi,\tilde \chi$, and $G$
take values respectively in the line bundles $\CO$, $\CO(-1)$,
$\CO(-2)$, $\CO(-3)$, and $\CO(-4)$.  These fields, geometrically,
are $(0,1)$-forms on $\Bbb{CP}^3$ with values in those line
bundles. The fields $(A,\chi,\phi,\tilde\chi,G)$ also have charges
$(0,-1,-2,-3,-4)$ for the charge $S$, defined above, that assigns
the value $+1$ to $\psi$.

The classical action describing the open strings is the same as
\nolfun, except that the field $A(Z,\bar Z)$ must be replaced by
$\CA(Z,\bar Z,\psi)$: \eqn\bolfun{I={1\over 2}\int_{Y}\Omega\wedge
\Tr\left(\CA \bar\partial\CA+{2\over
3}\CA\wedge\CA\wedge\CA\right).} Recall that in this
supersymmetric case, $\Omega$ is a measure for the holomorphic
variables $Z$ and $\psi$, locally taking the form $d^3Zd^4\psi$.
The product of $\Omega$ with the Chern-Simons $(0,3)$-form is thus
a measure on $Y$ that can be integrated to get the action \bolfun.
In terms of the expansion \duggo, the action becomes \eqn\holfun{
\eqalign{I=\int_{\Bbb
{CP}^3}\Omega'\wedge\Tr&\biggl(G\wedge(\bar\partial A+A\wedge
A)+\tilde\chi^A\wedge \bar D\chi_A\biggr.\cr&\left. +{1\over
4}\epsilon^{ABCD}\phi_{AB}\wedge\bar D\phi_{CD}+{1\over
2}\epsilon^{ABCD}\chi_A\wedge\chi_B\wedge\phi_{CD}\right).\cr}}
Here, $\bar D$ is the $\bar\partial$ operator with respect to the
connection $A$; for any field $\Phi$, $\bar
D\Phi=\bar\partial\Phi+A\Phi$. Also, $\Omega'={1\over
4!}\epsilon_{IJKL}Z^IdZ^JdZ^KdZ^K$ is a $(3,0)$-form on
$\Bbb{CP}^3$ that is homogeneous of degree 4 and so takes values
in the line bundle $\CO(4)$. It is obtained by integrating out the
$\psi^A$ from the measure $\Omega$ on $\Bbb{CP}^{3|4}$. On the
other hand, the trace in \holfun\ is a $(0,3)$-form with values in
$\CO(-4)$. So the product of the two is an ordinary $(3,3)$ form,
which can be integrated over $\Bbb{CP}^3$ to give an action. This
action clearly has definite charge $S=-4$, confirming that the
charge $S$ is not a symmetry of the $B$ model in twistor space.

The classical equations of motion obtained by varying the fields
in \holfun\ are \eqn\bucon{0=\bar\partial \CA+\CA\wedge \CA} or in
components \eqn\ucocn{\eqalign{0 & = \bar\partial A+A\wedge A\cr 0
& =\bar D\chi \cr 0 & =\bar
D\phi_{AB}-\chi_A\wedge\chi_B\cr 0& =\bar
D\tilde\chi^A-{1\over 2}\epsilon^{ABCD}\left(\chi_B\wedge \phi_{CD}+\phi_{CD}\wedge
\chi_B\right)\cr 0&=\bar
DG+\chi_A\wedge\tilde\chi^A-\tilde\chi^A\wedge \chi_A+{1\over
4}\epsilon^{ABCD}\phi_{AB}\wedge\phi_{CD} .\cr}}

If we linearize these equations around the trivial solution with $\CA=0$,
they tell us simply that $0=\bar\partial \CA$, or in components
\eqn\jeqn{0=\bar\partial\Phi,}
 where $\Phi$ is any of
$(A,\chi_A,\phi_{AB},\tilde\chi^B,G)$.  Because of the gauge invariance
\dosoggo,
which reduces to \eqn\keqn{\delta\Phi=\bar\partial\alpha} for each
component $\Phi$, the fields $\Phi$ define elements of
appropriate cohomology groups.
  To find the right ones, recall that each field $(A,\chi,\phi,
\tilde\chi,G)$ has charge  $S=-k$ for some $k=0,-1,-2,-3,-4$ and
is a  $(0,1)$-form with values in $\CO(-k)$. The equations \jeqn\
and gauge invariance \keqn\ mean that such a field determines an
element of the sheaf cohomology group $H^1(\Bbb{PT}', \CO(-k))$.
Here $\Bbb{PT}'$ is whatever portion of  twistor space
$\Bbb{PT}=\Bbb{CP}^3$ we choose to work with.

Now we come to a central point.  According to the Penrose
transform \refs{\penroseb,\revminus-\atiyah}, reviewed briefly in
the appendix, the sheaf cohomology group $H^1(\Bbb{PT}',\CO(-k))$
is equal to the space of solutions of the conformally invariant
 free massless wave equation
for a field of helicity $1-k/2$, on a suitable region $\Bbb U$ of complexified and
conformally compactified Minkowski
space (which depends on the choice of $\Bbb{PT}'$). These
conformally invariant equations are as follows: the anti-selfdual
Maxwell equations $F'{}^+=0$ for helicity 1, where $F'{}^+$ is the
selfdual part of the field strength $F'=dA'$ of an abelian gauge
field $A'$ in spacetime; the massless Dirac equation for helicity
$1/2$ and $-1/2$; the conformally coupled Laplace equation for
helicity 0; and finally, for helicity $-1$, the equation $dG'=0$
where $G'$ is a selfdual two-form.

So in this linearized approximation, the twistor space fields
$(A,\chi_B,\phi_{BC},\tilde\chi^C,G)$ correspond to spacetime
fields $(A',\chi_B',\phi_{BC}',\tilde\chi'{}^C,G')$ which are
respectively anti-selfdual gauge fields, positive chirality
spinors, scalars, negative chirality spinors, and a selfdual
two-form, all in the adjoint representation of $GL(N,\Bbb C)$. On-shell,
they describe particles of helicities $(1,1/2,0,-1/2,-1)$,
respectively. These are precisely the physical states of $\N=4$
super Yang-Mills theory, with just the familiar $SL(4)_R$ quantum
numbers.  Of course, this is largely determined by the manifest
$PSL(4|4)$ symmetry.

Moreover, the field content is almost recognizable (and will be
altogether recognizable to readers familiar with investigations by
Siegel of a chiral limit of super Yang-Mills theory \ref\siegel{W.
Siegel, ``The $\N=2(4)$ String Is Self-Dual $\N=4$ Yang-Mills,''
Phys. Rev. {\bf D46} (1992) R3235, ``Self-Dual $\N=8$ Supergravity
As Closed $\N=2(4)$ Strings,'' hep-th/9207043.}). $A'$ is the
gauge field of the $\N=4$ theory, while $\chi'$ and $\tilde \chi'$
are the usual positive and negative chirality fermions, and
$\phi'$ are the usual scalars. We still have to interpret $G'$, as
well as the anti-selfduality of $A'$.

Under the Penrose transform from twistor space fields
$(A,\chi,\phi,\tilde\chi,G)$ to spacetime fields $(A',\chi',
\phi',\tilde\chi',G')$, might the action \holfun\ magically turn
into the standard ${\cal N}=4$ action in spacetime, which has the
same superconformal symmetry?  The answer to this question is
``no,'' for a very instructive reason.  The action \holfun, in
addition to having the $PSL(4|4)$ symmetry of $\N=4$ super
Yang-Mills theory, is homogeneous of degree $-4$ with respect to
the ``anomalous'' $U(1)$ generator $S$ that assigns the values
$(0,-1,-2,-3,-4)$ to $(A,\chi_B,\phi_{BC},\tilde\chi^B,G)$.  When
we linearize around $\CA=0$, the Penrose transform is a linear
map, so we should assign the same quantum numbers
$(0,-1,-2,-3,-4)$ to $(A',\chi'_B,\phi'_{BC},\tilde \chi'{}^B,
G')$. With this assignment, the standard $\N=4$ Yang-Mills action
is a sum of terms most of which have $S=-4$ or $S=-8$.  The $S=-4$
terms include the kinetic energies $\tilde\chi^{\dot a}D_{a\dot
a}\chi^a$ and $(D_{a\dot a}\phi)^2$, as well as the Yukawa
coupling $\phi\chi^2$, while the other Yukawa coupling
$\phi\tilde\chi^2$ and the $\phi^4$ coupling have $S=-8$.  The
$\N=4$ theory also has a Yang-Mills kinetic energy $(F')^2$, with
$F'=dA'+A'\wedge A'$; it has $S=0$, but arises, in a description
with an auxiliary field, from another term with $S=-8$, as we
explain in section 4.4.

In short, the $\N=4$ action can be described as a sum of terms of
$S=-4$ and $S=-8$.
  Our proposal here is that the classical $B$ model of
$\Bbb{CP}^{3|4}$ gives the terms of $S=-4$, while the terms of
$S=-8$ will come from a $D$-instanton correction that will be
introduced later. The $S=-4$ part of the action is a
supersymmetric truncation of $\N=4$ super Yang-Mills theory that
has been studied in \siegel. According to the second paper in that
reference, where the supersymmetry transformations can also be
found, the supersymmetric action is \eqn\hucod{I=\int
d^4x\,\,\Tr\left( {1\over 2}G{}^{ a b}F_{ a b} +\tilde\chi^{A
a}D_{a\dot a}\chi^{\dot  a}_A+{1\over
8}\epsilon^{ABCD}\phi_{AB}D_{a\dot a}D^{a\dot a}\phi_{CD}+{1\over
4}\epsilon^{ABCD}\phi_{AB}\chi^{\dot a}_{C}\chi_{D \dot
a}\right).} Since confusion with twistor space fields seems
unlikely, we have here omitted the primes from the fields.

\bigskip\noindent{\it The $++-$ Amplitude And The Twistor Space Propagator}

Having identified the twistor fields $A$ and $G$ with spacetime
fields of helicities $1$ and $-1$, we can shed a little light on
one point from section 3.2. There we predicted, but did not quite
find, a local twistor space interaction with helicities $++-$.  In
complex twistor space $\Bbb{CP}^3$, this interaction exists; it is
simply the $AAG$ interaction that we can read off from \holfun.

Moreover, we can now understand the mysterious ``internal lines''
that appeared in section 2 in (for example) figures 3 and 5. The
fields propagating in these lines are $A$ and $G$. (For tree level
scattering of gluons, which was the main focus of section 2,
 the $SU(4)_R$ non-singlet fields $\chi,\phi,\tilde\chi$ do not contribute.)
 The kinetic energy of $A$ and $G$ is purely off-diagonal, of
the form $G\bar\partial A$, so the propagator is also purely
off-diagonal. This is why opposite ends of the internal lines are
labeled by opposite helicities. It still remains to explain later
the Riemann surfaces that the internal lines are attached to.

\subsec{The Auxiliary Field $G'$ And The Anti-Selfduality of $A'$}

By now, we have extracted from the twistor theory a spacetime
description that is much like conventional $\N=4$ super Yang-Mills
theory. The main differences are the appearance of a possibly
unexpected field $G'$ and the anti-selfduality of $A'$.

\nref\gcs{G. Chalmers and W. Siegel, ``The Self-Dual Sector Of QCD
Amplitudes,'' hep-th/9606061.}%
 To elucidate these points, supersymmetry is not really
essential, so we will start with a stripped down version with
fewer fields. We consider a $U(N)$ gauge field $A'$ in spacetime,
and a field $G'$ that is a selfdual antisymmetric tensor with
values in the adjoint representation of $GL(N,\Bbb C)$.
We define an $S$
quantum number under which $A'$ and $G'$ have charges $0$ and
$-4$. (The peculiar choice for $G'$ is of course motivated by the
supersymmetric example described above.) We begin with the action
\refs{\siegel,\gcs} \eqn\olpo{I=\int d^4x\,\Tr\left(G'\wedge
F'(A')\right)=\int d^4x\, \Tr\left( G'\wedge F'{}^+(A')\right).}
$F'^+$ is the selfdual part of $F'$.  The two expressions for the
action given in \olpo\ are equal, since $G'$ is selfdual. The
action has charge $S=-4$. The classical equations of motion are
\eqn\unco{F'{}^+(A')=0,~~{D\over D x^\mu}G'{}^{\mu\nu}=0,} where
$D_\mu$ is the covariant derivative including the gauge field
$A'.$ The first equation says that the nonzero part of the field
strength of $A'$ is the anti-selfdual part $F'{}^-$, which of
course obeys a Bianchi identity $D_\mu F'{}^-{}^{\mu\nu}=0$ that
is rather like the equation of motion for the selfdual field $G'$.

The fact that $G'$ and the nonzero part of $F'$ are respectively
selfdual and anti-selfdual  means that they describe particles of
opposite helicity.  In our conventions, $A'$ describes a particle
of helicity 1 and $G'$ describes a particle of helicity $-1$.  The
spectrum, at this linearized level, is thus that of conventional
Yang-Mills theory, with both helicities present.  The
interactions, however, are not the standard ones. Indeed, the
action \olpo\ has an $A'A'G'$ term, describing a vertex of three
fields with helicities $++-$, but in contrast to Yang-Mills
theory, it has no $--+$ vertex.  Indeed, that term would have
$S=-8$.

To cure this, we add a $(G')^2$ term (as was also discussed in
\gcs), to get an extended action \eqn\boggo{I_1=\int d^4x\,\,\Tr
\left(G' F'-{\epsilon\over 2}(G')^2\right).} Here $\epsilon$ is a
small parameter. The term we have added has $S=-8$.  It is nearly
the unique term that we can add to \olpo\ that is local, gauge
invariant, and conformally invariant. The only other possibility
is the topological invariant \eqn\oboggo{\Delta I=\int \Tr
F'\wedge F',} which has $S=0$ and is related in a familiar fashion
to instantons and the $\theta$ angle of four-dimensional quantum
gauge theory. As a topological invariant, this interaction has no
influence on Yang-Mills perturbation theory, but it is important
nonperturbatively.

We can integrate out $G'$ from \boggo\ to get an equivalent action
for $A'$ only.  It is \eqn\nuboggo{I_2={1\over 2\epsilon}\int d^4x
\,\,\Tr (F'{}^+)^2.} From the point of view of perturbation
theory, this is precisely equivalent to conventional Yang-Mills
theory. In fact, the topological invariant \oboggo\ is a multiple
of $(F'{}^+)^2-(F'{}^-)^2$.  So upon adding it with the right
coefficient,\foot{The right coefficient is imaginary.  For
example, if we are in Lorentz signature, the action should be
real, but because the selfdual and anti-selfdual conditions are
$F=\pm i *F$, $\Delta I$ is actually an imaginary multiple of
$(F'{}^+)^2-(F'{}^-)^2$.} we convert \nuboggo\ to
\eqn\hilfo{I_3={1\over 4\epsilon}\int d^4x\, \Tr (F')^2.}

If desired, we can also add the topological invariant \nuboggo\
with a real coefficient to incorporate the theta angle of quantum
gauge theory. This will play no role in the present paper, as we
will limit ourselves to trying to reconstruct perturbation theory
from twistor space. (We note, however, that the topological
invariant \oboggo\ is mapped by the twistor transform to the
second Chern class of the bundle $E$ over twistor space, and so
could be represented by a local interaction in twistor space.)

We have obtained our desired result.  \hilfo\ is equivalent to Yang-Mills
theory with the usual Yang-Mills coupling $g_{YM}$ being related to $\epsilon$
by $g_{YM}^2=\epsilon$.

What has happened?  Clearly, it is possible to take the $g_{YM}\to
0$ limit of Yang-Mills theory in such a way as to arrive at \olpo.
We are accustomed to taking the weak coupling limit of Yang-Mills
theory in a way that treats the two helicities symmetrically. But
it is possible instead to break this symmetry as $g_{YM}\to 0$ and
end up with \olpo.  Or one could make an opposite choice as
$g_{YM}\to 0$ and arrive at the parity conjugate of \olpo.  The
different choices differ by how the wavefunctions of states of
different helicities are scaled as $g_{YM}\to 0$.  We make this
more explicit momentarily in the context of the $\N=4$ theory.

\bigskip\noindent{\it Charges For $\N=4$}

We can now improve on an assertion that was made in section 4.3.
There we described the ${\cal N}=4$ super Yang-Mills action as the
sum of the $(F')^2$ term, of $S=0$, plus terms of $S=-4$ and
$S=-8$.  But now we see that, if the auxiliary field $G'$ is
included, then the $(F')^2$ term really comes from a $(G')^2$
interaction, which has $S=-8$.  So in this description, the ${\cal
N}=4$ super Yang-Mills action is of the form
\eqn\polo{I=I_{-4}+\epsilon I_{-8}.} where $I_{-4}$ is the sum of
terms of $S=-4$ and $I_{-8}$ is the sum of terms with $S=-8$.
$I_{-4}$ comes from the Penrose transform of \holfun\ to
spacetime, and $I_{-8}$ will arise as a one-instanton
contribution.  $I$ is the standard $\N=4$ action \schwarz, and
$I_{-4}$ was investigated in \siegel.

It is also interesting to understand how to express the action in
a standard form with all terms proportional to
$g_{YM}^{-2}=\epsilon^{-1}$. This is done simply by rescaling
every field that has $S=-k$ for some $k$ by a factor of
$\epsilon^{-k/4}$.  In the new variables, all terms in $I$ are
proportional to $\epsilon^{-1}$.  After this rescaling and
integrating out $G$, the action becomes manifestly invariant under
parity.  The different $\epsilon\to 0$ limits of $\N=4$ super
Yang-Mills theory thus arise from different ways to scale the
fields $\chi,\phi,\tilde\chi$ (or the corresponding helicity
states), as well as the gluon states of one helicity or the other,
as $\epsilon\to 0$.

\subsec{Relation To Perturbation Theory}

Now let us understand how the relation to Yang-Mills perturbation
theory must work, to recover the results of section 3.  We will
then look for an instanton construction that yields the right
properties.

For simplicity, we consider only the fields $A'$ and $G'$, with
the action \boggo, which takes the general form \eqn\noggo{I\sim
G'( dA'+(A')^2)-\epsilon (G')^2.}

\ifig\graphs{(a) A tree level Feynman diagram with $k$ vertices of
type $AAG$, connected by $AG$ propagators, leads to an $A^{k+1}G$
interaction, as sketched here for $k=2$. (b) Replacing an $AG$
propagator by an $AA$ amplitude adds a power of $\epsilon$ and
replaces an $A$ by a $G$ in the amplitude.  For $k=2$, we generate
in order $\epsilon$ an $\epsilon A^2G^2$ interaction, as sketched
here.} {\epsfbox{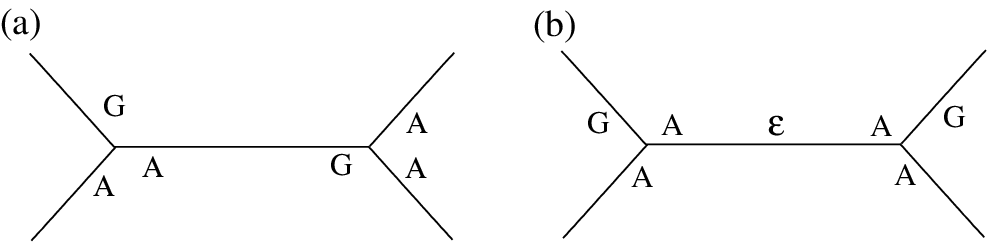}}

 We first consider the
theory at $\epsilon=0$.  The perturbation theory in this case has
already been analyzed in \gcs.
 The only interaction vertex is the
$G'A'A'$ vertex, which we identify with a configuration of
helicities $-++$. To form a Feynman diagram, we can start with any
number of $G'A'A'$ vertices, and then contract some fields with
propagators. Because of the off-diagonal nature of the $G'dA'$
kinetic energy, the propagator in the basis given by $(A',G')$ has
the general form \eqn\leftgo{\left(\matrix{0 & d^{-1}\cr d^{-1} &
0\cr}\right).} The only non-zero matrix element of the propagator
is $\langle G'~ A'\rangle$. As illustrated in figure 6(a), to make
a tree diagram we start with an arbitrary number $k$ of $G'A'A'$
vertices, and connect them by $k-1$ propagators, in the process
``contracting out'' $k-1$ factors of $G'$ and the same number of
factors of $A'$.  We are left with an amplitude $G'(A')^{k+1}$
with only one negative helicity field $G'$ and an arbitrary number
of positive helicity fields. (These amplitudes actually vanish for
$k>1$, after summing over diagrams, but this is not very apparent
in the present discussion.)

We could have predicted the same result without looking at Feynman diagrams
by noting that since (at $\epsilon=0$) the classical action is homogeneous
with $S=-4$, the tree level $S$-matrix elements, obtained by integrating out the
off-shell degrees of freedom, must have the same property.  So they are homogeneous
in $G'$ of degree 1.

We can extend that analysis to $\epsilon\not= 0$ simply by
assigning charge $S=4$ to $\epsilon$.  Then the whole action is
homogeneous with $S=-4$, so the generating function of scattering
amplitudes has the same property. Since the only objects carrying
the $S$-charge are $G'$ with charge $-4$ and $\epsilon$ with
charge 4, the generating functional of the tree level scattering
matrix elements must have the general form
\eqn\nibob{W=f_0(A')G'+\epsilon f_1(A')(G')^2+\dots
+\epsilon^{r-1}f_{r-1}(A')(G')^r+\dots.}

Of course, we can also reach the same conclusion by examining
Feyman diagrams (as in figure 6(b)). For this, we note that taking
$\epsilon\not= 0$ adds no vertices to the Lagrangian, but it does
add an additional term $\epsilon (G')^2$ to the kinetic energy.
The modification of the propagator is very simple.  Upon inverting
a $2\times 2$ matrix that is schematically
\eqn\rufo{\left(\matrix{ 0 & d\cr d&
\epsilon\cr}\right)} in the $(A',G')$ basis, we find that for
$\epsilon\not= 0$,  $\langle G'\,G'\rangle$ remains zero, but
$\langle A'\,A'\rangle $ is nonzero and of order $\epsilon$. Every
time that we replace the $\langle A'\,G'\rangle$ propagator in a
tree level Feynman diagram by an $\langle A'\,A'\rangle$
propagator, we multiply the amplitude by a factor of $\epsilon$
(from the propagator) and we retain an extra $G'$ field (which is
not contracted out).  This leads back to the structure found in
eqn. \nibob.

We interpret the $\epsilon^{r-1}f_{r-1}(A')(G')^r$ term in \nibob\
as the generating functional of tree level scattering processes
with precisely $r$ gluons of negative helicity.  As we also will
interpret $\epsilon$ as the instanton expansion parameter, it
follows that tree amplitudes with precisely $r$ negative helicity
gluons must arise from configurations with instanton number $r-1$.
(The instantons in a given configuration may be either connected
or disconnected, as we discussed in section 3.)  This reasoning
was the original motivation for the conjecture that was stated in
eqn. \ibob\ and explored in section 3.

We can straightforwardly extend this analysis to include loops.
After assigning charge $4$ to $\epsilon$, the whole action $I$ is
of charge $-4$.  If we introduce Planck's constant $\hbar$ with
charge $-4$ and define the rescaled action $I'=I/\hbar$, then $I'$
is invariant under $S$. On the other hand, an $l$-loop amplitude
is proportional to $\hbar^{l-1}$. As this factor has charge
$-4l+4$, it must multiply a function of $G'$ and $\epsilon$ of
total charge $4l-4$.  The allowed powers are $\epsilon^d(G')^q$
where $4d-4q=4l-4$, or \eqn\ribob{d=q-1+l.} This agrees with our
basic formula \ibob\ when we interpret the power of $\epsilon$ as
the instanton number and the power of $G'$ as the number of
negative helicity gluons in a scattering process.

Of course, we could alternatively have reached this conclusion
by counting the powers of $\epsilon$ in Feynman diagrams with loops.
We leave this to the reader.

\subsec{$D$-Instantons}

By now it should be clear that we need to enrich the $B$ model of
$\Bbb{CP}^{3|4}$ with instanton contributions that will introduce
additional violation of the quantum number $S$. But what kind of
instantons? The most obvious instantons  are worldsheet
instantons. However, one of the main claims to fame of the
$B$ model is that topological amplitudes in this model receive no
worldsheet instanton corrections. The $A$-model does have
worldsheet instanton contributions, but otherwise it falls badly
short of what we need.  For example, its space of physical states
is far too trivial, involving ordinary cohomology, which is
finite-dimensional for any reasonable subspace $\Bbb{PT}'$ of
twistor space. By contrast, the $B$ model leads to the far richer
$\bar\partial$ cohomology, and, via the Penrose transform, to
massless fields in Minkowski spacetime.  Somehow, we need a model
that combines the virtues of the $A$ model and the $B$ model.
 Another obvious shortcoming of the $A$ model is that, as it
requires no Calabi-Yau condition for the target space, it would
not explain the special role of $\N=4$ supersymmetry.

A clue comes by considering duality between heterotic and Type I
superstrings. The $B$ model of $\Bbb{CP}^{3|4}$ with $U(N)$ gauge
fields incorporated via Chan-Paton factors is a kind of
topological version of the Type I model. Suppose that, at least
for some values of $N$, the model has a heterotic string dual.
Then we would expect worldsheet instanton contributions to the
topological amplitudes.  Under duality between heterotic and Type
I superstrings, the heterotic string worldsheet instantons turn
into Type I $D$-instantons, which represent submanifolds in the
target space $\Bbb{CP}^{3|4}$ on which open strings may end. And
accordingly, in physical Type I superstring theory, $D$-instantons
do contribute to chiral amplitudes. All of this suggests that we
should incorporate $D$-instantons in the $B$ model of
$\Bbb{CP}^{3|4}$.  To preserve the topological symmetry of the $B$
model, these instantons must come from $D$-branes wrapped on some
holomorphic submanifold of $\Bbb{CP}^{3|4}$.  These holomorphic
submanifolds must be of complex dimension one, since we learned in
section 3 that perturbative Yang-Mills theory is related to curves
in twistor space.

A $D$-instanton carries a $U(1)$ gauge field, so whenever we
consider a $D$-instanton wrapped on a curve $C$, a holomorphic
line bundle ${\cal L}$ will be part of the discussion.  As we
discuss presently, if $C$ has genus $g$, ${\cal L}$ should have
degree $g-1$.  (In section 3, we did not notice that our curves in
twistor space were endowed with line bundles, because they all
were of genus zero, so ${\cal L}$ had no moduli.) Scattering
amplitudes with $g\geq 2$ should also receive contributions from
$D$-instantons of multiplicity $k>1$, that is, from a collection
of $k$ $D$-instantons wrapped on the same curve $C$.  In this
case, the gauge group supported on $C$ is $GL(k,\Bbb C)$, and $C$
will be endowed with a rank $k$ holomorphic vector bundle $F$.
When $F$ is irreducible, the $k$ $D$-instantons cannot separate; a
cluster of $k$ $D$-instantons with an irreducible vector bundle is
a component of instanton moduli space that needs to be included.
(But we will not do any computations that are nearly sophisticated
enough to see such components.)

The massless modes on the worldvolume of  a $D$-instanton, apart
from the gauge fields mentioned in the last paragraph, are just
the modes that describe the motion of the $D$-instanton.  So the
moduli space ${\cal M}$ of $D$-instanton configurations
parameterizes holomorphic curves $C\subset \Bbb{PT}$ endowed with
a holomorphic line bundle (or a holomorphic vector bundle in the
situation considered in the last paragraph).  $C$ may have several
disconnected components (possibly with different multiplicities)
as in some examples encountered in section 3.

To construct scattering amplitudes, we need, roughly speaking, to
integrate over ${\cal M}$.  But in the topological $B$ model, as
we recalled in section 4.2, the action is a holomorphic function
of complex fields, and all path integrals are contour integrals.
Thus, the integral will really be taken over a contour in ${\cal
M}$, that is, a middle-dimensional real cycle.  To be able to
integrate over such a contour, ${\cal M}$ must be endowed with a
holomorphic measure $\Upsilon$.  For an ordinary complex manifold
of dimension $n$, a holomorphic measure would be a holomorphic
$n$-form; for a supermanifold, a holomorphic measure is a
holomorphic section of the Berezinian of the tangent bundle.

We will give two choices of contour in ${\cal M}$, using a method
explained in section 4.2.  We pick a real structure $\tau $ on
twistor space $\Bbb{PT}$, that is, a $\Bbb Z_2$ symmetry of
$\Bbb{PT}$ that reverses its complex structure.\foot{Since we only
need to determine the homology class of the contour, it is enough
to have a symmetry that reverses the complex structure of the
bosonic reduction of supertwistor space.}  $\tau$ automatically
acts on ${\cal M}$, and we take the integration contour (for the
bosons in ${\cal M}$) to consist of the fixed points of $\tau$.

There are two possible choices of $\tau$. We can define $\tau$ to
simply act by complex conjugation on each of the homogeneous
coordinates $Z^I$ of $\Bbb{PT}$.  This choice is natural for
signature $--++$ in Minkowski spacetime, where the $Z^I$ can all
be real. This is the choice we will make in computing MHV
amplitudes, because the definition of twistor amplitudes is
simplest for that signature.  Alternatively, we can consider
the symmetry $\hat\tau(Z^1,Z^2,Z^3,Z^4)=(\bar Z{}^2,-\bar
Z{}^1,\bar Z{}^4,-\bar Z{}^3)$. This choice is natural for
studying Yang-Mills theory in signature $+++\,+$, because, in
acting on complexified Minkowski spacetime $\Bbb M$, $\hat \tau$
leaves fixed a real slice that has Euclidean signature.
(Unfortunately, I do not know how to pick a contour that is
naturally adapted to Lorentz signature in spacetime.)  One hopes
that the theories constructed using $\tau$ or $\hat\tau$ to pick
the contour are equivalent, but it is not clear how to prove this.

\bigskip\noindent{\it Construction Of The Measure}

How can we construct the holomorphic measure $\Upsilon$ on the
moduli space ${\cal M}$ of $D$-instantons?  In the topological
$B$ model with target space an ordinary (bosonic) Calabi-Yau
manifold, such a measure arises from the determinant of the
massless fields on the $D$-brane (whose zero modes are the
moduli).  I do not know technically how to do this when the target
space is a Calabi-Yau supermanifold, so I will just
construct the measure by hand for $D$-instantons of genus zero and
arbitrary degree.  As we will see, for these cases ${\cal M}$ is a
Calabi-Yau supermanifold, and the measure is uniquely determined
by holomorphy, up to a multiplicative constant.  The choice of
this constant for degree one determines the Yang-Mills coupling
constant, and the normalization of the measure for higher degrees
can be determined by factorization or unitarity.  (The
normalization given below is presumably compatible with
factorization, though this will not be proved.)

To construct a genus zero curve of degree $d$, we let $C_0$ be a
copy of $\Bbb{CP}^1$ with homogeneous coordinates $(u^1,u^2)$.
Then we describe a holomorphic map $\Phi:C_0\to \Bbb{CP}^{3|4}$
that maps the homogeneous coordinates $(Z^I,\theta^A)$ of
$\Bbb{CP}^{3|4}$ to homogeneous polynomials of degree $d$ in
$(u^1,u^2)$: \eqn\mormo{\eqalign{Z^I&=c^I_{i_1\dots
i_d}u^{i_1}\dots u^{i_d},\cr \psi^A&=\beta^A_{i_1\dots
i_d}u^{i_1}\dots u^{i_d}.\cr}} The map $\Phi$ is determined by the
coefficients $c^I_{i_1\dots i_d}$ and $\beta^A_{i_1\dots i_d}$,
which are, of course, respectively bosonic and fermionic.  The
coefficients $c$ and $\beta$ parameterize a linear space $\Bbb
L\cong \Bbb{C}^{4d+4|4d+4}$. On $\Bbb L$, there is a natural
holomorphic measure,
\eqn\tinon{\Upsilon_0=\prod_{I,\{i_1,\dots,i_d\}}dc^I_{i_i\dots
i_d}\prod_{A,\{i_1,\dots,i_d\}}d\beta^A_{i_i\dots i_d}.} The space
of maps $\Phi$ is parameterized by the $c$'s and $\beta$'s modulo
the scaling $(c,\beta)\to(t c,t \beta)$, $t\in \Bbb C^*$.  The
measure $\Upsilon_0$ is invariant under this scaling, since $c$
and $\beta$ have the same number of coefficients.  The space of
maps is thus a Calabi-Yau supermanifold
$\Bbb{PL}=\Bbb{CP}^{4d+3|4d+4}$.

We really are not interested in maps from $C_0$ to
$\Bbb{CP}^{3|4}$ but in holomorphic curves in $\Bbb{CP}^{3|4}$.
Two maps have as their images the same curve if and only if they
differ by the action of $SL(2,\Bbb C)$ on $(u^1,u^2)$.  The moduli
space $\CM$ of curves in $\Bbb{CP}^{3|4}$ of genus zero and degree
$d$ is thus $\CM=\Bbb{PL}/SL(2,\Bbb C)$. As $\Upsilon_0$ is
$SL(2,\Bbb C)$-invariant as well as being invariant under scaling,
$\Upsilon_0$ descends to an everywhere nonzero holomorphic volume
form $\Upsilon$ on $\CM$.   Thus, $\CM$ is a Calabi-Yau
supermanifold of dimension $4d|4d+4$.

For a genus zero instanton of degree 1, $Z^I$ and $\psi^A$ are
linear in $u^1,u^2$.  Writing as usual $Z=(\lambda,\mu)$, we can
generically use the $SL(2,\Bbb C)$ symmetry and scaling to put the map $\Phi$
in the form $\lambda^1=u^1$, $\lambda^2=u^2$, whereupon  the other
coordinates $\mu^{\dot a},\psi^A$ become linear functions of
$\lambda$.  After renaming the coefficients, the curve takes the familiar form
\eqn\enons{\eqalign{&\mu_{\dot a} + x_{a\dot a}\lambda^{ a}=0\cr
                   & \psi^A + \theta_a^A\lambda^a=0.\cr}}
The measure $\Upsilon$ reduces to the familiar measure
$d^4x\,d^8\theta$ that we used in section 3.1.

\bigskip\noindent{\it The $S$ Charge Of The $D$-Instanton Measure}

We introduced $D$-instantons in the hope of finding a new source
of violation of the quantum number $S$ whose role in Yang-Mills
perturbation theory we have discussed above.  Now we can determine
if this program has a chance to succeed.

Since the $S$-charges of $(Z,\psi)$ are $(0,1)$, the charges of
the cofficients $(c,\beta)$ are likewise $(0,1)$.  The
differentials $(dc,d\beta)$ therefore have charges $(0,-1)$, so
the $S$-charge of $\Upsilon$ is $-4d-4$. So a genus zero instanton
of degree $d$ contributes to the effective action a term that
violates the $S$-charge by this amount.  This is exactly what we
want. Since each negative helicity gluon has $S=-4$, while
positive helicity gluons have $S=0$, an amplitude with any number
of positive helicity gluons and $q$ gluons of negative helicity
has $S$-charge $-4q$. So a connected genus zero instanton of
degree $d$, with no other sources of $S$-charge violation, can
contribute to such an amplitude if and only if $d$ and $q$ are
related by \eqn\unon{d=q-1.} We recognize the familiar formula
\ibob\ whose consequences were explored in section 3.

The result is the same for disconnected instantons, of a type that
we really have already described in section 3.  If we consider $r$
$D$-instantons, all of genus zero, and of degree $d_i$, with
$d=\sum_id_i$, then their total $S$-charge is
$\sum_{i=1}^r(-4d_i-4)=-4d-4r$.  However, for such a configuration
to contribute to a connected amplitude, the $D$-instantons must be
connected by fields propagating in twistor space, as represented
by the ``internal line'' in figure 3.  Each such internal line
increases the $S$-charge by $+4$ (the propagator is the inverse of
the kinetic energy, which has charge $-4$).  To make a connected
configuration without loops, which we regard as a degenerate case
of a configuration of genus zero, the number of internal lines
must be $r-1$, whereupon the total $S$-charge violation in the
amplitude is $-4d-4r+4(r-1)=-4d-4$, as expected.

To extend the agreement with \ibob\ to higher genus, we would like
the $S$-charge of the measure for a connected $D$-instanton of
genus $g$ to be \eqn\onons{\Delta S=-4(d+1-g).} Though we will not
prove this rigorously by properly understanding the appropriate
worldvolume determinants of the $D$-instantons, one can give a
heuristic explanation of where the formula comes from.  For a
generic curve $C$ of genus $g$ and degree $d$ in $\Bbb {CP}^3$,
one expects from the index theorem that $H^0(C,\CO(1))$ will be of
dimension $d+1-g$.  (Here $\CO(1)$ is the usual line bundle over
$\Bbb{CP}^3$.)  The four $\psi^A$ are each sections of $\CO(1)$,
so $C$ has a total of $4(d+1-g)$ fermionic moduli, all of
$S$-charge 1, leading to the formula \onons\ for the $S$-charge of
the $D$-instanton measure. For disconnected $D$-instantons of any
genus, connected by internal lines, the agreement is preserved
because of arguments similar to those in the last paragraph.

\bigskip\noindent{\it $D1-D5$ Strings}

The key ingredient in computing scattering amplitudes, as we will
see presently in computing the MHV amplitudes, is the effective
action of the $D1-D5$ strings.

We consider a $D1$-brane $C$ located at $\psi^A=0$; the
$\psi$-dependence will be restored when we integrate over moduli
of $C$.  The $D5$-branes are of course the usual stack of $N$
(almost) space-filling branes. In quantizing the $D1-D5$ strings,
$\psi$ and its bosonic partners and the bosons and fermions normal
to $C$ in $\Bbb{CP}^3$ have no zero modes, since they obey
Dirichlet boundary conditions at one end of the string and Neumann
boundary conditions at the other end. Bosons and fermions tangent
to $C$ do have zero modes; their quantization leads in the usual
way for the $B$ model to the space of $(0,q)$-forms on $C$, where
in the present problem (as $C$ is of complex dimension one),
$q=0,1$. The derivation of this is rather similar to the
quantization of $D5-D5$ strings, which we briefly explained in
section 4.2.

The $D1-D5$ strings are thus $(0,q)$-forms $\alpha$ on $C$, with values in
$E_C\otimes\CL$, where $E_C$ is the $D5$-brane
gauge bundle restricted to $C$ and $\CL$
is a line bundle on $C$ that depends on the $U(1)$ Chan-Paton gauge field on $C$.
The $D5-D1$ strings are similarly $(0,q)$-forms $\beta$ on $C$, but now
with values in $E_C^*\otimes \CL'$, where $E_C^*$ is the dual bundle to
$E_C$, and $\CL'$ is another line bundle. (Dual bundles $E_C$ and $E_C^*$ appear here
because $D5-D1$ strings transform in the antifundamental representation
 of the $D5$-brane gauge group, while $D1-D5$ strings transform
in the fundamental representation.)  When we want to make manifest
the $GL(N,\Bbb C)$ quantum numbers of $\alpha$ and $\beta$ (or the
fact that they take values in $E_C$ and $E_C^*$, respectively), we
write them as $\alpha^x$, $\beta_x$, $x=1,\dots,N$. The kinetic
operator for topological strings is the BRST operator $Q$, which
when we reduce to the low energy modes is the $\bar\partial$
operator, or its covariant version $\bar D$ to include a
background field $\CA$. The effective action for the $D1-D5$
strings is thus \eqn\ononp{I_{D1-D5}=\int_Cdz \,\beta_x\bar
D\alpha^x.} Here $z$ is an arbitrary local complex parameter on
$C$. We have incorporated a possible background gauge field $\CA$
(which will represent initial and final particles in a scattering
amplitude) by using the covariant $\bar\partial$ operator $\bar
D=d\bar z(\partial_{\bar z}+\CA_{\bar z})$, where $\CA_{\bar z}$
is the component of $\CA$ along $C$. For the action to make sense,
it must be that $\CL\otimes\CL'\cong K$, where $K$ is the
canonical line bundle of $C$.  (This result should ideally be
explained more directly by more carefully quantizing the zero
modes.) All choices of $\CL$ are allowed, depending on the choice
of gauge field on $C$.

Only the $(0,0)$-form components of $\alpha$ and $\beta$ actually appear in this
Lagrangian.  The $(0,1)$-form components may possibly play some role in understanding
$c$-number contributions to the measure (at some deeper level that we will not
reach in this paper), but they do not couple to the background field $\CA$.
For the rest of this paper, therefore, we simply take $\alpha$ and $\beta$ to
be $(0,0)$-forms.

The coupling of $\CA$ to the $D1-D5$ strings can be read off from \ononp.  It is
\eqn\ponon{\Delta I=\int_C \Tr \,J\CA_{\bar z}d\bar z,}
where we define $J^x{}_y=\alpha^x\beta_y\, dz$;
we include the factor of $dz$ in the current
and interpret $J$ as a $(1,0)$-form on $C$ that (because of the way $\alpha$ and
$\beta$ transform under a change in local
parameter) is independent of the choice of $z$.  $J$ takes values in the Lie
algebra of $GL(N,\Bbb C)$ (acting as endomorphisms of $E$), and the trace in
\ponon\ is taken over this Lie algebra.

\nref\smallinst{E. Witten, ``Small Instantons In String Theory,'' Nucl.
Phys. {\bf B460} (1996) 541, hep-th/9511030.}%
\nref\douginst{M. R. Douglas, ``Gauge Fields And $D$-Branes,
J. Geom. Phys. {\bf 28} (1998) 255, hep-th/9604198.}%
The model seems to
make more sense if we assume that the $D1-D5$ string fields
$\alpha$ and $\beta$ are fermions.  Under appropriate conditions, $\alpha$ and
$\beta$ will have zero modes.  If $\alpha$ and $\beta$ are bosons,
the zero modes will lead to flat directions
 which by analogy with phenomena in critical
string theory \refs{\smallinst,\douginst} will represent the deformation of
the $D1$-brane into a smooth holomorphic bundle on $\Bbb{CP}^{3|4}$ with second
Chern class nonzero (and Poincar\'e dual to $C$).
By the twistor transform of the anti-selfdual Yang-Mills equations, such bundles
correspond to instantons in spacetime and thus to nonperturbative contributions
in the Yang-Mills theory.  However, the $D1$-branes do not couple to spacetime
fields like Yang-Mills instantons; rather, we will argue in section 4.7 that
they contribute to perturbative scattering amplitudes.

If $\alpha$ and $\beta$
are fermions, there is no contradiction, as
we would not expect to relate the $D1$-brane to a spacetime instanton.
In the computation that we actually perform in section 4.7,
however, the statistics of $\alpha$
and $\beta$ only affect the overall sign of the single-trace interaction; our
computation is not precise enough to determine this sign.
In any event, for whatever it is worth, the action \ononp\ is more natural
for fermions.

In quantizing the $D1$-branes, one must sum and integrate
over the choice of line bundle ${\cal L}$.
However, unless ${\cal L}$ has degree $g-1$, where $g$ is the genus of $C$,
there is a non-trivial index because of which
 $\alpha$ or $\beta$ have zero modes that  are not lifted by the coupling
to the external gauge field $\CA$.  ${\cal L}$'s of degree other than $g-1$
hence will not contribute.  In the specific computation that we will
perform presently, $C$ has genus 0, so we take ${\cal L}$ to have degree $-1$.
In genus 0, ${\cal L}$ has no moduli.  The coupling to ${\cal L}$ just means
that the fields $\alpha$ and $\beta$ are ordinary chiral fermions of spin
$(1/2,0)$, which is how we will interpret them in section 4.7.

\subsec{Computation Of MHV Amplitudes}

Now let us discuss how to use $D$-instantons in twistor space to actually
compute a scattering amplitude in spacetime.

We will consider an $n$-particle scattering amplitude.  The $i^{th}$ external
particle, for $i=1,\dots,n$,
is represented by a wavefunction that is a $\bar\partial$-closed $(0,1)$-form
$w_i$
on $\Bbb{PT}'$ (the part of super twistor space $\Bbb{CP}^{3|4}$
with $\lambda\not=0$).  Each $w_i$ takes values in the Lie algebra of
$GL(N,\Bbb C)$ (the gauge group carried by the $D5$-branes),
and so represents a cohomology class that takes values in the tensor product with
this Lie algebra of the twistor space cohomology group
$H^1(\Bbb{PT}',\CO)$.

The coupling of $w_i$ to a $D$-instanton wrapped on a Riemann surface $C$ is
according
to \ponon\
\eqn\thecop{B_i=\int_C\Tr\,J\wedge w_i.}
This is found by simply regarding $w_i$ as a contribution to the external gauge
field $\CA$ in \ponon.

If $C$ had no moduli, its contribution to the scattering amplitude for $n$ particles
coupling via $B_1,\dots ,B_n$ would
be found by evaluating the corresponding
 expectation value $\langle B_1\dots B_n\rangle$
 in the $D$-instanton worldvolume
theory.  Concretely, this would be done
by integrating over the fields $\alpha$ and $\beta$.
In actual examples, $C$ is a point in a moduli space $\CM$ of holomorphic
curves in supertwistor space.  We must pick a real cycle $\CM_{\Bbb R}$
in $\CM$ and integrate over it using the holomorphic measure $\Upsilon$.
The scattering amplitude with the given external wavefunctions $w_i$ is consequently
\eqn\onxon{A(w_i)=\int_{\CM_{\Bbb R}}\Upsilon \langle B_1\dots B_n\rangle.}
Actually, to get the proper power of the Yang-Mills coupling $g$ multiplying
a scattering amplitude and a possible multiplicative constant,
we need to also include a few additional factors:
normalization factors for external wavefunctions and a factor of
$e^{-I}$, with $I$ the $D$-instanton action.  We will omit these factors.

We will now show how to use this formalism to recover the supersymmetric
tree level MHV amplitudes, as described in twistor space in eqn. \tyno.
The ability to recover these amplitudes gives our most detailed evidence
that the $B$ model of $\Bbb{CP}^{3|4}$ is equivalent at least in the planar limit
to $\N=4$ super Yang-Mills amplitudes.

For tree level MHV amplitudes, we take $C$ to be a straight line, that is a curve
of genus zero and degree one.  We recall that the lines in supertwistor
space are described by the equations
\eqn\xnon{\eqalign{\mu_{\dot a}+x_{a\dot a}\lambda^a=&0\cr
                     \psi^A+\theta_a^{A}\lambda^a=&0.\cr}}
Here $x^{a\dot a}$ and $\theta^{aA}$ are the moduli of $C$.  The measure
is the usual superspace measure $\Upsilon=d^4x\,d^8\theta$.
We will use the real slice that is natural for signature $++-\,-$ in
spacetime; a point in $\Bbb{CP}^3$ is considered real if $\lambda$ and $\mu$
are real, and the real slice of $\CM$ is defined by simply saying that
$x^{a\dot a}$ is real for $a,\dot a=1,2$.

The scattering amplitude is therefore \eqn\umbu{ A(w_i)=\int d^4x
\,d^8\theta \,\langle B_1\dots B_n\rangle.} Let us assume that the
wavefunctions $w_i$ take the form $w_i=v_iT_i$, where $T_i$ is an
element of the Lie algebra of $GL(N,\Bbb C)$, and $v_i$ is an
ordinary (not matrix-valued) $(0,1)$-form. The amplitude has a
term proportional to $I=\Tr\,T_1T_2\dots T_n$. Let us extract this
term.  We have to compute the appropriate term in the expectation
value of a product of currents $\langle \Tr\,T_1J(\lambda_1)
\Tr\,T_2J(\lambda_2)\dots \Tr\,T_nJ(\lambda_n)\rangle$, or
essentially $\langle\Tr\,T_1\alpha\beta(\lambda_1)
\Tr\,T_2\alpha\beta(\lambda_2)\dots
\Tr\,T_n\alpha\beta(\lambda_n)\rangle$. The term proportional to
$\Tr\,T_1T_2\dots T_n$ arises from contracting $\beta(\lambda_i)$
with $\alpha(\lambda_{i+1})$ for $i=1,\dots,n$. The computation is
done with free fields on $C=\Bbb{CP}^1$; the result is a function
only of the $\lambda_i$, since the equations \xnon\ that
characterize $C$ let us express the other variables in terms of
the $\lambda_i$. In fact, for doing this computation, we can just
think of $C$ as a copy of $\Bbb{CP}^1$ with homogeneous
coordinates $\lambda$. The result of computing the free field
correlation function is that the desired part of $\langle
J(\lambda_1) J(\lambda_2)\dots J(\lambda_n)\rangle$ is
\eqn\ponon{\prod_{i=1}^n\epsilon_{ab}\lambda_i^ad\lambda_i^b
\prod_{i=1}^n{1\over\langle \lambda_{i+1},\lambda_{i}\rangle}.}
This expression is completely determined by the following
properties: it is homogeneous of degree zero in each $\lambda_i$
(so it makes sense), it a $(1,0)$-form in each variable
$\lambda_i$ (because each current $J(\lambda_i)$ is a
$(1,0)$-form), it is $SL(2,\Bbb C)$-invariant, and it has a simple
pole at $\lambda_{i+1}=\lambda_i$ because of the contraction of
$\beta(\lambda_i)$ with $\alpha(\lambda_{i+1})$. Perhaps the
formula \ponon\ is more familiar if written in terms of
$z_i=\lambda_i^2/\lambda_i^1$.  It then takes the form
\eqn\bonon{dz_1\,dz_2\dots dz_n\prod_{i=1}^n{1\over
z_{i+1}-z_{i}},} where $1/(z_{i+1}-z_{i})$ is the usual
free-fermion propagator on the complex $z$-plane.  One can
calculate this readily by using homogeneity in the $\lambda_i$ to
set $\lambda_i=(1,z_i)$ for all $i$, whereupon
\eqn\huono{\eqalign{\epsilon_{ab}\lambda^ad\lambda^b&=dz\cr
\langle\lambda_i,\lambda_{i+1}\rangle&=z_{i+1}-z_i.\cr}} The
scattering amplitude is thus (with the gauge theory trace
$\Tr\,T_1\dots T_n$ suppressed, as usual) \eqn\jugv{ A(v_i)= \int
d^4x \,d^8\theta \prod_{i=1}^n\int_C v_i(\lambda^a_i, \mu^{\dot
a}_i,\psi_i^A)\epsilon_{ab}\lambda_i^ad\lambda_i^b
\prod_{i=1}^n{1\over \langle \lambda_i,\lambda_{i+1}\rangle}.}
Here $\mu_i$ and $\psi_i$ are functions of $\lambda_i$, $x$, and
$\theta$ (obeying \xnon), since $v$ is evaluated on $C$. What is
integrated over $C$ is a $(1,1)$-form in each variable, since
$v_i$ is a $(0,1)$-form and $\epsilon_{ab}\lambda^a d\lambda^b$ is
a $(1,0)$-form. Clearly, this result is closely related to the
desired answer of eqn. \tyno.  To finish the derivation, we need
to convert the formula from the language of $\bar\partial$
cohomology to a formalism more like that used in section 3.

This can be done using the link between $\bar\partial$ cohomology
and Cech cocycles that is explained at the end of the appendix. In
order to carry out the calculation, we again use the homogeneity
of the twistor space variables to set $\lambda^1=1$ for each
particle. We write $z$ for $\lambda^2/\lambda^1$, as we already
did in \bonon, and leave unchanged the names of the rest of the
twistor coordinates $\mu^{\dot a},\, \psi^A$.  The homogeneity can
be restored at the end of the computation, if one wishes, by
multiplying by suitable powers of $\lambda^1$ and reversing the
steps in \huono.

As in the derivation of eqn. A.21, we write $z=\sigma+i\tau$, with
$\sigma$ and $\tau$ real.  We saw in eqn. A.23 that we can pick
the external wavefunctions to be \eqn\oncono{v_k={i\over 2}f_k
\,d\bar z_k\delta(\tau_k),} where $f_k$ is a holomorphic function
(whose singularities are far away from $ \tau_k=0$). Upon
inserting this in \jugv, writing $(i/2)dz\wedge d\bar
z=d\sigma\wedge d\tau$, and doing the $\tau$ integrals with the
help of the delta functions, we get \eqn\ugv{A(f_i)=\int
d^4x\,d^8\theta\prod_{i=1}^n\int_{-\infty}^\infty d\sigma_i
f_i(\sigma_i,\mu_i^a,\psi_i^A)\prod_{i=1}^n{1\over
\sigma_{i+1}-\sigma_i}.} Again, $\mu_i$ and $\psi_i$ are functions
of $\sigma_i$, $x$, and $\theta$ in such a way that the integral
runs on the curve $C$.  We can make this explicit: \eqn\buhugv{
A(f_i)= \prod_{i=1}^n\int_{-\infty}^\infty d\sigma_i\,d^2\mu^{\dot
a}_i \,d^4\psi^A_i f_i(\sigma_i,\mu_i^{\dot a},\psi_i^A)\tilde
A(\sigma_i,\mu_i^{\dot a},\psi_i^A),} where \eqn\jujug{ \tilde
A(\sigma_i,\mu_{i\,\dot a},\psi_i^A)=\int d^4x\,d^8\theta
\prod_{i=1}^n\delta^2(\mu_{i\,\dot a} +x_{a\dot a}\lambda_i^a)
\delta^4(\psi_i^A+\theta_a^A\lambda_i^a) \prod_{i=1}^n{1\over
\sigma_{i+1}-\sigma_i}.} The integral in \buhugv\ is carried out
over real twistor space -- that is, the integration variables
$\sigma_i$ and $\mu_i$ are all real. In $\tilde A$, we recognize
the MHV tree level scattering amplitude of eqn. \tyno, written
(with the help of \huono) in a coordinate system with
$\lambda_i^1=1$.
 The integral in \buhugv\ is the pairing
(described in sections 2.5 and 2.6) by which one integrates over a
copy of twistor space for each initial and final particle to go
from $\tilde A$ to a scattering amplitude with specified initial
and final states.

  In our derivation, the integral $d\sigma\,d^2\mu$ over real twistor
space $\Bbb{RP}^3$ arose in two steps: $z$ became real
because of the particular choice of external wavefunctions, and $\mu$ became real
because, for curves of degree one,
with our choice of real slice $\CM_{\Bbb R}$, $z$ being real leads
to $\mu$ being real.

The attentive reader might ask why we need not include additional
contributions where two external particles join in twistor space
(to couple to a quantum $\CA$-field that then propagates to the
$D$-instanton), using the $\CA\wedge \CA \wedge \CA$ term in the
twistor space effective action.  With our gauge choice, this does
not occur because the wedge products of the wavefunctions in
\oncono\ all vanish, as those wavefunctions are all proportional
to $d\bar z$.

We have obtained the tree level MHV amplitudes in terms of correlation functions
 of chiral currents on $\Bbb{CP}^1$, as suggested
by Nair  \nair.  In
Nair's paper, this is an abstract $\Bbb{CP}^1$, while in our framework, it is
a curve in twistor space.  Correlators of chiral currents are what one often
gets from  heterotic string worldsheet instantons, but we have obtained
them from $D$-instantons.

\newsec{Further Issues}

Here we will take a brief survey of a few further issues.

\subsec{Closed Strings}

\nref\aaa{E. S. Fradkin and A. A. Tseytlin, ``Conformal Supergravity,''
Phys. Rept. {\bf 119} (1985) 233.}%
\nref\bbb{A. Salam and E. Sezgin, {\it Supergravities
In Diverse Dimensions} (World Scientific, 1989),  vol. 2.}%
The most serious outstanding issue may be to understand the closed
strings. In the topological $B$ model in general, closed string
modes describe deformations of the complex structure of the target
space.  In the present problem, the target space is supertwistor
space $\Bbb{PT}'$.  Deformations of the complex structure of
twistor space describe -- according to the original application of
twistor theory to nonlinear problems \nongrav\ -- conformally
anti-selfdual deformations in the geometry of Minkowski spacetime.
In the case of supertwistor space, one would presumably get some
sort of chiral limit (analogous to the $GF$ theory studied  in
section 4.4 for open strings) of $\N=4$ conformal supergravity,
perhaps extended to a more standard theory with the aid of
$D$-instanton contributions. (For some reviews of conformal
supergravity, see \refs{\aaa,\bbb}.)
 This remains to be properly understood.

The holomorphic anomaly of the $B$ model \cec, which usually obstructs the
background independence of the closed string sector of the $B$
model, presents a conundrum. As the closed strings in this problem
presumably describe  gravitational fluctuations in spacetime,
we need to maintain the background independence.  Possibly, the anomalous
$S$ symmetry, which eliminates most string loop effects, avoids the
holomorphic anomaly in the present context.

There actually is a sign of closed string contributions in the
calculation of tree level MHV amplitudes in section 4.7.  There,
we extracted a single-trace interaction, and found it to agree
with the standard tree-level result of Yang-Mills theory. However,
the underlying formula \umbu\ for the amplitude also gives rise to
multi-trace interactions.  Where can they come from? The most
likely explanation is that they arise from the exchange of closed
string states that, being singlets of the $GL(N,\Bbb C)$ gauge
group, naturally produce multi-trace interactions.

To support this idea, we will analyze the four-gluon multi-trace
interactions that arise from \umbu.  In doing so, we only consider
gluons in the $SL(N,\Bbb C)$ subgroup of $GL(N,\Bbb C)$; the gluons
that gauge the center of $GL(N,\Bbb C)$ are likely to mix with closed
string modes (by analogy with a familiar mechanism for the usual critical
string theories), and one would not expect to be able to understand
the resulting scattering amplitudes without understanding this mixing.
This being so, we assume that $\Tr\,T_i=0$ for all $i$.
This only allows, up to a permutation of the gluons, one possible
group theory factor in a multi-trace four-gluon amplitude;
we can assume the group theory factor to be
 $\Tr\,T_1T_2\,\Tr\,T_3
T_4$. There are two essentially different cases: the helicities may
be $++--$ or $+-+-$.  Other cases are related to these by the obvious
permutation symmetries (exchanging 1 with 2, 3 with 4, or 1,2 with 3,4).

  The momenta of the four gluons are denoted
as usual $p_i^{a\dot a}=\lambda_i^a\tilde\lambda_i^{\dot a}$.
We consider first the $++-\,-$ amplitude.
The amplitude extracted from \umbu\ is
\eqn\numbu{A=(2\pi)^4\delta^4(\sum_ip_i)\Tr\,T_1T_2\,\Tr\,T_3T_4
\langle\lambda_3,\lambda_4\rangle^4{1\over
\langle\lambda_1,\lambda_2
\rangle^2\langle\lambda_3,\lambda_4\rangle^2}.}  In contrast to
our usual practice, we have written the group theory factor, since
it is unusual. This amplitude is conformally invariant, by the
same analysis as in section 2.4. The derivation of \numbu\ goes as
follows.  The factor
$1/\langle\lambda_1,\lambda_2\rangle^2\langle\lambda_3,\lambda_4\rangle^2$
comes from the current correlation function that is needed to get
a group theory factor $\Tr\,T_1T_2\,\Tr\,T_3T_4$.  (The relevant
contribution to
$\langle\Tr\,T_1J(\lambda_1)\Tr\,T_2J(\lambda_2)\dots\Tr\,T_4J(\lambda_4)\rangle$
is the disconnected piece $ \langle\Tr\,T_1J(\lambda_1)
\Tr\,T_2J(\lambda_2)\rangle\langle\Tr\,T_3J(\lambda_3)\Tr\,T_4J(\lambda_4)\rangle$,
leading to a double contraction in both the
$\lambda_1$-$\lambda_2$ and $\lambda_3$-$\lambda_4$ channels.) The
factor $\langle\lambda_3,\lambda_4\rangle^4$ in the numerator
(which is also present in the numerator of the conventional
single-trace MHV amplitude \hutu, where it arises in the same way)
is one of the terms that comes from the $d^8\theta$ integral.  We
simply picked the term associated with the helicity configuration
$++--$. If we let $k=p_1+p_2$, and observe that
$k^2=(p_1+p_2)^2=2p_1\cdot
p_2=2\langle\lambda_1,\lambda_2\rangle[\tilde\lambda_1,
\tilde\lambda_2]$, we can write
\eqn\bumbu{A=4(2\pi)^2\delta^4(\sum_ip_i)\Tr\,T_1T_2\,\Tr\,T_3T_4
{[\tilde\lambda_1,\tilde\lambda_2]^2
\langle\lambda_3,\lambda_4\rangle^2\over (k^2)^2}.} We can
reproduce this amplitude from tree level exchange of a scalar
field $\phi$ with a propagator $1/(k^2)^2$ (as expected for a
scalar field in conformal supergravity, which is a nonunitary
theory with higher derivatives) and a coupling $\phi\,\,\Tr
F_{\mu\nu}F^{\mu\nu}$.   Indeed, the matrix element of $\Tr
F_{\mu\nu}F^{\mu\nu}$ to create two photons of momentum $p_1,p_2$
and $+$ helicity is $[\tilde\lambda_1,\tilde\lambda_2]^2$, while
the matrix element of the same operator to create two photons of
momentum $p_3,p_4$ and $-$ helicity is $\langle
\lambda_3,\lambda_4\rangle^2$.  (We do not need to include scalar exchange
in crossed channels, as this produces other group theory factors.)
Actually, to avoid generating $++++$ and
$----$ amplitudes that are not present in \umbu, we need a slight elaboration
of this mechanism: two scalars $\phi_+$ and $\phi_-$, with couplings
$\phi_+\Tr\,(F^+)^2$, $\phi_-\Tr\,(F^-)^2$ to gluons of one helicity or
the other, and a purely off-diagonal propagator $\langle\phi_+\phi_+\rangle
=\langle\phi_-\phi_-\rangle=0$, $\langle\phi_+\phi_-\rangle=1/k^4$.

The $+-+-$ amplitude can be understood similarly in terms of
graviton exchange.  The amplitude is read off from \umbu\ to be
\eqn\numbun{A=(2\pi)^4\delta^4(\sum_ip_i)\Tr T_1T_2\,\Tr T_3T_4
\langle\lambda_2,\lambda_4\rangle^4{1\over
\langle\lambda_1,\lambda_2
\rangle^2\langle\lambda_3,\lambda_4\rangle^2}.} Using identities
such as \guyy, this can be rewritten
\eqn\umbun{\eqalign{A=&(2\pi)^4\delta^4(\sum_ip_i)\Tr T_1T_2\,\Tr
T_3T_4
{[\tilde\lambda_1,\tilde\lambda_3]^2\langle\lambda_2,\lambda_4\rangle^2
\over
\langle\lambda_1,\lambda_2\rangle^2[\tilde\lambda_1,\tilde\lambda_2]^2}
\cr =&4(2\pi)^4\delta^4(\sum_ip_i)\Tr T_1T_2\,\Tr T_3T_4
{[\tilde\lambda_1,\tilde\lambda_3]^2\langle\lambda_2,\lambda_4\rangle^2
\over (k^2)^2}.\cr}} We now consider a traceless metric
fluctuation $h_{\mu\nu}$ that in spinor language is written
$h_{ab\dot a\dot b}$ (symmetric in $a,b$ and in $\dot a,\dot b$)
with propagator \eqn\toncon{\langle h_{ab\dot a \dot b}h_{cd\dot
c\dot d}\rangle ={1\over
4}{(\epsilon_{ac}\epsilon_{bd}+\epsilon_{ad}\epsilon_{bc})
(\epsilon_{\dot a\dot c}\epsilon_{\dot b\dot d}+\epsilon_{\dot
a\dot d}\epsilon_{\dot b\dot c})\over (k^2)^2}.} We assume that
$h$ couples to gluons via a coupling $h^{ab\dot a\dot b} T_{ab\dot
a\dot b}$, where $T$ is the stress tensor.  The matrix element of
$T_{ab\dot a\dot b}$ to create two gluons of momenta $p_1,p_2$ and
helicities $+,-$ is $\tilde\lambda_{1\,\dot
a}\tilde\lambda_{1,\dot b}\lambda_{2\,a} \lambda_{2\,b}$, and
similarly with $1,2$ replaced by $3,4$.  Combining this matrix
element with the propagator in \toncon, we recover the amplitude
\umbun.

The tentative conclusion is that the $B$ model of $\Bbb{CP}^{3|4}$ has a closed
string sector which describes some sort of ${\cal N}=4$ conformal supergravity.
  If so, this $B$ model describes ${\cal N}=4$ Yang-Mills theory only for
planar amplitudes, in which the closed strings decouple.

Consideration of anomalies raises numerous puzzles that will not
be addressed here. The world-volume determinants of $D1-D5$ and
$D1-D1$ strings appear potentially anomalous; anomaly cancellation
may well involve contributions of closed strings, as in the
Green-Schwarz mechanism of heterotic and Type I anomaly
cancellation.  The $c$-number conformal anomaly of $\N=4$ super
Yang-Mills theory raises the question of how it could possibly be
coupled to any version of conformal supergravity.  Perhaps there
is more to the story.

\subsec{Yangian Symmetry}

The planar limit of $\N=4$ super Yang-Mills theory seems to have
an  extended infinite-dimensional symmetry group  that can be
described as Yangian symmetry. This result was first found in
strong coupling in \ref\polchin{I. Bena, J. Polchinski, and R.
Roiban, ``Hidden Symmetries Of The ${\rm AdS}_5\times S^5$
Superstring,'' hep-th/0305116.} and has also been found for weak
coupling \ref\dnw{L. Dolan, C. R. Nappi, and E. Witten, ``A
Relation Among Approaches To Integrability In Superconformal
Yang-Mills Theory,'' JHEP 0310:017 (2003), hep-th/0308089. }. We
therefore should look for such symmetry in the present framework.

Along with many two-dimensional models \ref\whop{M. Luscher and K.
Pohlmeyer, ``Scattering Of Massless Lumps And Nonlocal Charges In
The Two-dimensional Classical Non-linear Sigma Model,'' Nucl.
Phys. {\bf B137} (1978) 46.}, the two-dimensional sigma model with
target space $ \Bbb{CP}^{M-1}$ has nonlocal symmetries that
generate a Yang-Baxter or Yangian algebra, as investigated in
\ref\devega{ H. J. de Vega, H. Eichenherr and J. M. Maillet,
``Yang-Baxter Algebras Of Monodromy Matrices In Integrable Quantum Field Theories,''
Nucl. Phys. {\bf B240}
(1984) 377.}. However (in contrast to similar models
in which the target space is, for example, a sphere), the quantum
version of this model is believed to not be integrable \ref\bw{B.
Berg and P. Weisz, ``Exact $S$ Matrix Of The Adjoint $SU(N)$
Representation,'' Commun. Math. Phys. {\bf 67} (1979) 241.};
presumably, the nonlocal symmetries are anomalous, as the local
ones appear to be \ref\goldw{Y. Goldschmidt and E. Witten,
``Conservation Laws In Some Two-Dimensional Models,'' Phys. Lett.
{\bf 91B} (1980) 392.}.

The supersymmetric $\Bbb{CP}^{M-1}$ model also has Yangian
symmetry classically.  Quantum mechanically, it is believed to be
integrable with a factorizable $S$-matrix,
and anomaly-free Yangian symmetry \ref\oab{
E. Abdalla, in {\it Non-linear Equations in Classical and Quantum
Field Theory}, ed. N. Sanchez,  Springer Lectures in Physics, vol. 226, p. 140;
 E. Abdalla, M C B Abdalla and M Gomes, Phys. Rev. {\bf  D25} (1982) 452,
{\bf D27} (1983) 825.}.
  Granted this, Yangian symmetry will
also hold for $\Bbb{CP}^{M-1|P}$, as the anomalies generated by Feynman
diagrams  really only depend on $M-P$.  If
we set $M-P=0$, even more anomalies (such as the beta function) cancel.
So the $\Bbb{CP}^{3|4}$ model can be expected to have Yangian symmetry
at the quantum level.

The Yangian symmetry, like the more obvious $PSU(4|4)$ symmetry, commutes
with spacetime supersymmetry.  It also has no anomaly with the $U(1)$ $R$-symmetry
current by which we ``twist'' to make the topological $B$ model.  So Yangian
symmetry is expected in the $B$ model of $\Bbb{CP}^{3|4}$.

\subsec{Other Target Spaces}

What models can we make by replacing $\Bbb{CP}^{3|4}$ with another target space?

We can certainly replace $\Bbb{CP}^3$, which is the twistor space of Minkowski
space, by the twistor space of a more general
conformally anti-selfdual four-dimensional spacetime $X$.
(This twistor space is the space of null selfdual
complex planes in $X$, generalizing the
$\alpha$-planes introduced in the appendix.)
The topological $B$ model of this twistor space (or rather of its extension
with $\N=4$ supersymmetry) will describe $\N=4$ super Yang-Mills theory on $X$,
in the same sense that the topological $B$ model of $\Bbb{CP}^{3|4}$ describes
$\N=4$ super Yang-Mills theory in Minkowski space.

A more interesting generalization is to consider the weighted projective
space $\Bbb W=\Bbb{WCP}^{3|2}(1,1,1,1|1,3)$.  This is the projective space
with four bosonic homogeneous coordinates $Z^I$, $I=1,\dots,4$, of weight one,
 and two
fermionic homogeneous coordinates $\psi,\chi$, of weights one and three.
The homogeneous coordinates are  subject to the equivalence
relation  $(Z^I,\psi,\chi)\cong (t Z^I,t \psi,t^3\chi)$, for $t\in\Bbb C^*$.
$\Bbb W$ is a Calabi-Yau supermanifold because the sum of bosonic weights equals the
sum of fermionic weights.  The holomorphic
measure $\Omega_0=dZ^1\dots dZ^4 d\psi d\chi$ is invariant under $\Bbb C^*$,
and descends to a holomorphic measure $\Omega$ on $\Bbb{W}$,
ensuring that one can define a topological $B$ model with this target space.
The supermanifold $\Bbb W$ admits $\N=1$ superconformal symmetry $SU(4|1)$,
acting on $Z^I$ and $\psi$.  In the presence of $N$ (almost) space-filling
$D5$-branes
like those studied in this paper, the spectrum of the model is the $U(N)$ vector
multiplet with $\N=1$ supersymmetry, as one can verify by repeating the analysis of
section 4.3 for this case.

However, the topological $B$ model with target $\Bbb W$ cannot reproduce
$\N=1$ super Yang-Mills theory, as one would have hoped, because it has too
much symmetry.  To have any hope of reproducing $\N=1$ super Yang-Mills theory,
one would have to modify the model to deal with two problems:
 (A) In this model, the $SU(4|1)$ symmetry will persist quantum
mechanically, while in $\N=1$ super Yang-Mills symmetry, there is a conformal
anomaly that breaks $SU(4|1)$ to a subgroup.  (B) The $B$ model with target
$\Bbb W$ has additional symmetries $\delta \chi=P_3(Z,\psi)$, with $P_3$
a homogeneous polynomial of degree three.  These have no analog in $\N=1$ super
Yang-Mills theory.

\bigskip\noindent{\it The Quadric}

One other possible model is worth mentioning here. First of all,
if $\Bbb A$ is a copy of $\Bbb{CP}^{M-1}$ with homogeneous
coordinates $Z^I$,  then there is a natural ``dual'' projective
space $\Bbb B$ whose points parameterize hyperplanes in $\Bbb A$.
The equation of a hyperplane is \eqn\orm{\sum_{i=1}^NW_IZ^I=0,}
for some constants $W_I$, not all zero.  Moreover, an overall
scaling of the $W_I$ would give the same hyperplane.  So we take
$W_I$ as homogeneous coordinates for $\Bbb B$.  The relation
between $\Bbb A$ and $\Bbb B$ is clearly symmetric: $\Bbb B$
parameterizes hyperplanes in $\Bbb A$, and vice-versa.  We can
regard the equation \orm\ in one more way: its zero set defines a
``quadric'' $\Bbb Q$ in the product $\Bbb A\times\Bbb B$.

Now let $\Bbb A$ be the complex supermanifold $\Bbb{CP}^{3|3}$,
with homogeneous coordinates $Z^I,\psi^A$, $I=1,\dots,4$,
$A=1,\dots,3$. Let $\Bbb B$ be the dual projective supermanifold
$\Bbb{CP}^{3|3}$, parametrizing hyperplanes in $\Bbb A$.  We write
$W_I,\chi_A$, for the homogeneous coordinates of $\Bbb B$.  The
equation via which $\Bbb B$ parameterizes hyperplanes in $\Bbb A$,
and vice-versa, is
\eqn\torm{\sum_{I=1}^4Z^IW_I+\sum_{A=1}^3\psi^A\chi_A=0.} The zero
set of this equation is a quadric $\Bbb Q$ in
$\Bbb{CP}^{3|3}\times \Bbb{CP}^{3|3}$.

$\Bbb A$ and $\Bbb B$ are not Calabi-Yau supermanifolds, but $\Bbb Q$ is one.
(This is so because the first Chern class of $\Bbb A\times \Bbb B$ is $(1,1)$,
which is also the degree of the equation defining $\Bbb Q$.)
The topological $B$ model with target $\Bbb Q$ therefore exists, and should
describe a theory with symmetry group containing $SU(4|3)$, which is the symmetry
group of $\Bbb Q$.  The only evident four-dimensional field theory with
symmetry $SU(4|3)$ is $\N=4$ super Yang-Mills theory, which has the larger
symmetry $PSU(4|4)$.

By an analog of the twistor transform \witten, a holomorphic vector bundle
on a suitable region of $\Bbb Q$ corresponds to a solution of the equations of
$\N=4$ super Yang-Mills
theory on a suitable region of complexified and compactified Minkowski spacetime
$\Bbb M$ (in a description in which only $SU(4|3)$ is manifest). Essentially
the same construction was also obtained in a bosonic language \iyg.
The equations that arise here
 are the full Yang-Mills equations, not the selfdual or anti-selfdual
 version.  The intuition behind the construction was that the dependence
on $Z$ encodes the gauge fields of one helicity, and the dependence on $W$ encodes
the other.

It is therefore plausible that the topological $B$ model of $\Bbb
Q$ might give another construction of $\N=4$ super Yang-Mills
theory.  In this model, no $D$-instanton contributions would be
needed, and the mechanism by which perturbative Yang-Mills theory
would be reproduced would be completely different from what it is
in the case of $\Bbb{CP}^{3|4}$.  The main difficulty in making
sense of this idea seems to be that it is hard to understand the
right measure for the bosonic and fermionic zero modes on a
space-filling $D$-brane on $Q$. A somewhat similar problem was
treated recently by Movshev and Schwarz \mov,
who showed how to construct an ``integral form''
that enables one to define a suitable Chern-Simons action  on
certain complex supermanifolds that are related to super
Yang-Mills theory in roughly the same way that $\Bbb Q$ is. Their
motivation was in part to understand the covariant quantization of
the Green-Schwarz superparticle and superstring via pure spinors
\ref\berk{N. Berkovits, ``Super Poincar\'e Covariant Quantization
Of The Superstring,'' JHEP 0004:018 (2000), hep-th/0001035,
``Covariant Quantization Of The Superparticle Using Pure
Spinors,'' hep-th/0105050, JHEP 0109:016 (2001).}, which in some
ways is a cousin of twistor constructions. Many of their examples
have nonzero first Chern class and hence no topological $B$ model,
but their method of construction of the measure may be relevant to
understanding the topological $B$ model of $\Bbb Q$.

\appendix{A}{A Mini-Introduction To Twistor Theory}

Though there are numerous introductions to twistor theory
\refs{\penrind,\mcpen,\revminus-\revtwo}, and its applications to
Yang-Mills fields \atiyah, we will here offer a mini-introduction
to a few facets of the subject, with the aim of making the present
paper more accessible.  We begin by explaining the twistor
transform of the anti-selfdual Yang-Mills equations, following
Ward \ward, who developed the analog for gauge fields of the
Penrose transform \nongrav\ of the anti-selfdual Einstein
equations.  Then we explain the twistor transform of the linear
massless wave equations \penroseb.

\bigskip\noindent{\it Self-Dual Yang-Mills Fields}

We will describe a one-to-one correspondence between the following
two types of object:

(1) A  $GL(N,\Bbb C)$-valued gauge field $A_{a\dot a}(x^{b\dot
b})$ that obeys the anti-selfdual Yang-Mills equations on
complexified (but not compactified) Minkowski spacetime $\Bbb M'$.
($A$ is a connection on a holomorphic vector bundle $H$ over $\Bbb
M'$, which is automatically holomorphically trivial as $\Bbb
M'\cong \Bbb C^4$.) Here the $x^{b\dot b}$ are complex variables,
the $A_{a\dot a}$ are entire holomorphic functions of $x^{b\dot
b}$, and the curvature of $A$, which we write as $F=dA+A\wedge A$,
is anti-selfdual.  In spinor notation, with $F_{ab\dot a\dot
b}=[D_{a\dot a},D_{b\dot b}]$, anti-selfduality means that
\eqn\gufo{F_{ab\dot a\dot b}=\epsilon_{a b}\Phi_{\dot a\dot b}}
for some $\Phi_{\dot a \dot b}$.

(2) A rank $N$ holomorphic vector bundle $E$ over $\Bbb{PT}'$
(defined as the region of $\Bbb{PT}=\Bbb{CP}^3$  with
$\lambda^a\not= 0$) such that $E$ is holomorphically trivial when
restricted to each genus zero, degree one curve in $\Bbb{PT}'$.

What is remarkable about this construction is that in (1) we
impose a nonlinear differential equation, the anti-selfdual
Yang-Mills equation (and the purely holomorphic structure is
trivial), but in (2) we only ask for holomorphy. In a sense,
therefore, the correspondence solves the anti-selfdual Yang-Mills
equations.

This correspondence has numerous analogs and important
refinements.  One important point that we will omit (referring the
reader to standard references such as \atiyah) is that in $++++$
or $++-\,-$ signature (that is, whenever the anti-selfdual
Yang-Mills equations are real), one can impose a reality condition
and reduce the gauge group to $U(N)$ rather than $GL(N,\Bbb C)$ on a real
slice of $\Bbb M'$.

We will make the correspondence between (1) and (2) in a
computational way, and then explain it more conceptually.

A central role in the correspondence is played by the twistor equation,
which by now should be familiar to the reader:
 \eqn\polyo{\mu_{\dot
a}+x_{a\dot a}\lambda^a=0.}
This equation can be read in two ways.  If $x$ is given, and \polyo\ is
regarded as an equation for $\lambda$ and $\mu$, then it defines a curve
in twistor space of genus zero and degree one that we will call
$\Bbb D_x$.  Complexified Minkowski space is the moduli space of such curves,
a fact that we have extensively used in this paper.

Alternatively, if $\lambda$ and $\mu$ are given, and \polyo\ is
regarded as an equation for $x$, then the solution set $\Bbb K$
(or $\Bbb K(\lambda,\mu)$) is a two-dimensional complex subspace
of complexified Minkowski space $\Bbb M'$.  It is completely null
(any tangent vector to $\Bbb K$ is a null vector) and in a certain
sense is selfdual.
  Penrose calls $\Bbb K$ an $\alpha$-plane.
Thus, twistor space is the moduli space of $\alpha$-planes.

 Translations within the
$\alpha$-plane are generated by the operators
\eqn\torno{\partial_{\dot a}=\lambda^a{\partial\over\partial
x^{a\dot a}},~~\dot a = 1,2.}  These translations take the form
\eqn\takeform{x^{a\dot a}\to x^{a\dot a}+\lambda^a\epsilon^{\dot
a},} for arbitrary $\epsilon^{\dot a}$.

The significance of anti-selfduality for our purposes is that it
means that when restricted to an $\alpha$-plane, the gauge field
becomes flat.  We can verify this straightforwardly.  We define
\eqn\orno{D_{\dot a}=\lambda^a{D\over Dx^{a\dot a}},} with
$D/Dx^{a\dot a}$ the covariant derivative with respect to the
anti-selfdual gauge field $A$.  Then \eqn\forno{[D_{\dot
a},D_{\dot b}]=\lambda^a\lambda^b\left[{D\over Dx^{a\dot
a}},{D\over Dx^{b\dot
b}}\right]=\lambda^a\lambda^b\epsilon_{ab}\Phi_{\dot a\dot b}=0,}
where \gufo\  has been used.

Now,  let $V_1$ be the region of $\Bbb{PT}'$ in which $\lambda^1\not=
0$, and let $V_2$ be the region with $\lambda^2\not= 0$.  In
$\Bbb{PT}'$, $\lambda^1$ and $\lambda^2$ are not allowed to both
vanish, so $V_1$, $V_2$ give an open cover of $\Bbb{PT}'$.  $V_1$
and $V_2$ are both copies of $\Bbb C^3$ (for example, $V_1$ can be
mapped to $\Bbb C^3$ using the coordinates $\lambda^2/\lambda^1$,
$\mu^{ 1}/\lambda^1$, $\mu^{ 2}/\lambda^1$).  So a holomorphic
vector bundle on $V_1$ or $V_2$ is automatically holomorphically
trivial.  A holomorphic vector bundle $E$ on $\Bbb{PT}'$ can
therefore be defined by giving a ``transition function'' on
$V_{12}=V_1\cap V_2$.  This is a holomorphic function $U:V_{12}\to
GL(N,\Bbb C)$.  Explicitly, $U$ is a $GL(N,\Bbb C)$-valued
holomorphic function $U(\lambda,\mu)$, that is homogeneous in
$\lambda$ and $\mu$ of degree zero  and singular only if
$\lambda^1=0$ or $\lambda^2=0$. (Given $U$, the bundle $E$ is
defined by using $U$ to glue a trivial rank $N$ complex bundle
$F_1$ on $V_1$ to a trivial rank $N$ bundle $F_2$ on $V_2$.)  Two
transition functions $U$ and $U'$ define isomorphic bundles on
$\Bbb{PT}'$ if and only if we can write
\eqn\wedcan{U'=U_1UU_2^{-1},} where $U_1:V_1\to GL(N,\Bbb C)$ is
holomorphic throughout $V_1$ and likewise $U_2:V_2\to GL(N,\Bbb
C)$ is holomorphic throughout $V_2$.  If this is the case, then
$U$ can be converted to $U'$ by making gauge transformations of
$F_1$ and $F_2$ via $U_1$ and $U_2$, prior to the gluing.

Now as long as $(\lambda,\mu)\in V_1$,  the $\alpha$-plane $\Bbb
K$ contains a unique point $P(\Bbb K)$ with $x^{1\dot a}=0$. To
prove this, we just observe that if $\lambda^1\not= 0$, we can use
the translations \takeform\ to set $x^{1\dot a}=0$ in a unique
manner. Likewise, for $(\lambda,\mu)\in V_2$, $\Bbb K$ contains a
unique point $Q(\Bbb K)$ with $x^{2\dot a}=0$. For
$(\lambda,\mu)\in V_{12}$, $P(\Bbb K)$ and $Q(\Bbb K)$ are both
defined and vary holomorphically with $\lambda$ and $\mu$, and we
can set \eqn\orfog{U(\lambda,\mu) =P\exp\int_{Q(\Bbb K)}^{P(\Bbb
K)}A.} The integral is taken over any contour in $\Bbb K$.  The
choice of contour does not matter, since the gauge field is flat
when restricted to $\Bbb K$. Since $U$ is defined throughout
$V_{12}$ and takes values in $GL(N,\Bbb C)$, we can use $U$ to
determine a holomorphic vector bundle $E$ over $\Bbb{PT}'$.

If in making this construction, we replace $A$ by a
gauge-equivalent field, via $(\partial+A)\to Y(\partial+A)Y^{-1}$
for some holomorphic $GL(N,\Bbb C)$-valued field $Y$ on spacetime,
then $U$ transforms to $\tilde U=Y_PUY_Q^{-1}$, where $Y_P$ and
$Y_Q$ denote the values of $Y$ at $P(\Bbb K)$ and $Q(\Bbb K)$. As
$Y_P$ is holomorphic and invertible throughout $V_1$, and $Y_Q$
throughout $V_2$, the holomorphic vector bundles defined by
$\tilde U$ and $U$ are isomorphic.

We have almost shown how, from an object of type (1), to produce
an object of type (2).  We still must show that $E$ is
holomorphically trivial when restricted to a genus zero, degree
one curve in $\Bbb{PT}'$.  These are precisely the curves $\Bbb
D_x$ for some $x\in \Bbb M$, as described by \polyo.  To show that
$E$ is holomorphically trivial when restricted to $\Bbb D_x$, we
first restrict $U$ to $\Bbb D_x$, which is done by regarding
$\mu_{\dot a}$ as a function of $x$ and $\lambda$ that obeys the
twistor equation: $\mu_{\dot a}=-x_{b\dot a}\lambda^b$.  The
transition function of $E$ restricted to $\Bbb D_x$ is thus simply
$W(\lambda^a,x_{b\dot b})=U(\lambda^a,-x_{b\dot a}\lambda^b)$. To
show that the restriction of $E$ is trivial, we must show that $W$
can be factored holomorphically as $W=W_1W_2^{-1}$, where $W_1$ is
singular only at $\lambda^1=0$ and $W_2$ is singular only at
$\lambda^2=0$.  We simply define \eqn\wedef{\eqalign{W_1 & =
P\exp\int_x^{P(\Bbb K)}A\cr
                    W_2 & = P\exp\int_x^{Q(\Bbb K)}A.\cr}}
For any given $\lambda$, the contours are taken within the
$\alpha$-plane $\Bbb K$ of that given $\lambda$ which contains
$x$. Clearly, $W=W_1W_2^{-1}$.  The ability to make this
factorization depends on choosing $x$; in general $U$ has no such
factorization, but $W$ does.

To establish the converse, we start with a holomorphic bundle $E$
on $\Bbb{PT}'$ that is trivial on each $\Bbb D_x$.  We can assume
that $E$ is defined by a holomorphic transition function
$U(\lambda,\mu):V_{12}\to GL(N,\Bbb C)$, which is homogeneous of
degree zero. Now we reverse the above construction. We define
$W(\lambda^a,x_{b\dot b})=U(\lambda^a,-x_{b\dot a}\lambda^b)$. For
any fixed $x$, $W$ is homogeneous in $\lambda$ of degree zero.
From the definition of $\partial_{\dot a}$ and the chain rule, we
learn immediately that \eqn\urf{\partial_{\dot a}W=0.} The
holomorphic triviality of $E$ when restricted to each $\Bbb D_x$
means that $W$ can be factored
\eqn\wedef{W(\lambda,x)=W_1W_2^{-1},} where the $W_i$, $i=1,2$, are
singular only at $\lambda^i=0$.  If we plug this factorization
into \urf, we learn that \eqn\uncon{W_1^{-1}\partial_{\dot a}W_1=
W_2^{-1}\partial_{\dot a}W_2.} (This is understood as a
differential operator, via $W^{-1}\partial W = \partial
+W^{-1}[\partial, W]$.)  The left hand side of \uncon\ can only be
singular at $\lambda^1=0$. The right hand side can only be
singular at $\lambda^2=0$.  As they are equal, there can be no
singularity at all.  We define $D_{\dot a}$ to equal the left or
right hand side of \uncon.  It is homogeneous in $\lambda$ of
degree 1, since $\partial_{\dot a}$ has this property, and it is
clearly of the form $D_{\dot a}=\partial_{\dot a}+A_{\dot
a}(\lambda,x)$, where $A_{\dot a}$ is some function of $\lambda$
and $x$ valued in the Lie algebra of $GL(N,\Bbb C)$. Moreover, as
$A_{\dot a}$ is homogeneous in $\lambda$ of degree one and is
non-singular, it takes the form $A_{\dot a}=\lambda^a A_{a\dot
a}(x)$, where $A_{a\dot a}$ is a function only of $x$. Hence
\eqn\norko{D_{\dot a}=\lambda^a\left({\partial\over\partial
x^{a\dot a}}+A_{a\dot a}(x)\right).} Since the $\partial_{\dot a}$
commute, and their covariant versions $D_{\dot a}$ are conjugate
to $\partial_{\dot a}$ (via either $W_1$ or $W_2$), it follows
that the $D_{\dot a}$ also commute: \eqn\orko{[D_{\dot a},D_{\dot
b}]=0.}  When this is expanded out using \norko, we discover that
$\lambda^a\lambda^bF_{ab\dot a\dot b}=0$, where $F_{ab\dot a\dot
b}=[D_{a\dot a},D_{b\dot b}]$.  This implies, as promised, that
the gauge field $A_{a\dot a}(x)$ obeys the anti-selfdual
Yang-Mills equations \gufo.  We have thus completed the converse
step of obtaining an object of type (1) from an object of type
(2).

I leave it to the reader to show that if $U$ is replaced by an
equivalent transition function $\tilde U=U_1UU_2^{-1}$ in twistor
space, then $A$ is replaced by a gauge-equivalent connection in
Minkowski spacetime, and further to show that the two operations
that we have defined are indeed inverse to one another.

\bigskip\noindent{\it More Abstract Version}

A more conceptual version of the above proof -- not strictly
needed for the present paper -- goes as follows. We start with an
anti-selfdual connection $A$ on a $GL(N,\Bbb C)$ bundle $H$ over
spacetime. Given an $\alpha$-plane $\Bbb K$, we let $E_{\Bbb K}$
be the space of covariantly constant sections of $H$ restricted to
$\Bbb K$.  The $E_{\Bbb K}$ vary holomorphically with $\Bbb K$,
and fit together, as $\Bbb K$ is varied, to the fibers of a
holomorphic vector bundle $E$ over $\Bbb{PT}'$, which
parameterizes the space of $\Bbb K$'s.

To prove that the bundle $E$ is holomorphically trivial when
restricted to any $\Bbb D_x$, we note that if $T$ passes through
$x$, then $E_T$ can be canonically identified with $H_x$, the
fiber of $H$ at $x$.  Indeed, a covariantly constant section of
$H$ over $T$ is uniquely determined by its value at any point
$x\in T$; that value can be any element of $H_x$.   So the restriction
of $E$ to $\Bbb D_x$ is canonically the product of $\Bbb D_x$ with
the constant vector space $H_x$. This completes the more abstract
explanation of how to construct an object of type (2) from an
object of type (1).

Conversely, suppose we are given a rank $N$ holomorphic bundle $E$
over $\Bbb{PT}'$ that is holomorphically trivial on each $\Bbb
D_x$. Since it is trivial on $\Bbb D_x$, it has, when restricted
to $\Bbb D_x$, an $N$-dimensional space of holomorphic sections
which we call $H_x$.  As $x$ varies, the $H_x$ fit together to a
holomorphic vector bundle $H$ over $\Bbb M'$ (which is
holomorphically trivial as $\Bbb M'\cong \Bbb C^4$).

We wish to define a connection on $H$.  Suppose $x$ and $x'$ are
two points in $\Bbb M'$ that are at lightlike separation.  Then
they are contained in a unique $\alpha$-plane $\Bbb K$.  Once $x$
and $\Bbb K$ are given, $H_x$ is canonically isomorphic to
$E_{\Bbb K}$, the fiber of $E$ at $\Bbb K$.  This is so because
$\Bbb D_x$ can be regarded as the space of all $\alpha$-planes
that pass through $x$; $\Bbb K$ is one of those.  An element of
$H_x$ is a holomorphic section of the trivial bundle obtained by
restricting $E$ to $\Bbb D_x$; it can be identified with its value
at $\Bbb K$.  Likewise, when  $x'$ and $\Bbb K$ are given,
$H_{x'}$ is canonically isomorphic to $E_{\Bbb K}$. Combining
these isomorphisms of $H_x$ and $H_{x'}$ with $E_{\Bbb K}$, we get
a natural map from $H_x$ to $H_{x'}$ that we interpret as parallel
transport from $H_x$ to $H_{x'}$ along the light ray that connects
$x$ and $x'$.

Knowing parallel transport along light rays is enough to uniquely
determine a connection $A$.  To show that $A$ obeys the
anti-selfdual Yang-Mills equations, it suffices to show flatness
on $\alpha$-planes, which follows from the following:  if $x,x'$,
and $x''$ are contained in a common $\alpha$-plane $\Bbb K$, then
parallel transport around a triangle of light rays  from $x$ to
$x'$ to $x''$ and back to $x$ gives the identity.  This can be
readily proved using the above definitions.

Apart from concision and manifest gauge invariance, the advantage
of this abstract proof is that it generalizes to regions of
complexified, conformally compactified Minkowski space $\Bbb M$
other than $\Bbb M'$. Instead of starting with a solution of the
anti-selfdual Yang-Mills equations on $\Bbb M'$, we could start
with a solution defined on any open set $\Bbb U$ in complexified,
conformally compactified Minkowski space $\Bbb M$. (Actually, we
want a mild restriction on $\Bbb U$: its intersections with
$\alpha$-planes should be connnected and simply-connected.)  Let
$\Bbb G$ be the region of $\Bbb{PT}$ that parameterizes
$\alpha$-planes that have a non-empty intersection with $\Bbb U$.
Then the conceptual proof of the twistor transform extends
immediately to a correspondence between anti-selfdual Yang-Mills
fields on $\Bbb U$ and holomorphic vector bundles on $\Bbb G$ that
are trivial on $\Bbb D_x$ for all $x\in \Bbb U$. (Likewise, our
discussion later of the linear wave equations of helicity $h$
extends to a correspondence between solutions of the wave
equations on $\Bbb U$ and sheaf cohomology on $\Bbb G$.)

Here is a standard  application of this  generalization.
Yang-Mills instantons on $\Bbb S^4$ are automatically real
analytic (since the equation is elliptic) and so extend to a small
complex neighborhood $\Bbb U$ of $\Bbb S^4$ in $\Bbb M$. Using the
fact that every $\alpha$-plane in $\Bbb M$ has a non-empty
intersection with $\Bbb U$, one can then show that an instanton on
$\Bbb S^4$ corresponds to a holomorphic vector bundle defined on
all of $\Bbb{PT}$, and trivial on the generic $\Bbb D_x$ (and on
all of the ``real'' ones that correspond to points $x\in \Bbb
S^4$), and obeying a certain reality condition assuming that the
original instanton is real. For a systematic exposition, see
\atiyah.

\bigskip\noindent{\it Free Wave Equations}

In most of this paper, more than the nonlinear anti-selfdual
Yang-Mills equations, we really need  the twistor transform of the
linear wave equations for helicity $h$, for various values of $h$.
According to the Penrose transform, solutions of this wave
equation in $\Bbb M'$ are equivalent to elements of the sheaf
cohomology group $H^1(\Bbb{PT}',\CO(2h-2))$.  Here, following
Penrose \penroseb, we explain this correspondence in the context
of Cech cohomology, using the open cover $V_1,V_2$ of $\Bbb{PT}'$
and a holomorphic cocycle.  Then we convert the statement to
$\bar\partial$ cohomology, used in the rest of the paper.

Concretely, an element of $H^1(\Bbb{PT}',\CO(2h-2))$
is given by a ``cocycle,'' a
holomorphic function $f(\lambda^a,\mu^{\dot a})$ on $V_{12}$ that
is homogeneous of degree $2h-2$, and so is a section of
$\CO(2h-2)$. It may have singularities at $\lambda^1=0$ or at
$\lambda^2=0$. It is subject to the equivalence relation $f\to
f+f_1-f_2$, where $f_1$ is holomorphic on $V_1$ (and so may only
be singular at $\lambda^1=0$), and $f_2$ is holomorphic on $V_2$
(and so may only be singular at $\lambda^2=0$).

The first step is to define a function of $x$ and $\lambda$ by
setting $g(\lambda^a,x_{b\dot b})=f(\lambda^a,-x_{b\dot
a}\lambda^b)$. Rather as in the above discussion of the Yang-Mills
case,  a simple use of the chain rule and definition of
$\partial_{\dot a}$ gives \eqn\yugu{\partial_{\dot
a}g(\lambda,x)=0.}  For fixed $x$, $g$ can be regarded as a
cocycle defining an element of $H^1(\Bbb D_x,\CO(2h-2))$.
 We first consider the case that $h\geq 1/2$.
For this case, $H^1(\Bbb D_x,\CO(2h-2))=0$,
 so  \eqn\gongon{g(\lambda,x)=g_1(\lambda,x)-g_2(\lambda,x),}
where for $i=1,2$, $g_i$ is nonsingular except perhaps at
$\lambda^i=0$. From \yugu, we have \eqn\olop{\partial_{\dot
a}g_1=\partial_{\dot a}g_2.} The left hand side may be singular
only at $\lambda^1=0$, and the right hand side only at
$\lambda^2=0$; so in fact, there are no singularities at all.  We
write $\phi_{\dot a}$ for the left or right hand side.

If $h=1/2$, $\phi_{\dot a}(\lambda,x)$ is homogeneous in $\lambda$
of degree zero, and so, being nonsingular, is a function only of
$x$.  We claim that it obeys the Dirac equation:
\eqn\incin{{\partial\over\partial x^{a\dot a}}\phi^{\dot a}=0.} In
fact, as $\partial_{\dot a}\partial^{\dot a}=0$ and $\phi_{\dot
a}=\partial_{\dot a}g_1$, we have $0=\partial_{\dot a}\phi^{\dot
a}$, so, from the definition of $\partial_{\dot a}$,
$0=\lambda^a\partial_{a\dot a}\phi^{\dot a}$. As $\phi_{\dot a}$
is independent of $\lambda$, this does imply the Dirac equation.
This is the $h=1/2$ case of the Penrose correspondence.

If $h=1$, then $\phi_{\dot a}$, being homogeneous in $\lambda$ of
degree one, is of the form $\phi_{\dot a}=\lambda^aC_{a\dot
a}(x)$, where $C_{a\dot a}$ depends on only $x$ and not $\lambda$.
A linearized version of the same analysis that we gave in
discussing the anti-selfdual Yang-Mills equations shows that
$C_{a\dot a}$ obeys the linear anti-selfdual equation. For $h>1$,
which we do not need in the present paper, we refer to the
literature.

For $h\leq 0$, one instead uses a contour integral method due to
Penrose. If $h=0$, then $g$ is homogeneous in $\lambda$ of degree
$-2$. On $\Bbb D_x$, there is the holomorphic differential
$\epsilon_{ab}\lambda^ad\lambda^b$ that is homogeneous in $\lambda$
of degree 2.  Letting $C$ be any contour that surrounds one of
the two singularities in $g$, for instance the one at $\lambda^1
=0$, we define a function $\phi(x)$ by the integral
\eqn\cunon{\phi(x)={1\over 2\pi i}\oint_C
\left(\epsilon_{ab}\lambda^ad\lambda^b \right)g(\lambda,x).}  The
integral makes sense since the form being integrated is homogenous
in $\lambda$  of degree zero. The integral only depends on the
cohomology class represented by $f$; it is invariant under $g\to
g+g_1-g_2$ (where $g_1$ and $g_2$ each have only one singularity),
since if $g$ is replaced by $g_1$ or $g_2$,
 the contour can be deformed away,
shrinking it to $\lambda^1=0$ or $\lambda^2=0$. $\phi$ depends on
$x$ only as we have integrated over $\lambda$.  Finally, a simple
use of the chain rule shows that $\phi$ obeys the scalar wave
equation $\partial_{a\dot a}\partial^{a\dot a}\phi=0$.

For $h<0$, say $h=-k$, we define a field of helicity $h$ via
\eqn\bunon{\phi_{a_1a_2\dots a_{2k}}={1\over 2\pi
i}\oint_C\left(\epsilon_{cd}
\lambda^cd\lambda^d\right)\lambda_{a_1}\dots\lambda_{a_{2k}}g.}
This obeys the free wave equation $\partial_{a\dot
a}\phi^a{}_{b_1\dots b_{2k-1}}=0$, as one can again verify by a
simple application of the chain rule.

\bigskip\noindent{\it Relation To $\bar\partial$ Cohomology}

So far we have identified the space of solutions of the massless
linear wave equation for helicity $h$ with the sheaf cohomology
group $H^1(\Bbb{PT}',\CO(2h-2))$ defined in Cech cohomology. In
section 4, instead of Cech cohomology, we encountered the
$\bar\partial$ cohomology group
$H^1_{\bar\partial}(\Bbb{PT}',\CO(2h-2))$.  By general arguments
in complex geometry, $H^1$ and $H^1_{\bar\partial}$ are naturally
isomorphic.

In the present example, because $\Bbb{PT}'$ can be covered by two
such simple open sets, we can be completely explicit about this
isomorphism.  Before doing so, we will make a minor change of
notation.  This will enable us to obtain formulas that are more
convenient in section 4.  We make an $SL(2)$ transformation of the
$\lambda^a$ to move the singularities from
$\lambda_2/\lambda_1=\infty,0$ to $\lambda_2/\lambda_1=i,-i$.
Henceforth, $V_1$ is the portion of $\Bbb{PT}'$  with
$\lambda_2/\lambda_1\not= i$, $V_2$ the portion with
$\lambda_2/\lambda_1\not= -i$, and $V_{12}$ remains the
intersection.

Let $z=\lambda^2/\lambda^1$, and also let $z=\sigma+i\tau$, where
$\sigma$ and $\tau$ are real.  Let $\theta(\tau)$ be the function
that is 1 if $\tau>0$ and $0$ for $\tau<0$.  We have
\eqn\barp{\bar\partial\theta(\tau)=d\bar
z{\partial\over\partial\bar z}\theta(\tau)={i\over 2}d\bar
z\delta(\tau),} since $\tau=(z-\bar z)/2i$, $\partial_{\bar z}\bar
z=1$, $\partial_{\bar z}z=0$, and
$\partial_\tau\theta(\tau)=\delta(\tau)$. Consider an element
$\omega\in H^1(\Bbb{PT}',\CO(k))$, for some $k$, that is
represented by a cocycle $f$.  Thus, $f$ is a section of $\CO(k)$
that is holomorphic throughout $V_{12}$, and subject to the
equivalence \eqn\incon{f\to f+f_1-f_2,} where $f_i$, $i=1,2$, is
holomorphic throughout $V_i$.  The $\bar\partial$ cohomology class
corresponding to $\omega$ can be represented by the $(0,1)$-form
\eqn\snon{v=f\bar\partial\theta(\tau)={i\over 2}f\delta(\tau)d\bar
z.} The form $v$ is defined globally throughout $\Bbb{PT}'$, since
the singularities of $f$ are disjoint from the delta function. On
$V_{12}$, it can be written $\bar\partial(f\theta(\tau))$, and so
is trivial, but this representation is not valid everywhere
because of the singularity of $f$ at $z=i$, which is in the
support of $\theta(\tau)$.  The $\bar\partial$ cohomology class of
$v$ is invariant under the transformation \incon, since we have
$(f_1-f_2)\bar\partial
\theta(\tau)=\bar\partial\left(f_1(\theta(\tau)-1)-f_2\theta(\tau)\right)$,
a formula which is valid everywhere, as the $f_i$ multiply
functions that vanish near their singularities. Thus, we have
defined a mapping from $H^1$ to $H^1_{\bar\partial}$. This mapping
actually is an isomorphism.

\bigskip

I am indebted to N. Berkovits for numerous helpful discussions of
some of these ideas and pointing out a number of significant
references, to F. Cachazo for extensive assistance with computer
algebra, to L. Dixon for answering many queries about
perturbative Yang-Mills theory, and to M. F. Atiyah and R. Penrose for
mathematical consultations. This work was supported in part by NSF Grant
PHY-0070928.

\listrefs
\end